\documentclass[longauth]{aa}
\usepackage{euclid}
\usepackage[colorlinks=true,linkcolor=blue,citecolor=blue]{hyperref}
\usepackage[dvipsnames]{xcolor}
\usepackage{graphicx}
\graphicspath{{graphics}}
\usepackage{txfonts}
\usepackage[separate-uncertainty=true, list-units=repeat]{siunitx}
\usepackage{subcaption}
\usepackage{hyperref}
\usepackage[inline]{enumitem}
\usepackage{physics} 	
\usepackage{multirow}
\usepackage{bm}
\usepackage[flushleft]{threeparttable}
\usepackage{pdflscape}
\usepackage{graphicx}

\newcommand{\LePHARE}{\textsc{LePHARE}}
\newcommand{\Cigale}{\textsc{Cigale}}

\newcommand{\hst}{\textit{HST}}
\newcommand{\JWST}{\textit{JWST}}

\newcommand{\UVISTA}{UltraVISTA}

\newcommand{\NIRCAM}{NIRCam}
\newcommand{\MIRI}{MIRI}

\DeclareSIUnit{\Msun}{M_\odot}
\DeclareSIUnit{\year}{yr}
\DeclareSIUnit{\pc}{pc}
\DeclareSIUnit{\mag}{mag}
\DeclareSIUnit{\mas}{mas}
\DeclareSIUnit{\dex}{dex}
\DeclareSIUnit{\jansky}{Jy}
\DeclareSIUnit{\kpc}{kpc}

\newcommand{\zphot}{z_{\rm phot}}
\def\photoz{\mbox{photo-\textit{z}}}
\def\photoz{\mbox{photo-\textit{z}}}

\begin{document} 

   \title{COSMOS-Web: stellar mass assembly in relation to dark matter halos across $0.2<z<12$ of cosmic history}
   \subtitle{}  

   \author{M.~Shuntov\inst{\ref{DAWN},\ref{NBI}}\fnmsep\thanks{\email{marko.shuntov@nbi.ku.dk}} \and
O.~Ilbert \inst{\ref{LAM}} \and%
S.~Toft\inst{\ref{DAWN}, \ref{NBI}} \and%
R.~C.~Arango-Toro\inst{\ref{LAM}} \and
H.~B.~Akins\inst{\ref{UAT}} \and
C.~M.~Casey\inst{\ref{UAT}, \ref{DAWN}} \and
M.~Franco\inst{\ref{UAT}} \and
S.~Harish\inst{\ref{Rochester}} \and
J.~S.~Kartaltepe\inst{\ref{Rochester}} \and
A.~M.~Koekemoer\inst{\ref{STScI}} \and
H.~J.~McCracken\inst{\ref{IAP}} \and
L.~Paquereau\inst{\ref{IAP}} \and
C.~Laigle\inst{\ref{IAP}} \and
M.~Bethermin\inst{\ref{OAS}},
Y.~Dubois\inst{\ref{IAP}} \and
N.~E.~Drakos\inst{\ref{HawaiiHilo}} \and
A.~Faisst\inst{\ref{Caltech}} \and
G.~Gozaliasl\inst{\ref{Finland}} \and 
S.~Gillman\inst{\ref{DAWN}, \ref{DTU}} \and
C.~C.~Hayward\inst{\ref{NYC}} \and
M.~Hirschmann\inst{\ref{Lausanne},\ref{INAF}}
M.~Huertas-Company\inst{\ref{IAC}, \ref{LERMA}, \ref{Paris-Cite}, \ref{laLaguna}} \and
C.~K.~Jespersen\inst{\ref{Princeton}} \and
S.~Jin\inst{\ref{DAWN}, \ref{DTU}} \and
V.~Kokorev\inst{\ref{UAT}} \and
E.~Lambrides\inst{\ref{NASA-Goddard}}
D.~Le Borgne\inst{\ref{IAP}} \and
D.~Liu\inst{\ref{Nanjing}} \and
G.~Magdis\inst{\ref{DAWN},\ref{DTU}} \and
R.~Massey\inst{\ref{Durham}} \and
C.~J.~R.~McPartland\inst{\ref{DAWN}, \ref{NBI}} \and
W.~Mercier\inst{\ref{LAM}} \and
J.~E.~McCleary\inst{\ref{NoE}} \and
J.~McKinney\inst{\ref{UAT}}
P.~A.~Oesch\inst{\ref{Geneva},\ref{DAWN}} \and
J.~D.~Rhodes\inst{\ref{NASA}} \and
R.~M.~Rich\inst{\ref{LA}} \and
B.~E.~Robertson\inst{\ref{UnivCalifornia}} \and
D.~Sanders\inst{\ref{HawaiiManoa}}
M.~Trebitsch\inst{\ref{Sorbonne}} \and
L.~Tresse\inst{\ref{LAM}} \and
F.~Valentino\inst{\ref{ESO},\ref{DAWN}} \and
A.~P.~Vijayan\inst{\ref{DAWN}, \ref{DTU}} \and
J.~R.~Weaver\inst{\ref{UMASS}} \and
A.~Weibel\inst{\ref{Geneva}} \and
S.~M.~Wilkins\inst{\ref{Brighton}, \ref{Malta}}
    }

   \institute{ 
   Cosmic Dawn Center (DAWN), Denmark \label{DAWN} 
   \and
   Niels Bohr Institute, University of Copenhagen, Jagtvej 128, 2200 Copenhagen, Denmark \label{NBI}%
    \and
   Aix Marseille Univ, CNRS, CNES, LAM, Marseille, France \label{LAM} 
   \and
   The University of Texas at Austin, 2515 Speedway Blvd Stop C1400, Austin, TX 78712, USA\label{UAT} 
   \and
   Laboratory for Multiwavelength Astrophysics, School of Physics and Astronomy, Rochester Institute of Technology, 84 Lomb Memorial Drive, Rochester, NY 14623, USA \label{Rochester} 
   \and
   Institut d’Astrophysique de Paris, UMR 7095, CNRS, Sorbonne Université, 98 bis boulevard Arago, F-75014 Paris, France \label{IAP} 
   \and
   Space Telescope Science Institute, 3700 San Martin Drive, Baltimore, MD 21218, USA \label{STScI} 
   \and
   Université de Strasbourg, CNRS, Observatoire astronomique de Strasbourg, UMR 7550, 67000 Strasbourg, France\label{OAS}%
    \and%
   Department of Physics and Astronomy, University of Hawaii, Hilo, 200 W Kawili St, Hilo, HI 96720, USA \label{HawaiiHilo} 
   \and
   Caltech/IPAC, 1200 E. California Blvd. Pasadena, CA 91125, USA\label{Caltech}
   \and
   Department of Computer Science, Aalto University, P.O. Box 15400, FI-00076 Espoo, Finland \label{Finland}
   \and
   DTU Space, Technical University of Denmark, Elektrovej, Building 328, 2800, Kgs. Lyngby, Denmark \label{DTU} 
   \and
   Center for Computational Astrophysics, Flatiron Institute, 162 Fifth Avenue, New York, NY 10010, USA \label{NYC}
   \and
   Institute of Physics, GalSpec, Ecole Polytechnique Federale de Lausanne, Observatoire de Sauverny, Chemin Pegasi 51, 1290 Versoix, Switzerland \label{Lausanne}
   \and
   NAF, Astronomical Observatory of Trieste, Via Tiepolo 11, 34131 Trieste, Italy \label{INAF}
   \and
   Instituto de Astrofísica de Canarias (IAC), La Laguna, E-38205, Spain \label{IAC} 
   \and
   Observatoire de Paris, LERMA, PSL University, 61 avenue de l’Observatoire, F-75014 Paris, France \label{LERMA} 
   \and
   Université Paris-Cité, 5 Rue Thomas Mann, 75014 Paris, France \label{Paris-Cite} 
   \and
   Universidad de La Laguna, Avda. Astrofísico Fco. Sanchez, La Laguna, Tenerife, Spain \label{laLaguna}%
   \and
   Department of Astrophysical Sciences, Princeton University, Princeton, NJ 08544, USA \label{Princeton} 
   \and
   NASA-Goddard Space Flight Center, Code 662, Greenbelt, MD, 20771, USA \label{NASA-Goddard}
   \and
   Purple Mountain Observatory, Chinese Academy of Sciences, 10 Yuanhua Road, Nanjing 210023, China \label{Nanjing} 
   \and
   Centre for Extragalactic Astronomy, Durham University, South Road, Durham DH1 3LE, UK\label{Durham}%
   \and
   Department of Physics, Northeastern University, 360 Huntington Avenue, Boston, MA 02115, USA \label{NoE}
   \and
   Department of Astronomy, University of Geneva, Chemin Pegasi 51, 1290 Versoix, Switzerland \label{Geneva} 
   \and
   Jet Propulsion Laboratory, California Institute of Technology, 4800 Oak Grove Drive, Pasadena, CA 91109 \label{NASA} 
   \and
   Department of Physics and Astronomy, UCLA, PAB 430 Portola Plaza, Box 951547, Los Angeles, CA 90095-1547 \label{LA}
   \and
   Department of Astronomy and Astrophysics, University of California, Santa Cruz, 1156 High Street, Santa Cruz, CA 95064 USA \label{UnivCalifornia} 
   \and
   Institute for Astronomy, University of Hawai’i at Manoa, 2680 Woodlawn Drive, Honolulu, HI 96822, USA \label{HawaiiManoa}
   \and
   Sorbonne Université, Observatoire de Paris, PSL research university, CNRS, LERMA, 75014 Paris, France \label{Sorbonne} 
   \and
   European Southern Observatory, Karl-Schwarzschild-Str. 2, 85748 Garching, Germany \label{ESO}
   \and
   Department of Astronomy, University of Massachusetts, Amherst, MA 01003, USA \label{UMASS} 
   \and
   Astronomy Centre, University of Sussex, Falmer, Brighton BN1 9QH, UK \label{Brighton} 
   \and
   Institute of Space Sciences and Astronomy, University of Malta, Msida MSD 2080, Malta \label{Malta} 
   }

   \date{Received ; accepted }

 
  \abstract
  {
We study the stellar mass assembly of galaxies via the stellar mass function (SMF) and the co-evolution with dark matter halos via abundance matching in the largest redshift range to date $0.2<z<12$. We use the $0.53 \, {\rm deg}^2$ imaged by \JWST{} from the COSMOS-Web survey, in combination with ancillary imaging in over 30 photometric bands to select highly complete samples (down to log$\, M_{\star}/M_{\odot} = 7.5-8.8$) in 15 redshift bins.
\
Our results show that the normalization of the SMF monotonically decreases from $z=0.2$ to $z=12$ with strong mass-dependent evolution.
At $z>5$, we find increased abundances of massive (log$\, M_{\star}/M_{\odot}>10.5$) systems compared to predictions from semi-analytical models and hydrodynamical simulations. These findings challenge traditional galaxy formation models by implying integrated star formation efficiencies (SFE) $\epsilon_{\star}\equiv M_{\star}\, f_{\rm b}^{-1} M_{\rm halo}^{-1} \gtrsim 0.5$.
We find a flattening of the SMF at the high-mass end that is better described by a double power law at $z>5.5$, after correcting for the Eddington bias. At $z \lesssim 5.5$ it transitions to a Schechter law which coincides with the emergence of the first massive quiescent galaxies in the Universe, indicating that physical mechanisms that suppress galaxy growth start to take place at $z\sim5.5$ on a global scale.
\
By integrating the SMF, we trace the cosmic stellar mass density (SMD) and infer the star formation rate density (SFRD), which at $z>7.5$ agrees remarkably with recent \JWST{} UV luminosity function-derived estimates. This agreement solidifies the emerging picture of rapid galaxy formation leading to increased abundances of bright and massive galaxies in the first $\sim 0.7$ Gyrs. 
However, at $z \lesssim 3.5$, we find significant tension ($\sim 0.3$ dex) with the cosmic star formation (SF) history from instantaneous SF measures, the causes of which remain poorly understood. 
\
We infer the stellar-to-halo mass relation (SHMR) and the SFE from abundance matching out to $z=12$, finding a non-monotonic evolution. The SFE has the characteristic strong dependence with mass in the range of $0.02 - 0.2$, and mildly decreases at the low mass end out to $z\sim3.5$. At $z\sim3.5$ there is an upturn and the SFE increases sharply from $\sim 0.1$ to approach high SFE of $0.8-1$ by $z\sim 10$ for log$(M_{\rm h}/M_{\odot})\approx11.5$, albeit with large uncertainties. Finally, we use the SHMR to track the SFE and stellar mass growth throughout the halo history and find that they do not grow at the same rate -- from the earliest times up until $z\sim3.5$ the halo growth rate outpaces galaxy assembly, but at $z>3.5$ halo growth stagnates and accumulated gas reservoirs keep the SF going and galaxies outpace halos.

  }
   


   \maketitle


%

\section{Introduction}

The galaxy stellar mass function (SMF) is one of the most fundamental statistical measurements, quantifying the cosmic stellar mass assembly. It is shaped by all the physical processes that contribute to the growth of galaxies, and as such it reflects the cumulative history of star formation, mergers, and feedback processes. 
This is tightly linked to the dark matter halos in which galaxies form and evolve \citep{WhiteRees1978}. They provide the potential well in which gas can accrete and remain bounded, thus driving star-formation. By studying the SMF, we can gain insights into the connection between galaxies and dark matter halos, which is essential for constraining theoretical models of galaxy formation and evolution. 

Another importance in studying the SMF is that it quantifies the stellar mass density (SMD) assembled in the Universe at a given epoch. This, in turn, is a result of the integrated instantaneous cosmic star formation history (SFH) after correcting for stellar evolution processes \citep{Madau1998}. Comparisons between the SMD measured from SMF and the one obtained from integrating the cosmic SFH have shown a persistent discrepancy \citep[e.g.,][]{Hopkins2006, Wilkins2008, Leja2015, Wilkins2019}, with the integrated instantaneous SFH overpredicting the SMD. The reason for this tension remains poorly understood, and reconciling the two requires a detailed understanding of the various assumptions and systematics that enter the measurement of both quantities.

Historically, the SMF has been studied extensively in numerous works in the literature, from the very local universe \citep[e.g.,][]{Cole2001,Bell2003}, across vast cosmic time intervals \citep[e.g.,][]{Fontana06,Pozetti2007,Muzzin2013, Ilbert13,Davidzon2017,Wright2018,McLeod2021,Weaver2023}, to the earliest times that stellar masses could be robustly estimated \citep[e.g.,][]{Grazian2015, Song2016, Bhatawdekar19, Stefanon21}. These works rely on rest-frame optical photometric data either from ground-based imaging, limited by resolution and sensitivity in the near-infrared (NIR), from the \textit{Hubble Space Telescope} (\hst{}) limited in wavelength coverage to $<1.6\, \si{\micro \meter}$, or from \Spitzer{} that suffers from low resolution, source blending and confusion in the IR. 
Furthermore, studies of the SMF in the early Universe ($z>6$) have been carried out over very deep but small, pencil-beam surveys. These can be subject to significant cosmic variance relating to sampling over- and under-dense regions of large scale structure \cite[e.g.,][]{Steinhardt2021_cosmicvariance, Jespersen2024_mostmassive}, and are unable to probe significant samples of some of the rarest and most massive galaxies at different epochs.



The launch of \JWST{} has revolutionized our view of the early Universe with the unprecedented resolution and sensitivity at $1-5$ \SI{}{\micro \meter} from the Near Infrared Camera (NIRCam, Rieke et al. 2005).
\JWST{} observations confirmed that our $z>4$ samples based on optical selection have been considerably incomplete of typically red, dust-obscured and massive (log$M_{\star}/M_{\odot} > 10.5$) systems \citep{Barrufet2023, Gottumukkala2023, Weibel2024}.
Furthermore, early studies revealed surprisingly high abundances of massive galaxies in the early Universe \citep{Labbe2023Natur, Xiao2023, Casey2024}. These have challenged contemporary galaxy formation theories \citep{Boylan-Kolchin2023} by approaching the number density limit set by the dark matter halos and the universal baryonic reservoir.

Using \JWST{} observations, several works have already investigated the SMF at $z>4$ \citep{Navarro-Carrera2023, Weibel2024, Harvey2024, TWang2024} and agree on the enhanced abundances of massive (log$\, M_{\star}/M_{\odot} > 10$) systems at $z>6$, compared to theory predictions. They have also revealed a significant contribution in the number density from sources hosting active galactic nuclei (AGN), a subpopulation of which appear extremely red and compact \citep[so-called little red dots, LRD,][]{Labbe2023, Greene2024, Matthee2024, Kokorev2024, Akins2024}. These can boost the rest-frame optical flux and bias high the resulting stellar mass if the AGN component is not taken into account; therefore, they need to be carefully handled to avoid biases in the analysis. However, these works on the SMF are still limited to small areas and unable to probe the highest masses with high statistical significance.

The observed overabundances of massive galaxies can have important implications for baryon-to-star conversion efficiency in the early Universe and thus on our galaxy formation models. Since this is related to the host dark matter halo, a systematic study of the relation between stellar and halo mass is necessary and still lacking in order to investigate the star-formation efficiency in a large redshift regime and into the very early Universe. 
Several theories have been proposed to explain the overabundance of massive and galaxies. For example, Feedback-Free Bursts (FFB) \citep{Torrey2017, Grudic2018, Dekel2023, Li2023}, stochastic star formation \citep{PallottiniFerrara2023, Sun2023} and positive feedback from AGN \citep{Silk2024} in the early Universe, can cause increased star formation efficiencies that can rapidly assemble high stellar masses.
On the other hand, systematic biases and effects such as non-universal initial mass function \citep[e.g.,][]{Steinhardt2022} or outshining \citep[e.g.,][]{Gimenez-Arteaga2023, Narayanan2024} can lead to erroneously high or low inferred stellar masses respectively. Furthermore, even modifications to the standard $\Lambda$ cold dark matter ($\Lambda$CDM) cosmology have been suggested \citep{Boylan-Kolchin2023, Liu2024}. 



In this work, we use the largest \JWST{} survey COSMOS-Web to consistently measure the SMF for the first time in 13.4 Gyr, or $97\%$ of cosmic history. The NIRCam imaging at $1-4\, \mu$m allows us to compile highly complete rest-frame optical selected samples from $z=0.2$ to $z\sim10$ (and rest-frame UV selected out to $z=12$), and combined with 30 other photometric bands available in COSMOS spanning from the UV to the NIR, resulting in highly accurate photometric redshifts and stellar masses \citep{Weaver2022cosmos2020}. The large survey area ($0.53\,{\rm deg}^2$) allows us to study the assembly of some of the most massive and rarest systems. The large volume probes a range of environments and thus allows us to improve both the sampling statistics and the cosmic variance biases compared to previous work, especially at $z\gtrsim6$. Using abundance matching, we also make the connection to dark matter halos and study the evolution of the integrated star formation efficiency out to $z\sim12$.

This paper is organized as follows. Section~\ref{sec:data} presents the dataset we use to carry out our analysis, while Sect.~\ref{sec:sample-selection} details the sample selection for our SMF measurements. In Sect.~\ref{sec:measurements} we describe the methodology to measure the SMF, its associated sources of uncertainty, and the functional forms that we adopt to describe the SMF. The results are presented in Sect.~\ref{sec:results} and compared with the literature. We discuss our results in Sect.~\ref{sec:discussion} and put them into a broader perspective with respect to the cosmic stellar mass assembly and its connection to dark matter halos. Section~\ref{sec:conclusions} summarizes and concludes this work.

We adopt a standard $\Lambda$CDM cosmology with $H_0=70$\,km\,s$^{-1}$\,Mpc$^{-1}$, $\Omega_{\rm m,0}=0.3$, where $\Omega_{\rm b,0}=0.04$, $\Omega_{\Lambda,0}=0.7$, and $\sigma_8 = 0.82$. All magnitudes are expressed in the AB system \citep{1983ApJ...266..713O}. Stellar masses are obtained assuming  a \cite{Chabrier03} initial mass function (IMF) and when comparing to the literature, stellar masses are rescaled to match the IMF adopted in this paper.

\section{Data}\label{sec:data}
 
\subsection{Space- and ground-based observations of COSMOS}\label{sec:bands}
This work relies on space- and ground-based multi-band imaging data in the COSMOS field \citep{Scoville2007}. The cornerstone of our dataset is the multi-band imaging from the \JWST\ Cycle 1 program COSMOS-Web \citep[GO\#1727, PI: Casey \& Kartaltepe]{Casey2023}, which we use to carry out the principal scientific analysis in this paper. 
Additionally, we also rely on imaging in the COSMOS field from the deeper but smaller-area program PRIMER \citep[GO\#1837]{PrimerDunlop2021}, mainly for validation purposes and completeness estimation.

COSMOS-Web is a photometric survey that consists of imaging in four \NIRCAM{} (F115W, F150W, F277W, F444W) and one \MIRI{} (F770W) filters. The \NIRCAM{} (\MIRI{}) filters reach a 5$\sigma$ depth of AB mag $27.2 - 28.2$ ($25.7$), measured in empty apertures of $0.15\arcsec$ ($0.3\arcsec$) radius \citep{Casey2023}. Data reduction was carried out in the following way. For the January 2023 observational epoch we used  version 1.8.3 of the \JWST{} Calibration Pipeline \citep{Bushouse2022}, Calibration Reference Data System (CRDS) pmap-1017 and a \NIRCAM{} instrument mapping imap-0233. Subsequent data obtained in April 2023 were processed with an updated version of the JWST pipeline, version 1.10.0, alongside CRDS version 1075 (imap 0252). Finally, for data obtained in January 2024, we used the JWST pipeline 1.12.1 alongside CRDS version 1170 (imap 0273) for the version v0.6 and JWST pipeline 1.14.0 alongside CRDS version 1223 (imap 0285).
Mosaics are created at \SI{30}{\mas} for the short wavelength and \SI{60}{\mas} for the long wavelength and MIRI filters. The \NIRCAM{} and \MIRI{} image processing and mosaic making will be described in detail in Franco et al. (in prep.) and Santosh et al. (in prep.). This work uses the complete survey, imaged over three main epochs (January 2023, April 2023, January 2024), with missing visits ($\sim$5\%) completed in the April 2024 epoch.


We leverage the wealth of legacy data in COSMOS by including the multi-band imaging from ground- and space-based observatories, described in detail in Shuntov et al. (in prep.; see also \citealt{Weaver2022} and \citealt{dunlop_ultravista_2016}). These data tightly sample the SED of galaxies from ultraviolet to mid-infrared wavelengths in over 30 photometric bands. The $u$ band imaging comes from the CFHT Large Area $U$-band Deep Survey (CLAUDS; \citealt{Sawicki19}) at a 5$\sigma$ depth of 27.7 mag as measured from empty apertures of $\ang{;;2}$ diameter size. For the optical data, we use The Hyper Suprime-Cam (HSC) Subaru Strategic Program (HSC-SSP; \citealt{Aihara18}) DR3  \citep{Aihara21} in the ultra-deep HSC imaging region. This consists of five broad bands ($g,r,i,z,y$) and three narrow bands, with a sensitivity ranging 26.5-28.1 mag (5$\sigma$, $\ang{;;2}$ aperture size). In addition, we include the reprocessed Subaru Suprime-Cam images with 12 medium bands in optical \citep{Taniguchi07,Taniguchi15}. In the NIR, we also use the ground-based imaging from the \UVISTA{} survey \citep{McCracken12} in four broad bands $Y,J,H,K_s$ and one narrow band (\SI{1.18}{\micro\meter}). We use the latest and final \UVISTA{} DR6, which provides homogeneous coverage over the entire field at the depth of the Ultra Deep stripes. Even though these ground-based data are at a lower resolution and sensitivity than \NIRCAM{}, they are complementary in terms of wavelength coverage. Crucially, the \UVISTA{} bands fill in the wavelength gaps in the NIR left by the relatively sparse (4 band) \JWST{}/NIRCam coverage. Finally, we also include the \hst{}/ACS F814W band \citep{Koekemoer07} over the entire COSMOS area that is also covered by \NIRCAM{}. 

\subsection{Galaxy catalogues of photometry and physical parameters}
To consistently measure photometry in ground– and space–based data that have very different resolutions, we use a model-fitting approach to extract source photometry in all 33 bands. We use \textsc{SourceXtractor++} \citep{kummel20,bertin20} to fit single S\'ersic models convolved by the point-spread function (PSF) in each band for all sources detected in a $\sqrt{\chi^2}$ combination \citep{Szalay1999, Drlica-Wagner2018} of F115W, F150W, F277W and F444W, PSF matched to F444W. The structural parameters of the S\'ersic models are fitted on all NIRCam bands simultaneously, while the flux is fitted for each band independently. The details of the photometry extraction and catalog making in COSMOS-Web are described in detail in Shuntov et al. (in prep). We use the model-derived total photometry throughout the paper, unless stated otherwise.

\textbf{\textsc{LePHARE} photo$-z$ and physical parameters.}
We use the template-fitting code \LePHARE{} \citep{Arnouts02, Ilbert06} to measure photometric redshifts and physical parameters on the 33 photometric bands spanning $0.3-8.0 \, \si{\micro\meter}$. We fit a set of templates extracted from \citet{BC03} models that assume 12 different star formation histories (SFH; exponentially declining and delayed), as described in \citet{Ilbert15}. For each SFH, it generates templates at $43$ different ages going from 0.05 to \SI{13.5}{\giga\year}. Dust attenuation is implemented by varying the $E(B-V)$ in the range of $0 - 1.2$ from three attenuation curves \citep{Calzetti00, Arnouts2013, Salim18} (we use the high-$z$ analog curve from \citealt{Salim18}). We add emission lines by adopting the recipe in \citet{Saito20} and following \citet{Schaerer09}. We fit the normalization of the emission line fluxes by varying them by a factor of two (using the same ratio for all lines). Finally, the intergalactic medium (IGM) absorption is accounted for by using the analytical correction of \citet{Madau95}. \LePHARE{} provides the redshift likelihood distribution for each object, after a marginalization over the galaxy templates and the dust attenuation. We use it as the posterior redshift probability density function (PDF), assuming a flat prior. We adopt the median of the PDF($z$) as our point-estimate for the photometric redshifts. The physical parameters are derived in a second \LePHARE{} run. We use the same configuration as the one used to compute the photometric redshifts, except that we set the redshift to the point-estimate previously established. We also associate s PDF to the mass by summing all the probabilities at a given mass by marginalizing over the templates and dust. These mass PDFs do not include the redshift uncertainties.  

For calibrating \textsc{LePHARE} and assessing the photo-$z$ performance (see Table~\ref{table:photoz-performance}), we use a sample of about 12,000 spectroscopic redshifts with a high ($>97\%)$ confidence level out to $z=8$. These are compiled from most spectroscopic programs in COSMOS \citep[both public and private, e.g.,][]{lilly09,kartaltepe10,kartaltepe15b,silverman15,kashino19,lefevre15,casey12,capak11,kriek15,hasinger18}. The construction of the compilation, the individual surveys, and the properties/distribution of the galaxies will be presented in Khostovan et al., in prep. 

Stellar mas estimate can be significantly affected by SED modelling assumptions, such as dust attenuation law and formation histories \citep[e.g.,][]{Michalowski2012, Mitchell2013, Hayward2015, Haskell2023, Pacifici2023, Haskell2024}. To investigate this in our data, we also ran \Cigale{} \citep{Boquien19} with a non-parametric star formation history modeling and different dust attenuation law. The \Cigale{} SED modelling and comparison with \LePHARE{} is described and discussed in more detail in Appendix~\ref{appdx:cigale-comparison}.

\section{Sample selection}\label{sec:sample-selection}

\subsection{Selection function}

The unprecedented combination of sensitivity and resolution in the near-infrared of NIRCam, coupled with wide and deep observations of COSMOS-Web allow us to compile complete galaxy samples in a large dynamic range of stellar mass. 
The deep imaging ($\sim 28.0$ mag in F444W) in the wavelength range $\sim 1-5 \, \mu$m ensures that galaxies are detected in the rest-frame optical down to $z\sim10$. 
An advantage of the increase in depth in the NIR is that it leads to improved stellar mass completeness down to very low masses (see \S\ref{subsec:completeness-limits}), and the detection of obscured and red populations, especially at the high-mass end that have likely been missed by previous studies (e.g., \citealt{Weaver2023} for hints pre-\JWST{} and e.g., \citealt{Barrufet2023, Gottumukkala2023} for explanations using \JWST{})
However, the NIR selection function at $\sim 1-5 \, \mu$m can lead to some incompleteness of redshift $\lesssim2$ galaxies, especially those with young stellar populations and blue SEDs. We do not quantify this possible incompleteness, since the main interest is the high redshift regime.

The increased depth, coupled with the reduction of source blending thanks to the high resolution from space and the tight sampling of the SED by including ground-based bands, reduces photometric uncertainties and leads to more accurate estimates of physical parameters from SED fitting. 
The quality of the photo-$z$ and physical parameters in our catalog is presented in detail in Shuntov et al. in prep. The photo-$z$ show excellent performance which is summarized in Table~\ref{table:photoz-performance} with the standard metrics ($\sigma_{\rm MAD}$, outlier fraction, bias). This allows us to select complete samples in 15 redshift bins from $z=0.2$ to $z=12$. 

For the stellar mass estimates, we assess their performance with respect to different SED modelling assumptions by comparing them with the \Cigale{} results (Appendix~\ref{appdx:cigale-comparison}). We find largely consistent results, albeit with a scatter of about $0.1-0.3$ dex towards higher masses by \Cigale{}.

\begin{table}[h!]
\begin{threeparttable}
\centering
\caption{Photo-$z$ performance estimated using high-quality spectroscopic redshifts with $>97\%$ confidence.}
\begin{tabular}{c|c|c|c}
\hline
\hline
Mag range & $\sigma_{\text{NMAD}}$\tnote{a}  & Outliers\tnote{b} & Bias\tnote{c}  \\
\hline
$17 < m_{\rm F444W} < 23$ & $0.011$ & $1.44\%$ & $0.007$ \\
$23 < m_{\rm F444W} < 24$ & $0.014$ & $2.05\%$ & $0.005$ \\
$24 < m_{\rm F444W} < 25$ & $0.015$ & $3.94\%$ & $0.007$ \\
$25 < m_{\rm F444W} < 26$ & $0.020$ & $9.00\%$ & $0.005$ \\
$26 < m_{\rm F444W} < 28$ & $0.030$ & $9.52\%$ & $0.009$ \\
\hline
\end{tabular}
\begin{tablenotes}
\item[a] $\sigma_{\text{NMAD}} = 1.48 \times \text{ median}[(|\Delta z - \text{median}(\Delta z)|)/(1+z_{\rm spec})]$;  $\Delta z = \zphot - z_{\rm spec}$. 
\item[b] Outliers = $|\Delta z| > 0.15 (1+z_{\rm spec})$
\item[c] Bias = median$(\Delta z)$.
\end{tablenotes}
\label{table:photoz-performance}
\end{threeparttable}
\end{table}

We select our sample over a total of $\sim 1926\, \si{arcmin}^2$ ($0.534\, \si{deg}^2$).
We apply bright star masks that are defined in the HSC images, which in our case are the same as in \cite{Weaver2022cosmos2020}. These masks remove a conservatively large area around bright stars that contaminate the flux of nearby sources in the ground-based images. Even though the high-resolution NIRCam data is the pillar of our work and has a significantly smaller area affected by the bright stars, these conservative masks are necessary since we include HSC and \UVISTA{} fluxes in our SED fitting, and these data are unusable within the mask. These allow us to capture the Lyman break more accurately at $z\lesssim5$, especially in COSMOS-Web where HST coverage is limited, and more tightly sample the SED of galaxies especially in the NIR, where NIRCam coverage is limited to 4 bands, thanks to the \UVISTA{} bands. Applying these masks results in an effective area of $\sim 1551\, \si{arcmin}^2$, or $0.431\, \SI{}{\deg}^2$.

We apply a cut in F444W magnitude that corresponds to $\sim80\%$ completeness ($m_{\rm F444W}=27.5$, see Fig.~\ref{fig:mag-completeness}  and \S\ref{subsec:completeness-limits}). This ensures that the flux in the rest-frame optical is robustly measured at $S/N > 5$. Additionally, this cut also ensures that there is no spatial selection bias due to the small heterogeneity in depth in COSMOS-Web. This is rather conservative, however, the main goal of this work is leveraging the relatively large area to probe both ends of the typical knee of the Schechter form of the SMF, with a focus on the high-mass end. We also remove sources ($0.13\, \%$) with poorly constrained \photoz\ that have $>75\%$ of their $P(z)$ outside the $\zphot \pm \Delta z$, where $\Delta z$ is the width of the redshift bin. These are predominantly distributed near the low-mass completeness limit and their redshift distribution follows that of the full sample. The former is taken into account by the completeness correction (\S\ref{subsec:completeness-limits}). Finally, we remove stars and brown dwarfs using the $\chi^2_{\rm star}$ of the star/brown dwarf SED templates fitted by \LePHARE{}, coupled with compactness criteria.

\subsection{Completeness limits} \label{subsec:completeness-limits}

\begin{figure}[t]
  \centering
    \includegraphics[width=1\columnwidth]{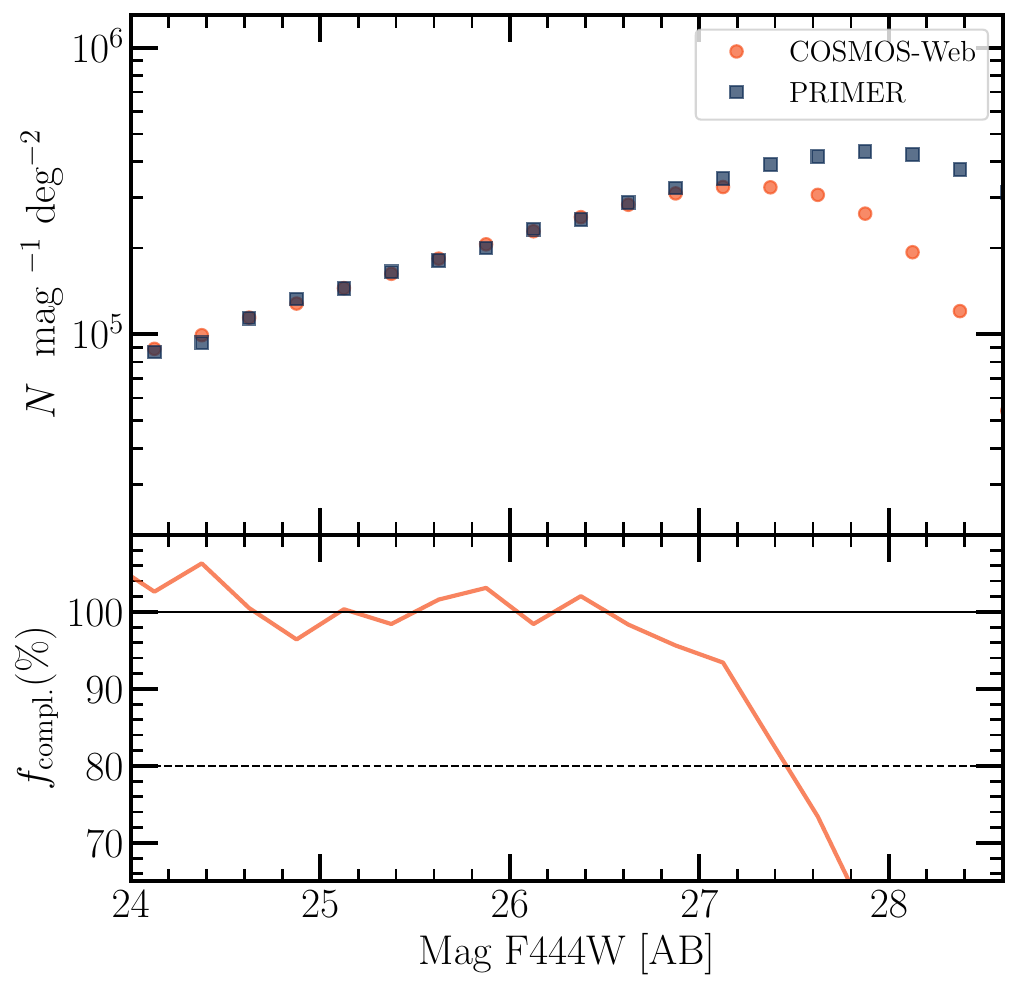}
 \caption{\textbf{Top:} total magnitude number counts as a function of F444W model magnitude in the COSMOS-Web and PRIMER-COSMOS catalogs. \textbf{Bottom:} Completeness fraction, defined as the ratio between the number counts of COSMOS-Web vs. PRIMER.
 }
  \label{fig:mag-completeness}
\end{figure}

Quantifying the completeness of statistical samples is crucial for unbiased interpretations of the stellar mass assembly throughout cosmic time. 
The limited depth of the survey means that faint and typically low stellar mass galaxies start to be missed beyond a certain limiting magnitude ($m_{\rm lim}$), imposing a completeness limit in stellar mass $M_{\rm lim}$, that we quantify in this section.

We compute the stellar mass completeness at the low-mass end following the method of \cite{pozzetti2010}. Briefly, at each $z$-bin, we take the 30$\%$ faintest galaxies as being the most representative of the population near the limiting stellar mass, and derive their mass ($M_{\rm resc}$) by scaling the F444W magnitude to the magnitude limit of the survey ($m_{\rm lim}$)
\begin{equation} \label{eq:mass-completeness}
    \text{log}_{10}(M_{\text{resc}}) = \text{log}_{10} (M_{\star}) + 0.4 ( m_{\text{F444W}} - m_{\rm lim} ).
\end{equation}
Then, we define the limiting stellar mass as the 95th percentile of the $M_{\rm resc}$ distribution. 
To compute the limiting magnitude, we rely on the F444W magnitude number counts. PRIMER-COSMOS is typically deeper than COSMOS-Web by about half a magnitude, and as such, PRIMER can serve us to estimate the magnitude limit of the COSMOS-Web catalog. The top panel of Fig.~\ref{fig:mag-completeness} shows the magnitude number counts in COSMOS-Web and PRIMER, along with a power-law fit in the range $23 < m_{\rm F444W} < 27$. We use model total magnitudes for both datasets, which is the reason why the number counts turn at brighter magnitudes than the image depth typically estimated from fixed-size apertures. The bottom panel shows the completeness fraction for COSMOS-Web defined as $f_{\rm compl.} = N_{\rm CWeb}/N_{\rm PRIMER}$. We then define the limiting magnitude $m_{\rm lim}$ the magnitude at which $f_{\rm compl.} = 80 \%$, which results in $m_{\rm lim} = 27.5$. We note that this is derived using the COSMOS-Web sample after applying the $P(z)$ criterion, therefore taking into account any incompleteness introduced by this.

\begin{figure}[t]
  \centering

    \includegraphics[width=1\columnwidth]{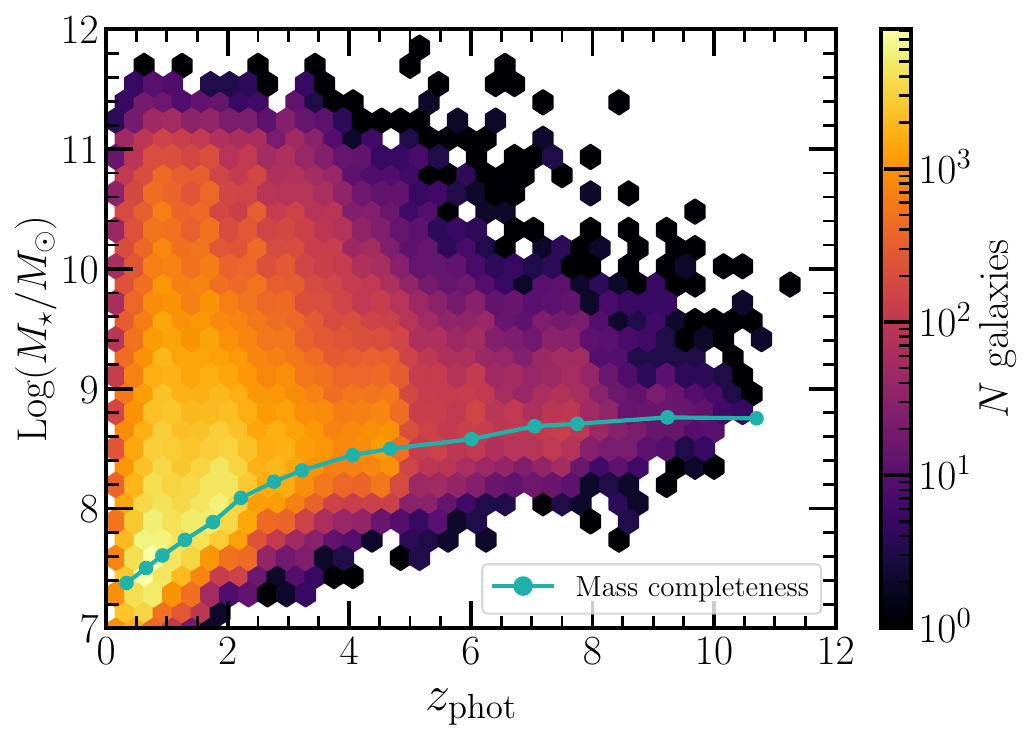}
 \caption{Photo-$z$ vs. stellar mass diagram showing the completeness limits for the COSMOS-Web catalog. The stellar mass completeness limits are derived by rescaling the stellar masses of the $30\%$ faintest sources to the limiting magnitude of the survey (Eq.~\ref{eq:mass-completeness}) and taking the $95$th percentile of this distribution.
 }
  \label{fig:photo-z_smass}
\end{figure}

Using these limiting magnitudes in Eq.~\ref{eq:mass-completeness}, we compute the mass completeness limits for each the $z$-bin. Fig.~\ref{fig:photo-z_smass} shows the photo-$z$ vs. stellar mass density histogram and the mass completeness limits for COSMOS-Web. We limit our analysis to samples more massive than these completeness limits. 

\subsection{Removal of potential AGN contamination}\label{sec:removal-of-lrd}

Recent work from JWST has shown a high abundance of galaxies at $4 \lesssim z \lesssim 9$ that exhibit active galactic nuclei (AGN) signatures identified from their broad H$\alpha$ and/or H$\beta$ lines \citep[e.g.,][]{Kocevski2023, Maiolino2023, Fujimoto2023, Harikane-AGN2023, Greene2024, Matthee2024}, A fraction of these ($\sim20-30\,\%$) show extremely compact (i.e., point-like) morphologies and extremely red colors in the \NIRCAM{} long wavelength bands, therefore dubbed `little red dots' \citep[LRDs,][]{Matthee2024}. $80\%$ completeness is reached at $m_{\rm F444w} = 27.5$.


The origin of the red rest-frame optical continuum emission of LRDs is still not solved, and their photometric features can create important degeneracies during SED fitting. 
They can mimic strong Balmer breaks resulting in high stellar masses from aged stellar populations up to 2 dex, depending on the relative contribution of the stellar and AGN components \citep{Wang2024-rub}. Alternatively, they can also be fitted by an SED with heavily dust-obscured components accounting for the rest frame optical emission, also resulting in high stellar mass. Therefore, unaccounted, these LRD can lead to biased measurements of high abundances of massive galaxies at high redshifts.
\LePHARE{}, being the SED fitting code of choice for this work, does not fit a composite template of stellar and AGN components. Therefore, our stellar masses are prone to be biased for the AGN/LRD population.  


We adopt a conservative approach of completely removing potential AGN/LRDs \citep[in line with other recent works on massive, $z>4$ falaxies, e.g.,][]{Chworowsky2023, Weibel2024, Harvey2024}. By applying the following criteria at $z>3.5$.
\begin{itemize}
    \item \textbf{AGN-SED as the best fit.} \LePHARE{} also fits AGN SEDs as part of the template set. However, this template does not distinguish the light originating from the stellar and AGN components and can not estimate the stellar mass. We use the corresponding $\chi^2_{\rm AGN}$ to identify sources whose photometry is best fit by a AGN SED, requiring that $\chi^2_{\rm AGN} < \chi^2_{\rm gal}$.
    \item \textbf{Compact.} We use the flux ratio in two different apertures \citep[following][]{Akins2024}, and the effective radius $R_{\rm eff}$ of the S\'ersic fit for the compactness criterion: $R_{\rm eff} < \ang{;;0.1} \, \lor \, 0.5<f(\ang{;;0.2})/f(\ang{;;0.5}) < 0.7$. We verify that the latter isolates clearly the locus of point-like sources, and the former corresponds to the FWHM of the F277W PSF. This is now possible in COSMOS-Web thanks to the high \NIRCAM{} resolution, compared to the lower resolution \UVISTA{} bands of the previous COSMOS catalog iteration -- COSMOS2020.
    \item \textbf{Red.} We use the two available NIRCam bands to identify the red colors of the LRD with the condition $m_{\rm F277W} - m_{\rm F444W} > 1.5$. 
    This criterion is also described in detail in the systematical study of LRDs in COSMOS-Web by \cite{Akins2024}.
    We individually inspected the stamps, photometry, and SED fits of many sources and verified that combining the three criteria efficiently identifies LRDs.
    \item \textbf{X-ray emission.} We also identify sources with an X-ray counterpart by cross-matching within $1\, \si{\arcsec}$ with the \cite{Civano_2016} catalog. We remove these regardless if they satisfy or not the previous three conditions.
    
\end{itemize}

The final condition to identify and remove sources dominated by AGN is: 
\textit{[(AGN-SED $\lor$ Red) $\land$ Compact] $\lor$ X-ray}. The main difference in our work is the inclusion of the AGN template criterion, whereas the \textit{(Red $\land$ Compact)} is an analogy of LRD condition from the literature \citep[e.g.,][]{Labbe2023}. 

We note that these conditions still do not capture a potential AGN-dominated population that has both point-like and extended components and no X-ray counterparts, that can still introduce biases. Identifying and dealing with these would involve a 2D decomposition, which we leave for future work.

Finally, in Appendix \ref{appdx:lrd-props}, Fig.~\ref{fig:lrd-agn_dist} we show the distribution of some of the properties of the AGN/LRD sample that we exclude from the analysis, and we discuss how they affect the SMF. 

\subsection{Summary of the selection criteria} 
Here, we summarize the criteria that we apply to select our samples in 15 redshift bins from $z=0.2$ to $z=12.0$.
\begin{itemize}
    \item $m_{\rm F444W}<m_{\rm lim}$, where $m_{\rm lim} = 27.5$.
    \item stellar mass selection above a completeness limit $M_{\star} > M_{\rm lim}(z)$.
    \item at $P(z) > 25\%$ within $\zphot \pm \Delta z$, where $\Delta z$ is the width of the bin.
    \item $\chi^2_{\rm gal} < \chi^2_{\rm star}$, coupled with compactness criteria to remove stars and brown dwarfs.
    \item Does not satisfy the AGN/LRD criterion 
    \textit{(AGN-SED $\lor$ Red) $\land$ Compact}, where
    AGN-SED : $\chi^2_{\rm AGN} < \chi^2_{\rm gal}$, 
    Compact : $R_{\rm eff} < \ang{;;0.1} \, \lor \, 0.5<f(\ang{;;0.2})/f(\ang{;;0.5}) < 0.7$,  
    Red : $m_{\rm F277W} - m_{\rm F444W} > 1.5$
    \item No X-ray counterpart.
    \item Outside of bright star masks defined in HSC.
    \item Finally, visually inspect sources of log$(M_{\star}/M_{\odot}) > 10.5$ at $z>5$, and remove possible artifacts.
\end{itemize}

Table \ref{table:selection-numbers} quantifies the number of sources remaining in our sample after applying the selection criteria.

\begin{table}[h!]
\begin{threeparttable}
\centering
\caption{Number of objects used in the analysis, after applying selection the criteria.}
\begin{tabular}{c|c|c}
\hline
\hline
Criterion & No. objects remaining & Fraction  \\
\hline
Initial\tnote{a} & $471,937$ & $100\%$ \\
$m_{\rm F444W}<27.5$ & $377,435$ & $78.98\%$\tnote{b} \\
$M_{\star}>M_{\rm lim}$ & $290,823$ & $77.05\%$\tnote{c} \\
$P(z)$ & $290,209$ & $99.79\%$\tnote{d} \\
AGN/LRD & $289,844$\tnote{e} & $99.66\%$\tnote{d} \\


\hline
\end{tabular}
\begin{tablenotes}
\item[a] Galaxy sample after applying star masks, removing stars and in the redshift interval we analyse in this paper $0.2 < z < 12.0$.
\item[b] Fraction of the initial sample.
\item[c] Fraction of the magnitude complete sample.
\item[d] Fraction of the mass complete sample.
\item[e] Final number of sources included in the scientific analysis.
\end{tablenotes}
\label{table:selection-numbers}
\end{threeparttable}
\end{table}


\section{Measurements}\label{sec:measurements}

\subsection{$1/V_{\rm max}$ estimator for the SMF}

We measure stellar mass functions in each redshift bin using the $1/V_{\rm max}$ estimator \citep{schmidt_space_1968}.  This estimator essentially corrects the \citet[][]{Malmquist1922} bias which refers to the fact that intrinsically faint galaxies can be observed within a smaller volume. With the $1/V_{\rm max}$ technique, each galaxy is weighted by the maximum volume in which it would be observed given the redshift range of the sample and magnitude limit of the survey. 
For the $i$-th galaxy the $V_{\text{max}, i}$ is computed as
\begin{equation}
    V_{\text{max}, i} = \dfrac{4\, \pi}{3} \dfrac{\Omega_{\rm survey}}{\Omega_{\rm sky}} \left( d_c(z_{\text{max},i})^3 - d_c(z_{\text{min},i})^3 \right),
\end{equation}
where $\Omega_{\rm survey} = 1520\, \rm \si{arcmin}^2$ and $\Omega_{\rm sky} = 41\,253\, \rm deg^2$ are the surface area of the survey and the full sky respectively, and $d_c(z)$ is the comoving distance at redshift $z$.
The comoving volume $V_{\rm max}$ is computed between $z_{\rm min}$ and $z_{\rm max}$, where  $z_{\rm min}$ is the lower redshift limit of the $z$-bin and $z_{\rm max} = \text{min}(z_{\rm bin,up}, z_{\rm lim})$, where $z_{\rm bin,up}$ is the upper redshift limit of $z$-bin and $z_{\rm lim}$ is the maximum redshift up to which a galaxy of a given magnitude can be observed given the magnitude limit of the survey $m_{\lim}$ in F444W.
Finally, the SMF is computed as
\begin{equation}
    \Phi(M_{\star}) \, \dd({\rm log} M_{\star}) = \displaystyle\sum_{i}^{N(M_{\star})}\dfrac{1}{V_{\rm max,i}\,f_{{\rm compl},i}},
\end{equation}
in the mass range starting from the mass completeness limit of each $z$-bin, with a bin size of $\Delta \text{log} M = 0.25$. Additionally, we correct for the magnitude incompleteness by applying $f_{{\rm compl},i}$ which the same completeness vs. magnitude function presented in \S\ref{subsec:completeness-limits}.

\subsection{Sources of uncertainty}\label{sec:sources-of-uncertainty}

\begin{figure*}[th!]
  \centering
    \includegraphics[width=1.0\textwidth]{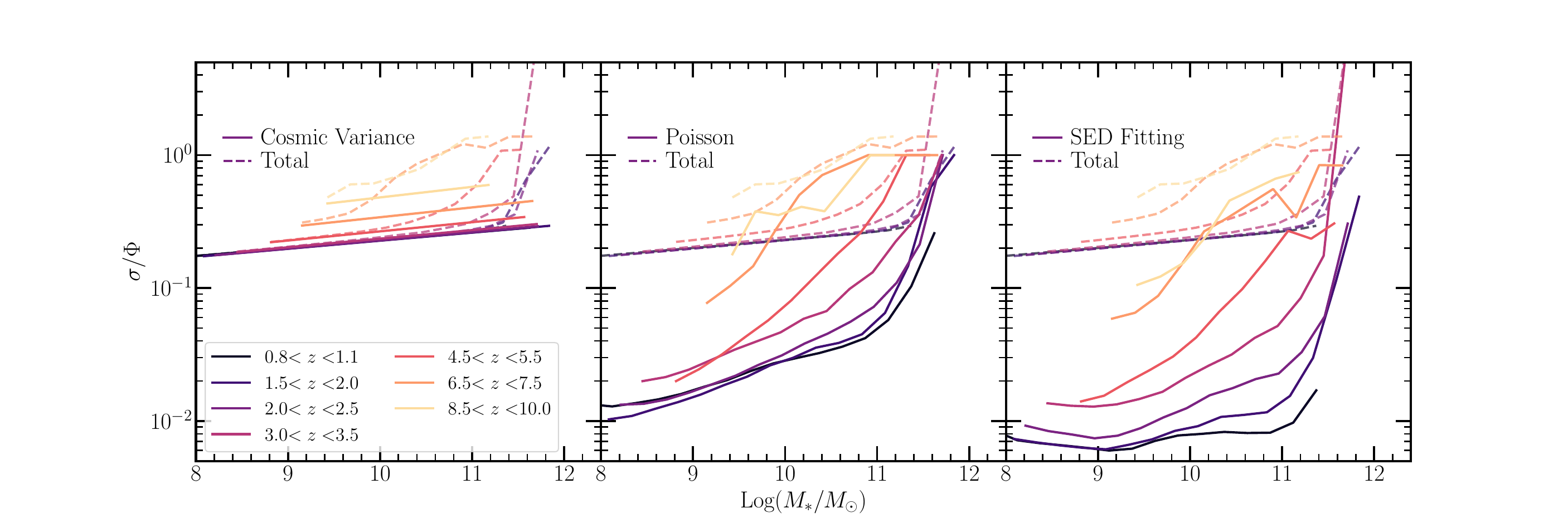}
 \caption{ The uncertainty budget of the SMF measurements. The three panels show the fractional uncertainty due to cosmic variance (left), Poisson (center) and SED fitting (right) in solid lines. The dashed lines, common to all panels, show the total uncertainty, obtained summing in quadrature the three contributions. For clarity, only a subset of our redshift bins are plotted.}
  \label{fig:smf-uncertainties}
\end{figure*}

SMF measurements are typically affected by three types of uncertainties of different origins, that need to be properly estimated. In the following, we describe how we account for these uncertainties in our measurements. 

\textbf{Poisson noise}.
Since measuring the SMF is essentially counting galaxies in bins of stellar mass, it conforms to the Poisson counting statistic. Taking into account the $1/V_{\rm max}$ estimator, the Poisson uncertainty $\sigma_{\rm Pois}$ is computed as $\sqrt{\sum{1/V_{\rm max}^2}}$ where the sum runs over the number of galaxies in the bin. Fig.~\ref{fig:smf-uncertainties} shows the contribution of the different sources of uncertainties to the SMF as a function of stellar mass. The middle panel shows the Poisson uncertainties that become more important and dominant when the number counts are low i.e., at the high-mass end.

\textbf{SED fitting uncertainties}. 
Due to photometric errors and degeneracies in the SED fits, there are uncertainties in $M_{\star}$ and photo-$z$ estimates that propagate to the SMF. We estimate the effects of stellar mass uncertainties from SED fitting on the SMF ($\sigma_{\rm fit}$), by using the ${\rm PDF}(M_{\star})$ computed by \LePHARE{}. 
We estimate $\sigma_{\rm fit}$ by drawing 1000 random samples from ${\rm PDF}(M_{\star})$, computing $\Phi(M_{\star})$ for each, and taking the standard deviation for each stellar mass bin. The right panel of Fig.~\ref{fig:smf-uncertainties} shows the relative uncertainty due to SED fitting that is modest for low-mass galaxies but becomes dominant at the high-mass end, mainly an effect of the Eddington bias. 

We note that these uncertainties do not capture a wider range of plausible star formation histories and other modelling assumptions than those we adopt for \LePHARE{}. In Appendix~\ref{appdx:cigale-comparison} we discuss how different SED modelling assumptions in \Cigale{} (such as non-parametric star formation histories and dust attenuation) result in about $0.1-0.3$ dex of scatter towards higher \Cigale{} masses. We do not propagate these uncertainties into our SMF measurements and caution that these should be considered as a lower limit.

\textbf{Cosmic variance.}
Since galaxies and halos are clustered, observing different fields implies a field-to-field variance in excess of Poisson noise, usually denominated \textit{cosmic variance} ($\sigma_{\rm cv}$, typically given as a fractional uncertainty, see \citealt{Robertson2010}, \citealt{Moster2011_cosmicvariancecookbook} for a definition). The final uncertainty is then $\sigma_{\Phi}^2 = \sigma_{\rm Pois}^2 + \sigma_{\rm fit}^2 + \sigma_{\rm cv}^2 $. Cosmic variance is naturally higher for smaller volumes/fields \citep{Steinhardt2021_cosmicvariance, Vujeva2024_jwst_cosmic_variance}, and also scales as a function of mass and redshift because galaxy bias scales as a function of mass and redshift \citep{Moster2011_cosmicvariancecookbook}. Although \cite{Moster2011_cosmicvariancecookbook} provided a cosmic variance calculator calibrated to observations, it was only calibrated in the low redshift regime and does not generalize to higher redshifts, where it provides significantly too high estimates of the cosmic variance, as discussed by \cite{Jespersen2024_mostmassive} and \cite{Weibel2024}. Here we instead follow \cite{Jespersen2024_mostmassive} and recalibrate the cosmic variance to the \texttt{UniverseMachine} simulation suite \citep{Behroozi2019_UniverseMachine}. This directly incorporates the scatter in the stellar-to-halo mass relation ignored by \cite{Moster2011_cosmicvariancecookbook}, which is on the order of 0.3 dex \citep{Jespersen2022_mangrove}. To fit the cosmic variance, we first sample the number counts in the same angular size, redshift, and mass bins as used in this work, and then fit a power-law in mass with a redshift-dependent normalization and slope. The fitting is done using \texttt{scipy.optimize}, incorporating the additional error terms identified by \cite{Jespersen2024_mostmassive}. The results are shown in Fig.~\ref{fig:smf-uncertainties}, and are significantly below what would be calculated with the \cite{Moster2011_cosmicvariancecookbook} calculator.

\textbf{Eddington bias}. The exponential cut-off of the SMF at $M_{\star}\gtrsim10^{11} \, M_{\odot}$ means that uncertainties in the stellar mass will tend to upscatter more galaxies towards the more massive end than vice versa. This can inflate the number densities of high-mass galaxies, known as the Eddington bias \citep{Eddington1913}. To account for the Eddington bias, we adopt the approach common in the literature \citep{Ilbert13, Davidzon2017, Weaver2022}, where we infer the intrinsic SMF by fitting a functional form convolved with a kernel, $\mathcal{D}(M_{\star})$, that describes the stellar mass uncertainty in bins of mass and redshift. Typically, this kernel takes the functional form of a product of Gaussian and Lorentzian components. However, in our work we build this kernel from the data itself for each redshift bin, by stacking the PDF($M_{\star}$), centered at the median of the distribution, of all galaxies in the $z$-bin. We verified that there is no appreciable change of the width $\mathcal{D}(M_{\star})$ in different mass bins, which is the reason why we use a single kernel per $z$-bin. The resulting $\mathcal{D}(M_{\star})$ for all $z$-bins are shown in Appendix~\ref{appdx:eddington-bias-kernels}.


\subsection{Functional forms to describe the SMF}\label{sec:functional-forms-smf}
The SMF is typically well described by a parametric function often referred to as the \cite{Schechter76} function. This function is a composite of a power law and an exponential cut-off function that describes the low- and high-mass ends respectively. We consider three different functional forms to describe our measurements.

\textbf{Single Schechter function}.
We fit our $1/V_{\rm max}$ measurements with the classical single Schechter function that is traditionally found to describe well the SMF at $z>2$ \citep{Ilbert15, Grazian2015, Davidzon2017, Weaver2022}. Written in terms of the logarithm of the stellar mass, it has a the following form \citep{weigel2016}
\begin{equation} \label{eq:schechter}
\begin{split}
    & \Phi\, \dd({\rm log}M_{\star}) = \\
    & = {\rm ln}(10) \, \Phi^*\, e^{-10^{{\rm log}M_{\star} - {\rm log}M^*}}  \left(10^{{\rm log}M_{\star} - {\rm log} M^*}\right)^{\alpha+1} \dd ( {\rm log}M_{\star}),
    \end{split}
\end{equation}
where $M^*$ is the characteristic stellar mass that marks the so-called `knee' of the SMF separating the power-law of slope $\alpha$ at the low-mass end and exponential cut-off at higher mass. $\Phi^*$ sets the overall normalization that corresponds to the number density at $M^*$. 

\textbf{Double Schechter function}.
For lower redshifts, however, numerous studies have shown that a double Schechter function better describes the galaxy number densities \citep[e.g.,][]{ Peng2010,pozzetti2010,Baldry2012,Ilbert13}, given by the following form
\begin{equation} \label{eq:double-schechter}
\begin{split}
    & \Phi\, \dd({\rm log}M_{\star}) = {\rm ln}(10) \,  e^{-10^{{\rm log}M_{\star} - {\rm log}M^*}} \times \\
    & \left[ \Phi^*_1  \left(10^{{\rm log}M_{\star} - {\rm log} M^*}\right)^{\alpha_1+1} + \Phi^*_2  \left(10^{{\rm log}M_{\star} - {\rm log} M^*}\right)^{\alpha_2+1} \right] \dd ( {\rm log}M_{\star}).
    \end{split}
\end{equation}
The two components share the same characteristic stellar mass $M^*$, but have different normalization $\Phi_1^*$ and $\Phi_2^*$ and low-mass slopes $\alpha_1$ and $\alpha_2$. In our work, we fit the double Schechter form out to $z\sim3$, finding that it provides a better fit to the observed SMF.

\textbf{Double power-law}. We also consider a double power-law (DPL) functional form to fit the SMF. The DPL has been increasingly used to describe the high-$z$ measurements of the UV luminosity function (UVLF) \citep[e.g.,][and references therein]{Bowler2014,Bowler2020,Finkelstein2022,Finkelstein2023}. Since our measurements indicate that the high-$z$ SMF also resembles a DPL (Fig.~\ref{fig:smf-cweb-onepanel},\ref{fig:smf-cweb-literature}), we carry out these fits and quantify how well this form describes the data. We adopt the following functional form of the DPL
\begin{equation} \label{eq:dpl}
    \Phi\, \dd({\rm log}M_{\star}) = \dfrac{\Phi^*}{\left[10^{-({\rm log}M_{\star} - {\rm log} M^*)({\alpha_1+1})} + 10^{-({\rm log}M_{\star} - {\rm log} M^*)({\alpha_2+1})}\right]},
\end{equation}
where $\Phi^*$ is the normalization at $M^*$ which is the characteristic stellar mass that separates the two power-law components described by the slopes $\alpha_1$ and $\alpha_2$.

\subsection{MCMC fits}
To fit the functional forms of the SMF, we define a Gaussian likelihood and employ the affine-invariant ensemble sampler implemented in the \texttt{emcee} code \citep{foreman-mackey_emcee_2013}, using $6\times n_{\rm param}$ walkers. To consider the chains converged, we use the auto-correlation time $\tau$ with the requirement that $\tau > 60$ times the length of the chain and that the change in $\tau$ is less than $5\%$. We discard the first $2\times\text{max}(\tau)$ points of the chain as the burn-in phase and thin the resulting chain by $0.5\times\text{min}(\tau)$. We impose flat priors on all parameters. For the double Schechter function fits we require $\alpha_1 < \alpha_2$ and log$(\Phi^*_2/\Phi^*_1) > 0$. We let all parameters be fitted at all $z$, except the low-mass slope at $z>5.5$ that we set at $\alpha = -2$ for both single Schechter and DPL. We discuss this choice in \S\ref{subsec:intrinsic-SMF}.

\section{Results}\label{sec:results}

\subsection{Galaxy stellar mass function since $z \sim 10$} \label{sec:results-5.1}

\begin{figure*}[t]
  \centering
    \includegraphics[width=0.98\textwidth]{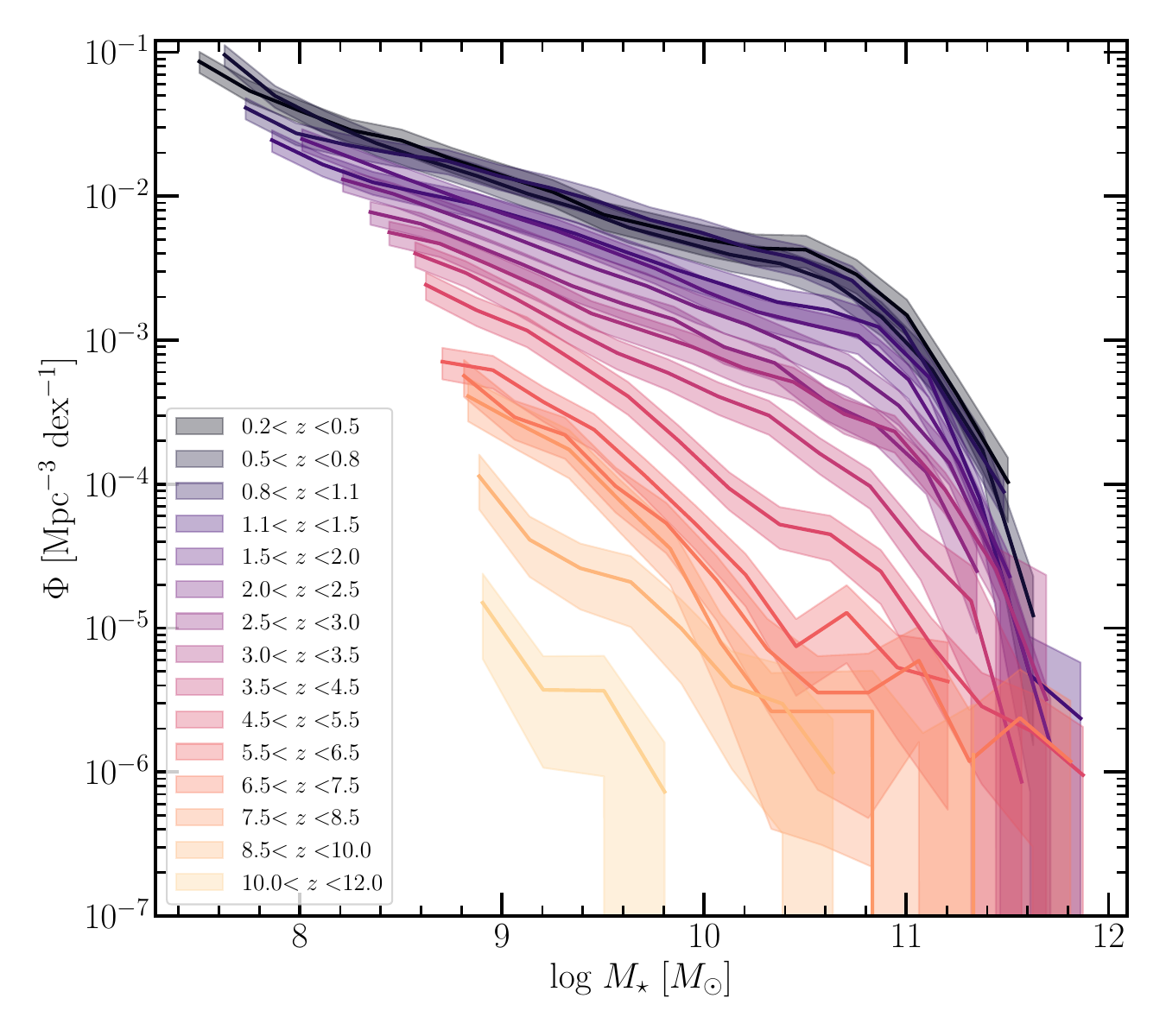}
 \caption{Measurements of the SMF in COSMOS-Web in all 15 redshift bins. Each color corresponds to a different redshift bin. The solid lines mark the measurements, while the filled areas envelop the $1\,\sigma$ confidence interval including Poisson, cosmic variance and SED-fitting errors. The SMF monotonically decreases at all redshifts, with a strong mass dependent evolution.}
  \label{fig:smf-cweb-onepanel}
\end{figure*}

Figures~\ref{fig:smf-cweb-onepanel} and \ref{fig:smf-cweb-literature}  present our measurements of the SMF in COSMOS-Web in each of the 15 redshift bins from $z=0.2$ to $z=12.0$\footnote{We provide our SMF measurements in tabulated form at \url{https://github.com/mShuntov/SMF_in_COSMOS-Web_Shuntov2024}}. The low mass cut is determined by the completeness limit estimated in \S\ref{subsec:completeness-limits}, and the $1/V_{\rm max}$ is measured in bins of $\Delta M_{\star} = 0.25\,M_{\odot}$. The downwards pointing arrows on Fig.~\ref{fig:smf-cweb-literature} represent the upper confidence limits of bins with zero galaxies, estimated as $1.841/V$ following \cite{Gehrels1986}, for the survey volume of COSMOS-Web.

Our work represents the most comprehensive set of measurements of the SMF in the largest redshift and stellar mass range to date.
From Fig.~\ref{fig:smf-cweb-onepanel} we can infer the evolution of the SMF since the first $\sim 360$ Myr. As the redshift increases, the overall number densities of galaxies (the normalization of the SMF) decreases. 
The evolution has been slow since cosmic noon, with the normalization and shape staying very similar since $z\sim2$, only about 0.3 dex change in the last $\sim 10$ Gyr. At earlier times, the evolution picks up pace and the normalization changes by 1.1-1.4 dex from $z\sim6$ to $z\sim2$. This is consistent with the known cosmic star formation history \citep[e.g.,][]{MadauDickinson2014}. 

The evolution of the SMF with redshift can be interpreted in two extreme scenarios: pure mass evolution and pure density evolution \citep[e.g.,][]{Johnston2011, Ilbert13}. In the pure mass evolution scenario, galaxies grow in mass only by star formation, resulting in a horizontal shift of the SMF towards higher masses with decreasing redshift. Our results show that this evolutionary scenario is highly mass-dependent -- a galaxy of $M_{\star}\sim 10^9 \, M_{\odot}$ would increase by $\sim0.9$ dex between $z\sim2.7$ and $z\sim0.3$, while a $M_{\star}\sim 10^{11} \, M_{\odot}$ would increase by $\sim0.4$ dex in the same time. This is consistent with the very efficient `mass quenching' of galaxies once they reach the same characteristic $M^*$ \citep{Peng2010}, which does not appear to evolve considerably with redshift (at least out to $z\sim5$, \S\ref{subsec:intrinsic-SMF}).
In the pure density evolution scenario, the number density of galaxies increases due to the creation of new galaxies, resulting in a vertical shift. Our results show that this scenario too is highly mass-dependent -- the low-mass end evolves faster than the high-mass end. However, the improved completeness at the low-mass end (by about 1 dex in $M_{\star}$) compared to previous work in COSMOS, reveals that the density evolution at masses just below the `knee' of the SMF is faster than the low-mass end, with $\sim 3$ dex and $\sim 2$ dex change in $\Phi$ since $z\sim7$ for log$(M_{\star}/M_{\odot})\sim 10.4$ and log$(M_{\star}/M_{\odot})\sim 9$ respectively. 
As the redshift increases, the low-mass slope steepens, and the `knee' flattens. The `knee' of the SMF, typically at log$(M_{\star}/M_{\odot}) > 10.5$ seems to be disappearing at $z>3.5$. Our measurements show that the number densities and their $1\,\sigma$ confidence intervals at the most massive end (beyond the knee) are within 1 dex since $z=5$ and 2 dex at all robustly probed redshifts. However, due to the rarity of these massive galaxies, the limited survey volume, and possibly the effect of Eddington bias, the `knee' of the SMF and the most massive end becomes difficult to determine robustly at high-$z$. Nonetheless, the flattening of the high mass end at $z>5$ indicates that a fraction of the most massive galaxies has assembled very efficiently in the first few Gyrs and did not grow significantly since.

In this work we can also leverage the large MIRI coverage ($\sim 0.2 \, {\rm deg}^2$) and investigate the resulting SMF for sources covered at $\sim 7.7 \, \SI{}{\micro \meter}$. MIRI is important in probing the rest-frame optical at $z>4$ and results in more robust stellar mass estimates \citep{TWang2024}. In Appendix \ref{appdx:smf-in-miri} and Fig.~\ref{fig:smf-miri} we show the SMF in 6 $z$-bins from $z>4.5$ measured in the MIRI-covered area and compared with the one from the full COSMOS-Web. We find a very good consistency between the two, which also serves as a validation of our measurements in the full COSMOS-Web area. The small differences between the two remain within the uncertainties and small number statistics near the empty bin limits. This is somehow contrary to the conclusions in \citep{TWang2024}, but this remains sensitive to the various ingredients that go into the SED fitting and different codes used in the two works.

Finally, we also compare with the SMF measured from \Cigale{}, as shown in Appendix~\ref{appdx:cigale-comparison}, Fig.~\ref{fig:cigale-compar-smfs}. This shows how different SED modelling assumptions propagate into the SMF and captures the variance and uncertainty arising from it. Overall, there is very good consistency between the two. However, the $0.1-0.3$ dex higher masses by \Cigale{} result in enhanced abundances of more massive galaxies.

\subsection{The $10<z<12$ SMF}\label{sec:the_10z12_smf}
At $10<z<12$, we carried out a more rigorous selection of galaxy candidates for our SMF measurement. In addition to the selection criteria outlined in \S\ref{sec:sample-selection}, we required that $P(z>9) > 68\%$. Furthermore, we visually inspected the median HSC $grizy$ stacks of each source and removed sources that show a counterpart, or are severely blended in this low-resolution image. This removed $25\%$ of the initial selection, and resulted in 27 galaxy candidates at $10<z<12$. We discuss their properties more in \S\ref{sec:zgt10_candidates}. We note that this conservative selection prioritizes purity, therefore the SMF measurement in this bin likely suffers from incompleteness. Another reason for a possible incompleteness are sources that have a photo-$z$ solution at $z<3$ but a secondary peak at $z>10$, that even though can be selected as $z>10$ dropouts, have a preferred SED solution at low redshifts. We do not investigate these further in this work, but they will be studied in detail in dedicated papers on the very high redshift population (e.g., \citealt{Casey2024}, \citealt{Franco2024}, Franco et al. in prep)

By virtue of the large area of COSMOS-Web this number of 27 objects that are brighter than AB mag 27.5 in F444W, allows us to carry out a statistical measurement such as the SMF at $10<z<12$. This results in an SMF that shows a power-law decrease with a slope consistent with $\alpha\approx2$ (see \S\ref{subsec:intrinsic-SMF}). We caution that this is only a tentative measurement, since at $z>10$ the rest-frame optical is no longer probed by the F444W filter. In such cases, stellar mass estimates are highly prone to a number of systematic biases that arise from not probing the rest-frame optical. Additionally, other systematics can also be important at these redshifts, including a top-heavy IMF \citep{Steinhardt2023}. In the MIRI-covered area, 10 sources have a MIRI F770W counterpart at S/N$>2$, and the SMF for these sources remains consistent with the full area (Appendix~\ref{appdx:smf-in-miri}, Fig.~\ref{fig:smf-miri}), however the MIRI area remains insufficient to robustly measure the number densities, and does not significantly help to remove low-$z$ outliers.

\subsection{Comparison with the literature} \label{sec:smf-lit-compar}

\begin{figure*}[p]
  \centering
    \includegraphics[height=0.91\textheight]{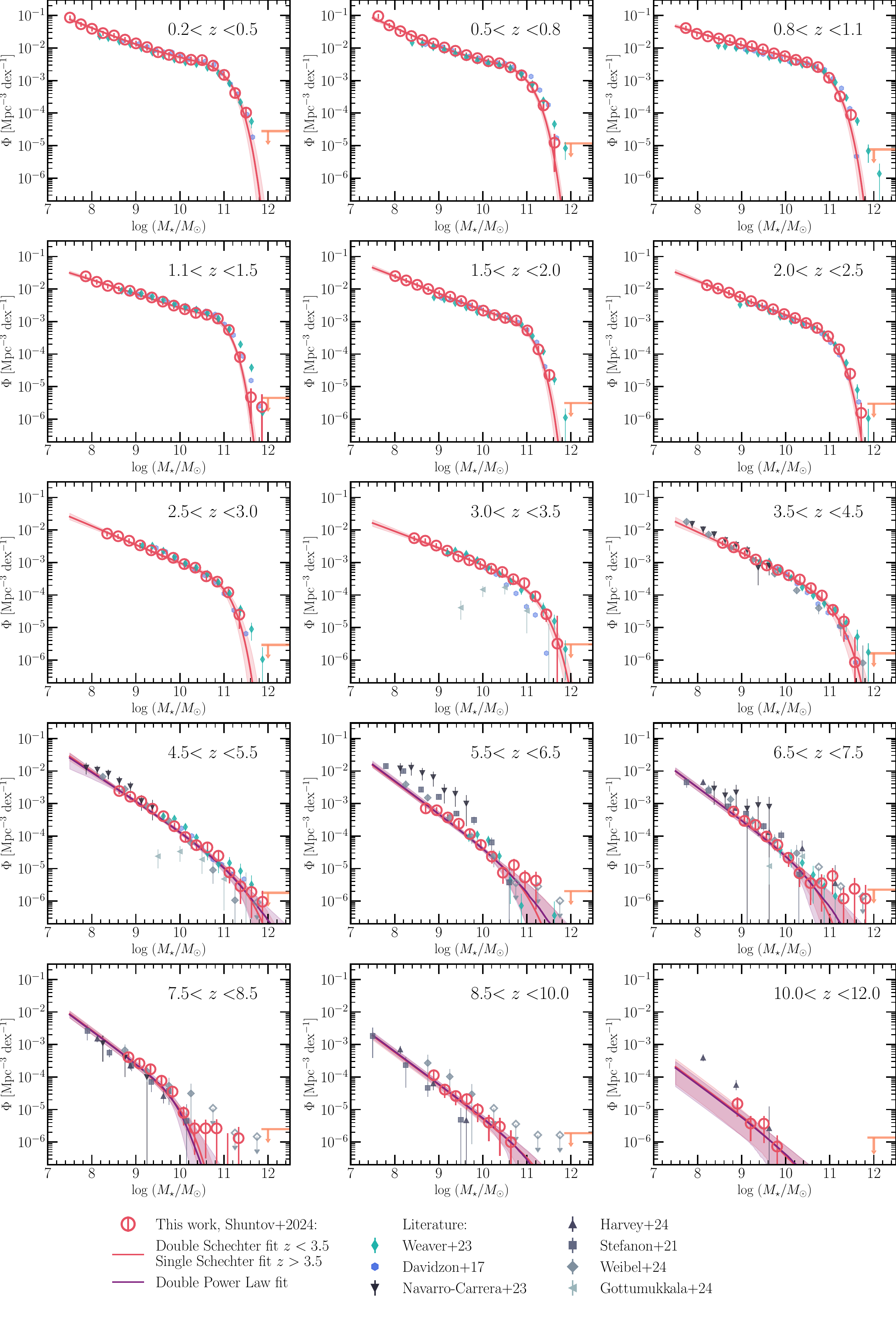}
 \caption{Measurements of the SMF and its evolution with redshift in COSMOS-Web. Each panel shows the SMF in a given redshift bin, along with measurements in the literature from \cite{Weaver2022, Davidzon2017, Stefanon21, Navarro-Carrera2023, Weibel2024, Harvey2024}. The upper limits for empty bins are shown for the COSMOS-Web volume. In each $z$-bin, we show the functional form that fits best the data (lowest Bayesian information criterion), while at $z>4.5$, we show both the single Schechter and DPL fits for illustration.  }
  \label{fig:smf-cweb-literature}
\end{figure*}

Figure~\ref{fig:smf-cweb-literature} also compares our measurement with a selection of recent results from the literature, namely those of \cite{Weaver2022, Davidzon2017, Stefanon21} pre-\JWST\ and \cite{Navarro-Carrera2023, Weibel2024, Harvey2024} using \JWST\ data. 

\textbf{Comparison with previous measurements in COSMOS (pre-\JWST).} In general, there is excellent agreement with previous measurements of the SMF in COSMOS by \cite{Davidzon2017} and \cite{Weaver2022} at all redshifts.
The latter two rely on the ground-based UltraVISTA and the low-resolution, space-based, and relatively shallower (compared to NIRCam) \Spitzer{}/IRAC data to sample the rest frame optical emission from $z\gtrsim2$. Therefore, given this difference in data and photometry extraction techniques, the excellent agreement at $z\lesssim2.5$ attests to the robustness of the measurements. Thanks to the substantially deeper selection in the NIR from \JWST, our work extends the $z<3.5$ SMF down to $\sim0.5-1$ dex lower masses. However, there is some discrepancy, especially at the high-mass end between $z\sim3$ and $z\sim5$ with our results showing lower number densities compared to \cite{Weaver2023}. By matching the samples in these $z$-bins used in COSMOS2020 and in our catalog, we find that the dominant source of this difference is the $M_{\star}$ solutions in our COSMOS-Web catalog that are lower than those in COSMOS2020, especially at the high-mass end. This is likely due to the considerably improved depth and resolution from NIRCam which allows superior deblending and more accurate flux measurements. On the other hand, the $M_{\star}$ solution in COSMOS2020 is mostly constrained by the confusion-limited IRAC data, which can introduce uncertainties and biases in the flux measurements, resulting in biased $M_{\star}$ solutions towards higher values.

\textbf{Comparison the deepest \hst{} and \Spitzer{} measurements (pre-\JWST{}).} \cite{Stefanon21} measure the SMF from $z\sim6$ to $z\sim10$ using the deepest available data from \hst{} and \Spitzer{} in the XDF/UDF, parallels and the five CANDELS (total of 731 arcmin$^2$). Their measurements show higher number densities in the $z\sim 6$ and $z\sim 7$ bins by about 0.2 dex and 0.1 dex respectively. The $z\sim 8$ and $z\sim 9$ bins are consistent with ours within the $1\, \sigma$ confidence intervals, albeit slightly lower at $z\sim9$. Their sample consists of exclusively Lyman-break galaxies, with $H_{1.6 {\mu}m}$ being the reddest detection band. Given the fact that in our work we are not limited to detecting only Lyman-break galaxies thanks to the $4.4 \, \mu$m detection band \citep[see e.g.,][]{Barrufet2023}, the increased number densities of \cite{Stefanon21} at $z\sim6-7$ is somehow surprising. The most likely reason for this difference might be cosmic variance due to the smaller volume. Indeed, spectroscopic redshift searches for overdensities in the GOODS fields (cornerstone of the \citealt{Stefanon21} study) have revealed several significant overdensities at $z\sim6-7$ \citep{Helton2023}, that can explain this difference in the SMF. Another potential reason is a difference in the redshift bin size and mean redshift of the sample, that we do not account for.

\textbf{Comparison with \JWST{} measurements.}
Several works have already reported measurements of the SMF leveraging the unprecedented sensitivity of \JWST{} in the NIR. In one of the earliest works, \cite{Navarro-Carrera2023} measured the low-mass end of the SMF at $3.5 < z < 8.5$ in the PRIMER-UDS and the HUDF fields, in a total of $\sim 20$ arcmin$^2$. \cite{Harvey2024}, assembled some of the deepest \JWST{} observations in numerous fields totaling $187$ arcmin$^2$ and measured the SMF at $6.5 < z < 13.5$. \cite{Weibel2024}, assembled the largest area (prior to our work) from \JWST{} programs in the CANDELS-EGS, -COSMOS, -UDS and -GOODS-S fields, totaling $460$ arcmin$^2$ to study the SMF at $z\sim 4-9$. All of these works probe the SMF down to lower masses by about 0.7-0.5 dex, thanks to the deeper observations compared to our work. However, our work probes more than three times the combined area of the former, and thus more robustly constrains the massive end of the SMF. In general, there is good agreement between the different JWST measurements. 

In the $z\sim6$ and $z\sim7$ bins, the \JWST{} results from \cite{Navarro-Carrera2023} show higher normalization of the SMF by about 0.5 dex. A likely explanation for this difference is cosmic variance, because there are the known overdensities in the GOODS-South field at these redshifts \citep{Helton2023}, that is a part of that measurement. Additional culprits might be uncertainties in the photometry and photometric calibration (given the relatively early release of that paper and the evolution of NIRCam calibration files since) and SED fitting systematics.

At $z\sim8$, $z\sim9$ and $z\gtrsim10$, our measurements are consistent with those of \cite{Harvey2024}, but ours show abundances of massive galaxies that are not probed by their $\sim 8$ times smaller volume. At the low mass end, their measurements are consistent with extrapolating the SMF from our work beyond the completeness limit (at least at $z<10)$. 

Compared to \cite{Weibel2024}, our measurements are fully consistent at all redshifts. At $z\sim9$, our SMF is lower but within the $1\,\sigma$ uncertainties. This difference can be due to the small number statistics in the smaller field of \cite{Weibel2024} ($54$ galaxies at $8.5<z<9.5$ compared to $209$ in our sample at $8.5<z<10.0$). Another likely contribution is the $\Delta z = 0.5$ difference in the redshift bins. However, at these redshifts all measurements should be interpreted cautiously because of the difficulty of measuring photo-$z$s and stellar masses with the wavelength coverage limited to $\lesssim 4.5 \, \SI{}{\micro \meter}$, especially with the limited NIRCam coverage of four bands in our work.

\subsection{The best-fit model of the SMF} \label{subsec:intrinsic-SMF}
In this section, we present the fitted functional forms of the $1/V_{\rm max}$ measurements of the SMF. The resulting best-fit functions represent the intrinsic SMF, with the effect of the Eddington bias removed. This is because during the fitting, we convolve the functions (\S\ref{sec:functional-forms-smf}) with the kernel describing the stellar mass uncertainties $\mathcal{D}(M_\star)$. We fitted the double Schechter out to $z=3.5$, single Schechter at $z>2.5$, and DPL at $z>4.5$, and used the Bayesian information criterion \citep[BIC,][]{Schwartz-BIC} computed for the median posterior values to quantify the best-fitting functional forms.
Table~\ref{table:fit_total} lists the median, 16th and 84th percentiles of the Schechter/DPL parameters, along with the BIC for each fit. In the remainder of this paper, for a given $z$-bin, we use the functional forms with lower BIC, unless stated otherwise. Fig.~\ref{fig:smf-cweb-literature} shows the fitted functions and their $1\, \sigma$ confidence interval. The solid line and the shaded regions are obtained by computing the Schechter/DPL functions for $1000$ randomly drawn samples of the posterior distribution and taking the median, 16th, and 84th percentiles. In this figure, we show the intrinsic functions i.e., not convolved with $\mathcal{D}(M_\star)$, which is the reason for the typical mismatch at high masses. 

At $z<3.5$, the data is better fit by a double Schechter function, which is a higher redshift than previous works that have done this out to $z=3.0$ \citep[e.g.,][]{Ilbert13, Davidzon2017}. This is likely because our improved depth, resolution, and wavelength coverage in the rest-frame optical compared to previous work allow us to 1) be more complete in detecting red quiescent and dusty systems and 2) more robustly determine the redshifts, stellar masses and SFR of this population at $z>3$. An upcoming paper will investigate this in detail (Shuntov et al., in prep.). Since the double Schechter results from massive populations with suppressed SFR\footnote{Which are not necessarily classified as quiescent given a sharp cut in sSFR or color. Instead, the double Schechter is a result of mixing populations with intermediate to low SF activity, and therefore can be observed even for the SMF of active galaxies \citep[e.g.,][]{Ilbert13,Davidzon2017,Weaver2023}} \citep{Peng2010}, this explains why we find it to be a better fit out to $z=3.5$. Additionally, the effects of the Eddington bias can have significant effect on the shape of the SMF \citep[e.g.,][]{Grazian2015, Davidzon2017}, so a narrower mass uncertainty kernel (see \S\ref{sec:sources-of-uncertainty}) can result in revealing the double Schechter. The double Schechter out to higher redshifts can also be a result of probing a wider range of environments -- \cite{Lovell2021} have shown in the \textsc{Flares} simulation that a double Schechter describes the SMF at all redshifts and becomes increasingly pronounced for denser environments.
However, given the fact that denser environments do indeed host more evolved galaxies with suppressed SFR, it is likely that the origin of the double Schechter form remains to be the rise of massive populations with lower SFR.
 
At $3.5 < z < 5.5$, the best fitting functional form is the single Schechter, and at $z>5.5$ it is the DPL. However, both models are virtually indistinguishable visually, and quantitatively when integrating them to compute the stellar mass density. Furthermore, at $z\gtrsim6$ both models fail to robustly capture the high mass end at ${\rm log} (M_{\star}/M_{\odot}) > 10.5$. This is because of several reasons: 1) the large errorbars of these bins containing only a few sources, 2) mass uncertainties that can upscatter the measured number densities, but are not accounted for by the Eddington bias kernel, 3) AGN component that can boost the flux but is unaccounted for in the SED fitting, and are not classified as AGN/LRD.
We note that the majority of these sources with SED fitting solutions at $z>6$ and ${\rm log} (M_{\star}/M_{\odot}) > 10.5$ are very difficult to constrain because they are red, highly dust attenuated and some have sub-mm counterparts; we discuss these further in \S\ref{sec:most-extreme}.

\begin{figure}[th!]
  \centering
    \includegraphics[width=0.72\columnwidth]{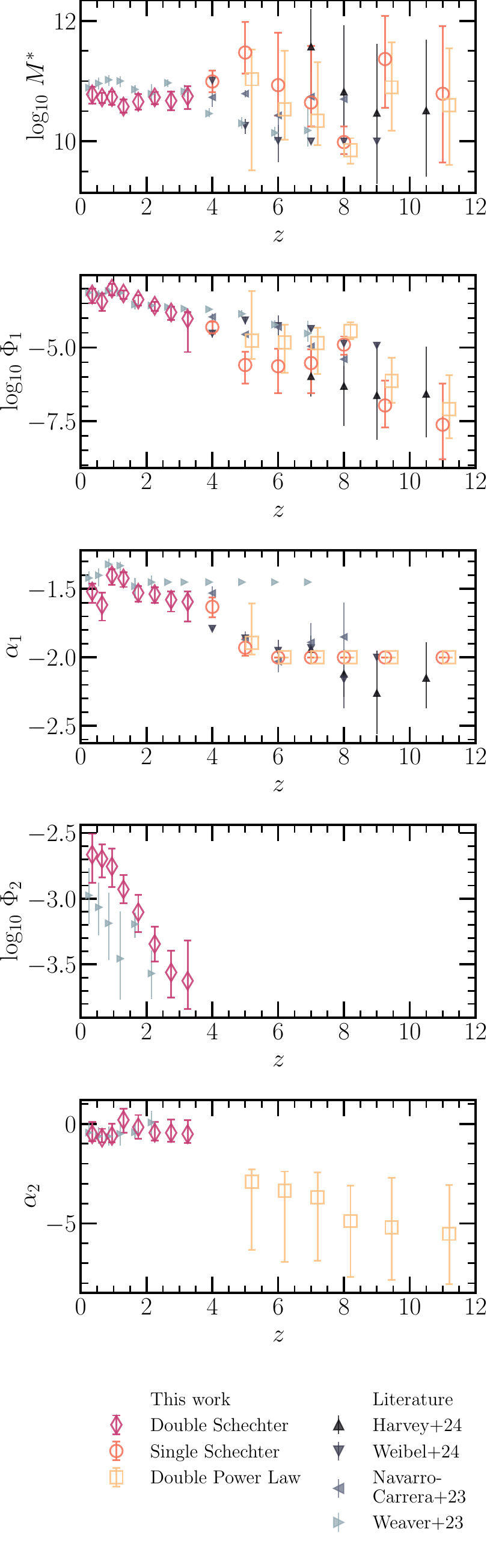}
 \caption{Redshift evolution of the best-fit parameters for the Schechter (double and single) and DPL, along with a literature comparison. The different empty symbols correspond to the best-fit function adopted at a given redshift using the BIC. For illustration, at $z>5.5$, we show the results from both Schechter and DPL fits.
 }
  \label{fig:fit-params-vs-z}
\end{figure}

\subsection{Evolution of the model parameters with redshift.} 
Fig.~\ref{fig:fit-params-vs-z} shows the evolution with redshift of the best-fit model parameters for the double/single Schechter and DPL fits, along with a compilation of literature results. At $z>5.5$, we show the results from both the Schechter and DPL fits for illustration. 

The characteristic mass log$(M^*/M_{\odot})$ is in the range of $\approx 10.6-11$ and does not show a significant evolution out to $z\sim4$, in agreement with \cite{Weaver2023}. However, the uncertainties are too large to infer an evolution meaningfully, and all the measurements in the literature are within the confidence intervals. This is because it is very difficult to constrain the log$(M_{\star}/M_{\odot})>10.5$ regime at $z>6$ with limited survey volumes, even with the $0.53$ deg$^2$ of COSMOS-Web. Constraints on this are poised to come from the next-generation Cosmic DAWN survey -- $59$ deg$^2$ survey of the \textit{Euclid} deep and auxiliary fields combining UV-IR data from \textit{Euclid}, CFHT, HSC, and \Spitzer{} \citep{CDS2024}.

The normalization of the low-mass Schechter component $\Phi_1$ (or $\Phi$ for the DPL) shows little evolution out to $z\sim 1$, after which it decreases rapidly, in agreement with previous works \citep[e.g.,][]{Weaver2023}. The DPL results show an increased normalization compared to the Schecher component by $\sim0.5$ dex, but within the uncertainties. At $z>5$ there is about 1 dex dispersion compared to the results from the literature \citep{Weaver2023, Harvey2024, Weibel2024, Navarro-Carrera2023}, but a general decreasing trend.
The normalization of the high-mass double Schechter component $\Phi_2$ decreases with redshift and shows relatively large uncertainties; compared to \cite{Weaver2023} the normalization is higher by about 0.3 dex, but the trend is in agreement.

The low-mass slope $\alpha_1$ (or $\alpha$ for the single Schechter) remains roughly constant out to $z\sim4$ with a slight peak at $z\sim1$, in agreement with \cite{Davidzon2017, Weaver2023}. At $z>4$, the slope drops to $\approx -2$, and at $z>5.5$, we set it to $-2$, because our data is not deep enough to constrain it. The choice of fixing $\alpha = -2$ is motivated by the fact that literature results deep enough to provide meaningful constraints on $\alpha$ appear to converge at this value \citep{Stefanon21, Navarro-Carrera2023, Weibel2024, Harvey2024}. Furthermore, extrapolating our measurements to lower masses with $\alpha = -2$ is consistent with deeper measurements in the literature.
We note that the low mass slope has an impact on the inferred stellar mass density, which we quantify by sampling $\alpha_1$ out of a normal distribution with a variance of $50\%$.
The low-mass slope of the high-mass component of the double Schechter function, $\alpha_2$, remains roughly constant, albeit with a slight increase, largely consistent with \cite{Weaver2023}. On Fig.~\ref{fig:fit-params-vs-z} we also plot $\alpha_2$ of the high-mass end of the DPL component, although it has a different physical meaning. It has values of $\approx -4$, with large uncertainties that prevent us from identifying any significant redshift trend.

\subsection{Cosmic evolution of the stellar mass density} \label{sec:smd}

The cosmic stellar mass density describes the total stellar mass content in a volume of the Universe at a given epoch. Since the stellar mass growth in the Universe is directly related to the star formation activity, galaxy formation models need to link the observations of both consistently. Therefore, accurate measurements of stellar mass density as a function of cosmic time are essential.

We derive the cosmic evolution of the stellar mass density $\rho_{\star}$ by integrating our SMF presented in \S \ref{sec:results} in each redshift bin
\begin{equation}
    \rho_{\star}(z) = \displaystyle\int_{10^8}^{10^{13}}\dd M_{\star} \, M_{\star} \, \Phi(M_{\star}, z).
\end{equation}
We use the best-fit models described in \S\ref{sec:functional-forms-smf} by taking the models with the lower BIC (double Schechter at $z<3.5$, single Schechter at $3.5<z<5.5$ and double power law at $z>5.5$. We take $10^{8}\, M_{\odot}$ as the lower integration limit, which is the most commonly used in the literature, facilitating comparisons. At $z>2$, our limiting mass is greater than $10^{8}\,M_\odot$, so we rely on the extrapolations of the best-fit functions. The $1\,\sigma$ uncertainty is derived by integrating the 16th and 84th percentiles of the best-fit functions presented in \S\ref{subsec:intrinsic-SMF}.

The left panel of Fig~\ref{fig:smd-sfrd} shows the stellar mass density from our work, along with a compilation of some recent measurements in the literature. Our results are shown in the solid orange line and the shaded region marking the $1\, \sigma$ confidence interval. The latter is obtained by integrating the lower and upper 68 percentiles of the SMF.
Our measurements show a constant increase of $\rho_{\star}$ with cosmic time, with a flattening at $z<1$, consistent with the peak and downturn of the cosmic star formation history. At $z>1$, $\rho_{\star}$ shows no significant change of slope within the uncertainties, at least out to $z\sim9$, indicating a steady buildup of the stellar mass density with time. 

We compare our results with the literature, including \cite{Moutard2016, Wright2018, Bhatawdekar19,Stefanon21,Weaver2023,Navarro-Carrera2023, Weibel2024} and \cite{Harvey2024}. 
Compared with the literature, generally, there is good agreement with our work, most notably with \cite{Weaver2023} at all redshifts.
Compared to \cite{Weibel2024}, there is a good agreement within the uncertainties at all redshifts, albeit with a $\sim 0.1 dex$ difference towards lower $\rho_{\star}$ from our measurements.
\cite{Harvey2024} shows a steeper drop going from $+0.5$ dex at $z\sim7$ to $-0.1$ dex $z\sim9$, and flat trend to the $z\sim11$ bin.
At the highest redshifts, the biggest difference is with the precipitous drop of the $\rho_{\star}$ by \cite{Stefanon21}; this can be because in the last two $z$-bins they fix both log$(M^*/M_{\odot}) = 9.5$ (which is lower than values fitted with our SMF) and $\alpha=-2$.

We also compare with the integrated SFRD function of \cite{MadauDickinson2014}, assuming a return fraction that depends on cosmic time $f_r(t-t') = 0.048\, \log(1 + (t - t') / 0.88\, {\rm Myr})$  (see \S\ref{sec:sfrd-from-smd} for justification), based on a \cite{Chabrier03} IMF, shown in the solid gray line. There is relatively good agreement in the general trend at $z<5$, while at $z>5$ our measurements are consistently lower with a slightly steeper slope. One reason for the discrepancy in this regime could be the fact that the SFRD by \cite{MadauDickinson2014} at $z>5$ is constrained with limited data from only two surveys \citep{Bouwens2012a,Bouwens2012b,Bowler2012} and is an extrapolation at higher redshifts. However, at all redshifts the \cite{MadauDickinson2014} SFRD integration is constantly higher than the direct measurements from our work. This constant difference towards higher $\rho_{\star}$ obtained from integrating the SFRD compared to direct SMD measurements is persisting since the first comparisons between the two \citep[e.g.,][]{Hopkins2006, Wilkins2008}. Some of the reasons discussed in the literature to reconcile this discrepancy are overestimated instantaneous SFR measurements due to overestimated dust attenuation, uncertain UV luminosity to SFR conversion factor, which in turn can be due to an evolving IMF. We discuss this further in \S\ref{sec:sfrd-from-smd}.

Finally, we also compare with the theoretical limits imposed by the dark matter halo evolution \citep{BehrooziSilk2018}. We scale\footnote{$\Phi_{\rm h}(M_{\star}\, f_{\rm b}^{-1}\, \epsilon_{\star}^{-1},z)$} the halo mass function \citep[by][]{2013MNRAS.433.1230W} with the universal baryonic fraction $f_{\rm b} = 0.16$ and with different values for the integrated star formation efficiency, $\epsilon_{\star}$ = [0.05, 0.1, 0.3, 0.6, 1] and integrate it from the same $10^8\, M_{\odot}$ mass limit. This comparison indicates relatively low cosmic star formation efficiencies below $5\%$ at $z>1.5$, with a trend towards increasing efficiencies at $z\lesssim4$ and $z\gtrsim 7$. This suggests that the assembly of halo and stellar mass has not happened at the same rate throughout cosmic history. This can be qualitatively explained by stagnating halo growth rate at $z \lesssim 5$ and a gas reservoir that can keep the star formation efficiency increasing \citep{Lilly2013}. We study in closer detail $\epsilon_{\star}$ and its evolution with both mass and redshift in \S\ref{sec:dm-connection-sfe}, where we discuss this trend.

\section{Discussion} \label{sec:discussion}


\subsection{Abundance of massive galaxies and transition from Schechter to a double power-law}

The form of the SMF and its evolution with redshift provide valuable constraints on physical models. The exponential cutoff of the Schechter function is thought to be a result of `mass~ quenching' \citep{Peng2010}, where AGN feedback in massive galaxies efficiently suppresses star formation and prevents the growth of more massive galaxies \citep{gabor_hot_2015}. The mass scale at which this happens is marked by the characteristic mass $M^*$ of the exponential cutoff. Therefore, robustly measuring the high-mass end of the SMF is important to shed light on the onset and efficiency of the AGN feedback in the early Universe. 

Our results show a transition from the Schechter form to a DPL at $z > 5.5$. In the DPL, the SMF decreases following a power law with increasing mass, in contrast to the exponential cutoff of the Schechter. Although both functional forms do not fully fit the observed SMF points at ${\rm log}(M_{\star}/M_{\odot}) > 10.5$, the DPL providing a lower BIC also qualitatively confirms the observation of overabundance of massive galaxies at $z\gtrsim5$.
The existence of such massive galaxies in excess of predictions from a Schechter law suggests very efficient growth at early times. This is likely due to efficient cooling, gas accretion and higher merger rates, that are shown to increase at earlier times \citep{Duan2024}.
The transition to a Schechter law at $z\lesssim5.5$ also coincides with the rise of the first quiescent galaxies in the Universe \citep[e.g.,][]{Carnall2023,Valentino2023}. This indicates that the physical mechanisms that suppress galaxy growth start to take place at $z\sim5.5$ on a global scale, at least out to masses that we can probe in the COSMOS-Web volume. This is likely due to the onset of negative AGN feedback at these redshifts.
However, we note that this remains only a tentative interpretation, since the relatively large uncertainties and potential systematics in photo-$z$ and mass estimates at the high-mass, high-$z$ end prevent us from drawing robust conclusions. Next-generation wide-area and deep NIR surveys from \textit{Euclid} \citep[e.g.,][]{CDS2024} and \textit{Roman} will be crucial in providing robust constrains on the evolution of the high mass end, $\log (M_{\star}/M_{\odot}) \gtrsim 10.2$, at $z>5$.

\begin{figure*}[th!]
  \centering
  \begin{subfigure}{0.49\textwidth}
    \includegraphics[width=\linewidth]{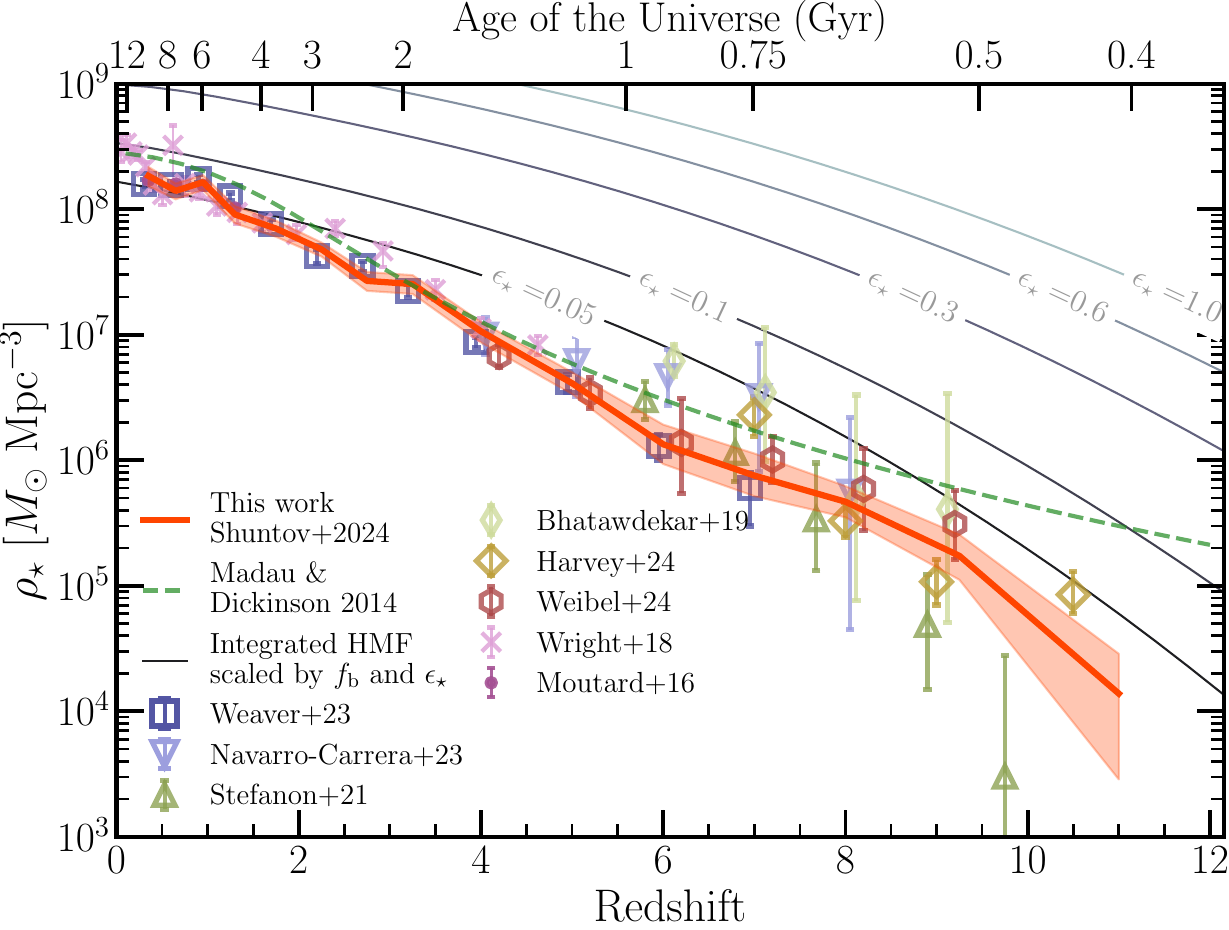}
    \label{fig:SMD}
  \end{subfigure}
  \hfill
  \begin{subfigure}{0.49\textwidth}
    \includegraphics[width=\linewidth]{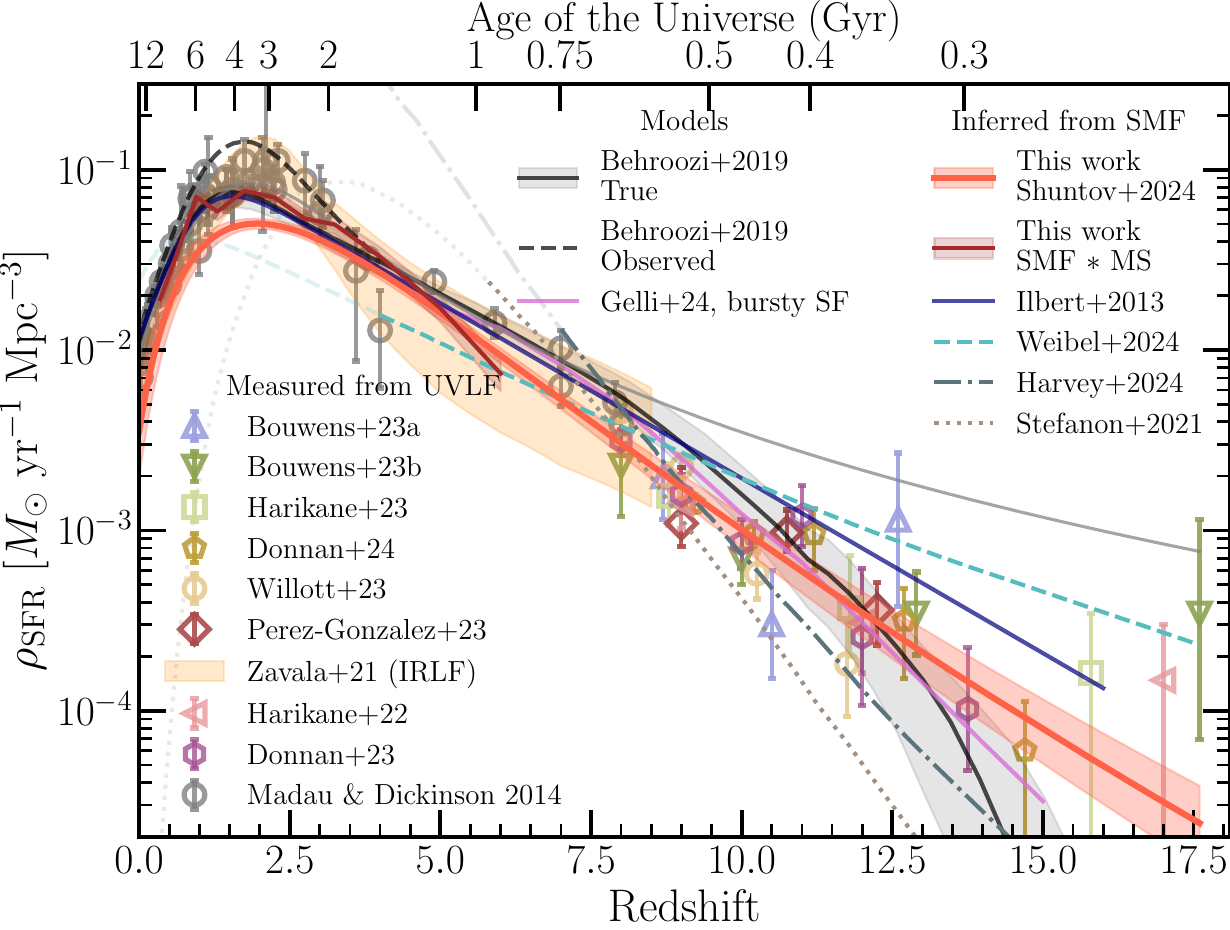}
    \label{fig:second_figure}
  \end{subfigure}
  \caption{{\bf Left:} Cosmic evolution of the stellar mass density, showing a steady increase with time, with no significant change in slope at $1<z<9$. This is computed by integrating best-fit SMF at a given redshift from a common lower limit of $10^8 \, M_{\odot}$. The comparison with some of the most recent literature results includes \cite{Moutard2016,Wright2018,Bhatawdekar19,Stefanon21,Weaver2023,Navarro-Carrera2023, Weibel2024} and \cite{Harvey2024}. The solid gray line shows the $\rho_{\star}$ obtained by integrating the \cite{MadauDickinson2014} SFRD function multiplied by a time-dependent return fraction based on Chabrier IMF (see \S~\ref{sec:smd}). The gray lines show the theoretical limits imposed by the halo mass function, scaled by the baryonic fraction $f_{\rm b}$ and different integrated star formation efficiencies $\epsilon_{\star}$, integrated from the same $10^8\, M_{\odot}$ limit.
  {\bf Right:} Inferred cosmic evolution of the star formation rate density. Comparison with the literature includes measurements of $\rho_{\rm SFR}$ from UVLF from \cite{Harikane2022, Harikane2023, Donnan2023, Donnan2024, Willott2024, Perez-Gonzale2023, Bouwens2023a} and \cite{Bouwens2023b} or from IRLF by \cite{Zavala2021}, as well as the compilation in \cite{MadauDickinson2014}. All are obtained using a common lower integration limit of the UVLF of $M_{\rm UV}=-17$, including \cite{Bouwens2023a, Bouwens2023b} that we rescale using a factor of 0.5 dex. All points are converted to a Chabrier IMF. The $\rho_{\rm SFR}$ inferred from SMF measurements is also shown for a compilation of the most recent literature works \citep{Ilbert13, Stefanon21, Weibel2024, Harvey2024} shown only for the best-fit function, without confidence intervals for clarity. The solid lines turn to transparent at lower $z$ that are not probed in the corresponding work. The solid brown line and shaded region show the SFRD obtained by integrating the SFRF obtained using our SMF and the \cite{Popesso2023} SF main sequence. The black solid (dashed) line and shaded region show the true (observed) SFRD from \cite{Behroozi2019_UniverseMachine}. Our results indicate lower inferred SFRD at $z<3.5$, in tension with instantaneous SFR indicators, while at $z>7.5$ we find remarkable consistency with recent \JWST{} UVLF results.
  }
  \label{fig:smd-sfrd}
\end{figure*}

\subsection{Implied cosmic star formation history}\label{sec:sfrd-from-smd}

The total stellar mass assembled at a given epoch in a volume of the Universe is a result of the integrated star formation activity until that epoch, times a factor accounting for the stellar mass loss due to material returned to the ISM via stellar winds and supernovae \citep{RenziniVoli1981}. Therefore, the stellar mass density can be related to the star formation rate density $\rho_{\rm SFR}$ as \citep{Wilkins2008}
\begin{equation} \label{eq:smd-from-sfrd}
    \rho_{\star}(t) = \displaystyle\int_{0}^{t} \dd t' \, \rho_{\rm SFR}(t') \, \left(1 - f_{r}(t-t') \right),
\end{equation}
where $f_r$ is the stellar mass loss that depends on the age of the stellar populations, but also on metallicity \citep{RenziniBuzzoni1986}. 

The interest in comparing the implied star formation history from stellar mass density measurements to the one derived from star formation measurements is that a solid understanding of the physical processes that affect galaxy evolution will yield a consistent picture with both measurements in agreement. This is because both are affected by complementary systematic uncertainties.
The star formation rate density is typically inferred from instantaneous indicators of star formation, such as rest-frame integrated UV emission, emission lines and IR luminosities, typically coming from young stellar populations \citep[e.g.,][]{Kennicut1998, Calzetti2007, MadauDickinson2014}. These instantaneous indicators can be subject to greater uncertainty due to dust attenuation, and due to various assumptions in the stellar population synthesis models in deriving luminosity-to-SFR calibration factors. On the other hand, stellar masses are inferred from the light of more evolved stellar populations and are subject to different uncertainties (e.g., the assumed star formation history in SED fitting). In this section, we infer the cosmic SFH from our SMD measurements and compare to a range of literature results on the $\rho_{\rm SFR}$, typically inferred from the instantaneous probe of SF -- the UV luminosity function (UVLF).

We adopt a parametrized relation for the stellar mass loss that is calibrated using the computations in \citet[][\S 2.4 and Appendix A in their paper]{Dubois2024}, and assuming a \cite{Chabrier03} IMF: $f_r (t-t') = 0.048 \ln(1 + (t-t')/0.88 {\rm Myr})$. We adopt a log-normal parametrization of $\rho_{\rm SFR}$ and fit our observed $\rho_{\star}$ using Eq.~\ref{eq:smd-from-sfrd}, finding
\begin{equation}\label{eq:sfrd-from-smd}
    \rho_{\rm SFR}(z) = 
    0.05_{-0.01}^{+0.01}\, {\rm exp}\left[ -\dfrac{1}{2} \left(\dfrac{{\rm log}(1+z) - {\rm log}(1+1.95^{+0.17}_{-0.18})}{0.47_{-0.03}^{+0.03}} \right) \right],
\end{equation}
The inferred $\rho_{\rm SFR}$ from $\rho_{\star}$ depends on the functional form assumed for the former. We also tested the parametrization of \cite{MadauDickinson2014}, \cite{Behroozi13}, as well as a double log-normal, but we found that Eq.~\ref{eq:sfrd-from-smd} provides the best fit with lower $\chi^2_{\rm red}$ and Bayesian information criterion.

For consistency check and validation, we obtain the SFRD with an alternative, and perhaps more direct approach. This relies on the star-formation main sequence (MS), as parametrized by \cite{Popesso2023}, to obtain the star-formation rate function (SFRF) from our SMF measurements, by simply converting the $M_{\star}$ to SFR: $\Phi_{\rm SMF} (M_{\star} \to {\rm SFR}) \equiv \Phi_{\rm SFR} ({\rm SFR})$. We also convolved the SFRF with a Gaussian kernel with $\sigma=0.3$, corresponding to the typical intrinsic scatter of the MS \citep[e.g.,][]{Schreiber2015}. Then, we get $\rho_{\rm SFR}$ by simply integrating the SFRF \citep{Picouet2023} from the SF limit corresponding to $M_{\star}=10^{8}\, M_{\odot}$, out to $z\sim6$ where the MS parameterization is calibrated.

The right panel of Fig.~\ref{fig:smd-sfrd} shows the inferred cosmic SFH from our work (Eq.~\ref{eq:sfrd-from-smd} in orange solid line and $1\, \sigma$ confidence interval). We show the SFH extrapolated to the highest redshifts probed by some of the work in the literature. The $z>12$ part can be reasonably well constrained with this approach because after integration it has to result in the measured SMD at $z<12$ from our work. The SFRD obtained from the SMF and the MS is shown in the brown line and shaded area.

\textbf{Comparison with the SFRD inferred from UVLF measurements.}
We compare the SFRD inferred from SMD measurements of our work to the one from instantaneous star formation measurements. These include the compilation from \cite{MadauDickinson2014} out to $z\sim7$ using various tracers from UV to IR, \cite{Harikane2022, Harikane2023, Donnan2023, Donnan2024, Willott2024, Perez-Gonzale2023, Bouwens2023a, Bouwens2023b} that are based on the UVLF, and \cite{Zavala2021} based on the IRLF. We rescale the results of \cite{Bouwens2023a, Bouwens2023b} by a factor of 0.5 dex to bring them to a common lower integration limit of the UVLF $M_{\rm UV}=-17$. We also convert all measurements to a Chabrier IMF using a conversion factor of $0.63$. We verify that the integration limits of $M_{\rm UV}=-17$ and $M_{\star} = 10^{8}\, M_{\odot}$ are consistent for this comparison by using observed $M_{\star}-M_{\rm UV}$ relations \citep[e.g.,][]{Song2016}. We tested this further by reintegrating the SFRF by considering galaxies with $M_{\star} > 10^{6}\, M_{\odot}$ and finding a negligible difference. 

At $z\lesssim3.5$, we find that the SFRD inferred from our SMD is $\sim0.3$ dex lower than the SFRD from instantaneous SFR measurements by \cite{MadauDickinson2014}, with the largest difference around the peak of the cosmic SFH $z\sim2$. However, the SFRD obtained from integrating the SFRF is higher than the one from SMD and in relatively good agreement with \cite{MadauDickinson2014} out to $z\sim4$, with a difference of about 0.1-0.2 dex around the peak of the cosmic SFH. This is perhaps unsurprising, because the SFR estimates from the \cite{Popesso2023} MS are calibrated similarly to \cite{MadauDickinson2014}.

One reason for the lower SFRD inferred from the SMD can be in the assumed stellar mass loss function that has stronger effect at lower redshift (due to the integrated effect over time, Eq.~\ref{eq:smd-from-sfrd}). We tested different mass loss functions \citep[e.g,][]{ConroyWechsler2009, Ilbert13} and found that although it can lower the difference and reconcile them $z\lesssim1.5$, it cannot fully account for the difference over a wider redshift range.

We compare our findings with \cite{Behroozi2019_UniverseMachine} \textsc{UniverseMachine} model, which infers both the `true' and the `observed' SFRD (shown in solid and dashed black lines respectively in Fig.~\ref{fig:smd-sfrd}). The latter is obtained by applying correction that accounts for redshift-dependent observational systematic offsets. 
The `true' SFRD is consistently lower that the `observed' with the biggest difference ($\sim0.4$ dex) around the peak of the SFRD ($z\sim2$). This is consistent with our results from the SMD, albeit with a difference of about $0.1-0.2$ dex at $z<2.5$.
The observational systematic offset applied in \cite{Behroozi2019_UniverseMachine} is calibrated exactly against this tension between the integrated SFRD and the SMD, that has been repeatedly reported and studied in the literature \citep[e.g,][]{Hopkins2006, Wilkins2008, Leja2015, Wilkins2019}. Our results corroborate this tension, the cause of which remains poorly understood, with several possible culprits (aside the stellar mass loss assumption). These include uncertain effects of dust attenuation on both stellar mass and SFR estimates, uncertain calibration for SFR indicators, IMF assumptions, and SFH and other modeling assumptions in SED fitting-derived stellar masses.

To shed some light on this tension, \cite{Wilkins2019} have investigated the effects of different (and more realistic) stellar population synthesis models on the luminosity-to-SFR calibration factors and on the resulting SFRD. They find that the recalibrated factors can lower the \cite{MadauDickinson2014} SFRD by $\sim0.2$ dex. This would alleviate part of the tension that we report in our work, and is one of the most likely explanations. Further investigation with a consistent dataset and methodology from both the SFR and stellar mass side, which is now possible with \JWST{}, is necessary in order to reconcile the two fundamental cosmic observables.

At $z>7.5$, the latest literature measurements from the UVLF, mostly from \JWST{} data \citep{Harikane2022, Harikane2023, Donnan2023, Donnan2024, Willott2024, Perez-Gonzale2023, Bouwens2023a, Bouwens2023b}, show a good agreement with the star formation history inferred from our work, albeit with a large scatter and uncertainties. This remarkable agreement in the SFH from two parallel approaches solidifies the emerging picture of rapid galaxy formation leading to increased abundances of bright and massive galaxies very early in the Universe.

\textbf{Comparison with the SFRD inferred from SMF measurements.}
We also infer $\rho_{\rm SFR}$ from some of the latest literature measurements of the $\rho_{\star}$ using the same procedure. These are shown in dashed/dotted/dash-dotted lines from \cite{Weibel2024, Stefanon21, Harvey2024}. The lines turn to transparent at the lowest $z$-range probed in the corresponding work. There are varying degrees of differences between these works, ours and the $\rho_{\rm SFR}$ from instantaneous SF tracers, with none in very good agreement in a large redshift range at $z>6$. This is mainly because these works do not measure the SMD at $z\lesssim5$, and the lower $z$ measurements are essential in constraining the full $z$ behavior. This is also shown by the result in \cite{Ilbert13} that constrains the SFH from SMD measurements at $0.2<z<4.0$, although in this case the lower maximum $z$ limit likely creates the difference at $z\gtrsim3$.

\subsection{The most massive galaxies in COSMOS}\label{sec:most-extreme}

In this work, we leverage the $\sim 4\times10^6$ Mpc$^3$ volume observed with COSMOS-Web to reveal some of the rarest and most massive objects in the Universe. Massive galaxies in the early Universe provide stringent constraints on the galaxy formation models and on the $\Lambda$CDM cosmology \citep{Steinhardt2016,BehrooziSilk2018,Boylan-Kolchin2023}. 
A number of prior studies based on \JWST{} have reported increased abundances of massive galaxies at $z\gtrsim5$ \citep{Labbe2023Natur, Furtak20230, Xiao2023, Akins2023, Chworowsky2023, Casey2024} that have posed challenges to our models of galaxy formation and cosmology \citep{Boylan-Kolchin2023,Lovell2023}. 

\begin{figure}[th!]
  \centering
    \includegraphics[width=1\columnwidth]{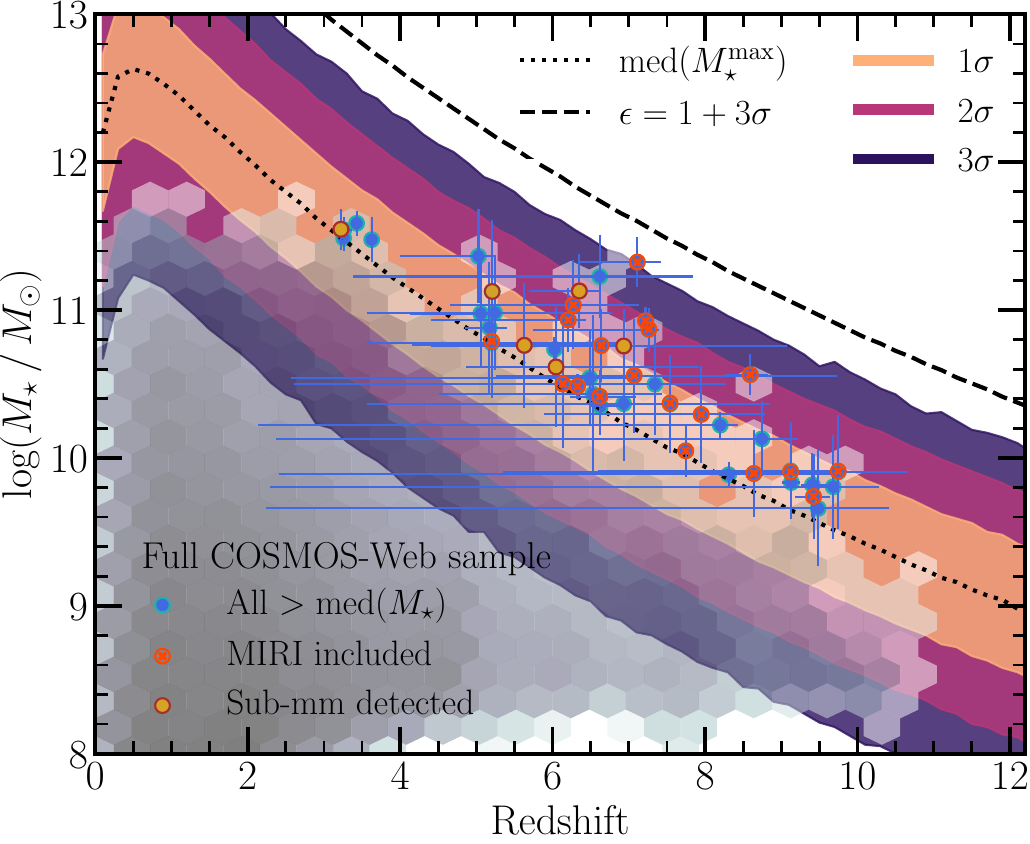}
 \caption{Theoretical limits imposed from halo abundances of the most massive plausible galaxies at a given epoch and volume within the extreme value statistics (EVS) formalism. The colored regions indicate the EVS confidence intervals for the COSMOS-Web survey area of $0.43$ deg$^2$. 
 The dotted line marks the median of the EVS distribution of the maximum plausible stellar mass, assuming a log-normal distribution of the star formation efficiency centered at $\sim 0.14$, while the dashed line shows the $3\,\sigma$ upper limit assuming a star formation efficiency of unity. The grayscale hex bin histogram shows the $M_{\star}-\zphot$ distribution of the full sample. Above the median $M_{\star}^{\rm max}$, the points mark individual galaxies in blue; those that have MIRI photometry are colored orange, and sub-mm detected sources are marked with yellow. The observed $M_{\star}$ are corrected for the Eddington bias using Eq.~\ref{eq:edd-bias-corr-mass}. We find several candidates that are within the 2 and 3$\sigma$ upper limits from the EVS as exceptionally massive for their epoch and volume. We mark sub-mm detected galaxies that are highly dust-attenuated with discrepant $z_{\rm phot}$ and $M_{\star}$ solutions when including FIR-to-radio data in the SED fitting.
 }
  \label{fig:evs-interpret}
\end{figure}

We interpret the most massive galaxies observed in COSMOS-Web within the extreme value statistics \citep[EVS,][]{Lovell2023} formalism. EVS is a probabilistic approach to estimating the PDF$(M_{\rm max})$ of observing the most massive object at a given redshift and given cosmological volume. The EVS PDF$(M_{\rm max})$ is first derived for halos using theoretical halo mass functions, and the conversion to the PDF of the most massive galaxy is done using the universal baryon fraction $f_{\rm b}$, and stellar fraction $f_\star$. The latter is equivalent to the integrated star formation efficiency $\epsilon_{\star}$ that we use in this work: $M_{\star} = M_{\rm h}\,f_{\rm b}\,\epsilon_{\star}$. The stellar fraction is modeled as a lognormal distribution PDF$(\epsilon_{\star}) =\ln N(\mu=e^{-2}, \sigma=1)$, truncated between 0 and 1. This distribution is centered at 0.14 and incorporates the range of values found in the literature for $\epsilon_{\star}$, as well as capturing a potential intrinsic scatter \citep[e.g.,][]{wechsler_connection_2018}. However, it assumes no dependence on redshift and halo mass, the latter of which is well-known to exist, and the former is increasingly hinted at by recent studies.

Figure~\ref{fig:evs-interpret} shows the distribution of stellar masses and redshifts of our COSMOS-Web sample, along with the EVS confidence intervals for observing the most massive objects. We corrected the observed stellar masses for the effect of Eddington bias following \cite{Lovell2023}
\begin{equation} \label{eq:edd-bias-corr-mass}
    \ln M_{\star, \rm Edd} = \ln M_{\star, \rm obs} + \dfrac{1}{2} \, \gamma \, \sigma_{M_{\star}}^2,
\end{equation}
where $\gamma$ is the slope of the HMF at $M_{\rm h} = M_{\star}/f_{\rm b}$, and $\sigma_{M_{\star}}^2$ are the uncertainties on the stellar mass. This correction lowers the observed stellar mass because the slope is negative.
The colored regions show the EVS $1\, \sigma, 2\, \sigma$ and $3\, \sigma$ confidence intervals that assume the lognormal distribution for $\epsilon_{\star}$, while the dashed line shows the $3\,\sigma$ upper limit assuming $\epsilon_{\star} = 1$. Exceeding this limit would mean that there is more stellar mass than available baryons (under the assumption of a universal $f_{\rm b}$) and represents a $3\,\sigma$ tension with the underlying $\Lambda$CDM cosmology. Above the median $M_{\star}^{\rm max}$, we mark individual galaxies with points on Fig.~\ref{fig:evs-interpret}.

None of the most extreme objects in our sample exceed the $3\,\sigma$ limit of $\epsilon_{\star} =1$. However, a handful of objects come in mild tension with the theory by approaching the $3\,\sigma$ limit of the lognormal $\epsilon_{\star}$ distribution centered at 0.14. In Fig.~\ref{fig:evs-interpret}, we also mark the sources having MIRI photometry that in principle results in more robust stellar mass estimates \citep{TWang2024}, but do not see a significantly preferred distribution in the $M_{\star}-z$ plane above the med$(M_{\star}^{\rm max})$ line.
In Appendix~\ref{appdx:extreme-gals}, Fig.~\ref{fig:extreme-properties} we show the F277W $-$ F444W color vs. F444W magnitude distribution of the extreme sample, color coded by $E(B-V)$. A significant fraction ($>50\%$) of our extreme sample is very red ($m_{\rm F277W} = m_{\rm F444W} > 1$) and dust attenuated with $E(B-V)>0.5$. The red colors and potentially high dust attenuation can make the derived photo-$z$ and stellar masses uncertain due to degeneracies in the SED fitting. This is captured in Fig.~\ref{fig:evs-interpret} by the large redshift error bars of some of the candidates, making their nature uncertain.

We check if our extreme sample, defined as $M_{\star} > {\rm med}(M_{\star}^{\rm max})$, have sub-mm counterparts by cross-matching with the SCUBA-2 selected sample of sub-mm galaxies (SMGs) by \cite{McKinney2024}. This sample is one of the most comprehensive and up-to-date compilation that combines all archival photometric data from optical through radio in COSMOS. About $12\%$ of our extreme sample are SMGs in the SCUBADive \citep{McKinney2024} catalog. Given their extreme red colors ($m_{\rm F277W}-m_{\rm F444W}>1.2$), these are highly dust attenuated and difficult cases to fit for our SED modeling that is configured to yield accurate results over all redshifts and populations (e.g., $E(B-V)$ limited to 1.2). Indeed, SMGs are known to be difficult cases for SED fitting, that depending on modelling assumptions and available data, can have mass estimates varying up to 1 dex \citep{Hainline2011, Michalowski2012}. We find a relatively large scatter between our photo$-z$ solutions and those by \cite{McKinney2024} that include FIR-to-radio data and more tailored SED modeling, which prefer lower redshift solutions by about $\Delta z = 1-2$. Fig.~\ref{fig:extreme-properties} shows that some of these sources are fitted with low $E(B-V)$ in our catalog, likely due to the dust attenuation-stellar mass degeneracy, and likely have overestimated stellar masses. We mark in yellow these sources in Fig.~\ref{fig:evs-interpret}, and caution their interpretation.

We also cross-match our extreme sample with spectroscopically confirmed SMGs and find two matches. One source has $z_{\rm spec} = 5.051$ from \cite{ShuowenJin2019}, consistent with our $z_{\rm phot} = 5.68^{+1.22}_{-1.19}$. This source has also been studied in \cite{Gentile2024} with the name ERD-1 and has a consistent stellar mass solution as ours (log$(M_{\star}/M_{\odot}) = 11.28$). 
The second is found at $z_{\rm phot}=5.38^{+0.84}_{-1.93}$ and log$(M_{\star}/M_{\odot}) = 11.61^{+0.24}_{-0.34}$ in our catalog, but has $z_{\rm spec}=3.68$ and log$(M_{\star}/M_{\odot}) = 11.09$ derived from spectroscopy \citep{LiuD2019b}. We do not exclude this source from our SMF which is in a $M_{\star}$ bin that has count one and is within the empty-bin confidence interval \citep[][see also \S\ref{sec:results-5.1}]{Gehrels1986}.

In Appendix~\ref{appdx:extreme-gals}, Fig.~\ref{fig:Stamps-of-extremes} we show one of the most extreme sources in our sample. The source ID=762429 is found at $z_{\rm phot} = 7.11^{+0.31}_{-0.75}$ and log$(M_{\star}/M_{\odot}) = 11.43^{+0.10}_{-0.07}$, with several modes in the $P(z)$ down to $z\sim3.5$ as plausible solutions (with $P(z<5) \approx 5\%$). It has rising flux densities out to F770W and is highly dust attenuated. The system shows an extended morphology and possibly clumpy as indicated by the residual of the smooth S\'ersic model. The rising flux densities can also be fitted with the AGN template but resulting in overall worse $\chi^2_{\rm AGN}$. However, from the clumpy morphology indicated by the residual images, it is possible that this is a merging system and therefore with components with significantly lower masses.

In summary, we find about 40 galaxies with masses that exceed the limits imposed from halo abundances and local-Universe baryon-to-star conversion factors. These massive systems at high redshifts tend to be red and highly dust-attenuated, with $12\%$ of them being detected at sub-mm and/or FIR. As such, they are very difficult cases for SED fitting and their photo-$z$ and stellar mass solution are uncertain. 
These sources are certainly very interesting and warrant further in-depth investigation, since their extreme nature is a powerful probe of the galaxy formation theory. This would need to involve NIR spectroscopy and further sub-mm observations.

\subsection{Galaxy candidates at $10<z<12$}\label{sec:zgt10_candidates}

We select 27 galaxy candidates at $10<z<12$ as described in \S\ref{sec:the_10z12_smf}; Fig.~\ref{fig:zgt10-properties} shows the distribution of some of their properties. Given the relatively shallow depth but over large area ($0.53 \, {\rm deg}^2$), these candidates are brighter than 27.5 AB mag and have a median of 26.7 AB mag in F444W. They have median F277W$-$F444W colors of $\sim 0.08$ and are bluer than 0.4, with one candidate having $m_{\rm F277W} - m_{\rm F444W} \sim 0.7$. Their median redshift is 10.3, and have $60\%$ of their stacked $P(z)$ at $z>10$. Several candidates show a secondary peak at $z\sim2.5$, that has $<32\%$ of the $P(z<9)$, while the stacked $P(z)$ has $2.5\%$ at $z<5$. In this paper, we only analyze these candidates within the statistical context of the SMF (and quantities derived from it). We do not find that their number densities are in tension with limits from $\Lambda$CDM (see also \S\ref{sec:model-comparison} and \S\ref{sec:dm-connection-sfe}), with the caution that they might suffer from incompleteness (cf. \S\ref{sec:the_10z12_smf}). However, these candidates are certainly interesting on their own, and warrant spectroscopic confirmation and in depth investigation.

Finally, we note that in \cite{Casey2024} we report 9 exceptionally luminous and massive galaxy candidates at $10\lesssim z \lesssim 12$ in COSMOS-Web over half of the survey footprint of this paper ($0.25 \, \si{deg}^2$). These are selected such that they have consistent $z_{\rm phot}$ solutions with three different codes: \textsc{LePHARE}, \textsc{eazy} and \textsc{Bagpipes}, as well as with different photometry measurements (e.g., model fitting from \textsc{SE++}, aperture photometry). We find that only one source from our $10<z<12$ sample is in common (source name COS-z10-4 in \citealt{Casey2024}). However, from the rest, all but one have broad and multimodal $P(z)$ that encompasses a $z>9$ solution. This difference is likely from the different \textsc{LePHARE} configuration in this work, tuned to yield robust results for all redshifts and populations (e.g., high allowed attenuation and strong emission lines that can prefer a lower-$z$ solution). Therefore, we take our results cautiously at $z>10$ and the true redshift and nature of these candidates remains to be confirmed with spectroscopy.

\subsection{Comparison to theoretical models} \label{sec:model-comparison}

\begin{figure*}[th!]
  \centering
    \includegraphics[width=1\textwidth]{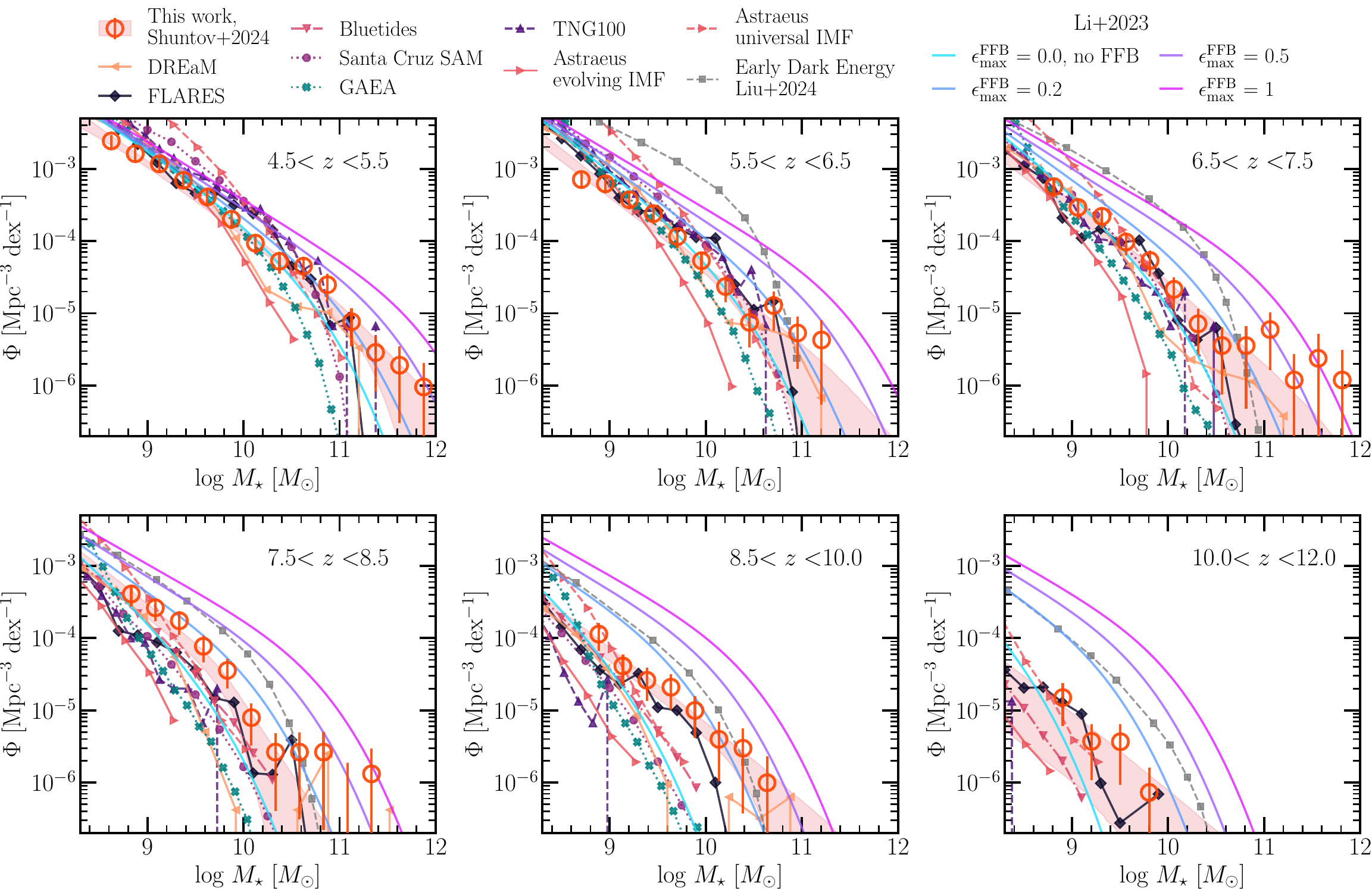}
 \caption{Comparison of the COSMOS-Web SMF to theoretical models. Measurements and best fits from this work are shown in the orange circles and filled region. We compare with the predictions from the feedback-free starbursts (FFB) model by \cite{Dekel2023,Li2023}, which are based of \textsc{UniverseMachine} \citep{Behroozi2019_UniverseMachine} corresponding to different maximum star formation efficiencies (SFE, $\epsilon_{\star}$) in the FFB regime, shown in the solid lines. We also show the SMF from a number of semi-analytical (\textsc{Astraeus}, \citealt{Hutter2021,Cueto2023}; \textsc{Santa Cruz SAM}, \citealt{Yung2019}; \textsc{DREaM}, \citealt{Drakos2022}; \textsc{GAEA} \citealt{DeLucia2024}) and hydrodynamical simulations (\textsc{Bluetides}, \citealt{Feng2016,Wilkins2017};
\textsc{FLARES}, \citealt{Lovell2021, Vijayan2022, Wilkins2023}; \textsc{TNG100}, \citealt{springel_first_2018,nelson_first_2018,pillepich_first_2018,marinacci_first_2018,naiman_first_2018}), and the predictions of the Early Dark Energy model by \cite{Liu2024}.
 }
  \label{fig:smf-compare-literature}
\end{figure*}

In this section, we compare to theoretical models and simulations to interpret some of the possible physical mechanisms behind our observations. Fig.~\ref{fig:smf-compare-literature} compares the SMF from our work to a compilation of semi-analytical (\textsc{Astraeus}, \citealt{Hutter2021,Cueto2023}; \textsc{Santa Cruz SAM}, \citealt{Yung2019}; \textsc{DREaM}, \citealt{Drakos2022}; \textsc{GAEA} \citealt{DeLucia2024}), hydrodynamical simulations (\textsc{Bluetides}, \citealt{Feng2016,Wilkins2017}; \textsc{Flares}, \citealt{Lovell2021, Vijayan2022, Wilkins2023}; \textsc{TNG100}, \citealt{springel_first_2018,nelson_first_2018,pillepich_first_2018,marinacci_first_2018,naiman_first_2018}), the analytical feedback-free starbursts (FFB) \citep{Dekel2023, Li2023}, and the Early Dark Energy (EDE) model \citep{Liu2024}. For brevity, we focus only on the SMF at $z>5$. We show our measurements of the SMF in the orange circles and the $1\,\sigma$ confidence region of the fitted parametric forms. The latter take into account the Eddington bias and therefore represent a more just comparison to the simulations. We note that for this compilation, we compare the SMFs at the median redshift that falls in the redshift bins from our work. This neglects potential differences in the redshift distributions. We note this caveat and do not attempt to correct for it.

\textbf{Comparison with simulations} that use different physical modeling assumptions can shed light on some of the mechanisms that result in the observed SMF. In Fig.~\ref{fig:smf-compare-literature} we compare to a non-exhaustive list of semi-analytical and hydrodynamical simulations. Overall, there is a relatively good agreement between SMFs from our work and the simulations, especially out to $z\sim7$. At $z>7.5$ our measurements are consistently above the simulations. The best agreement out to the highest $z\sim 11$ is with the \textsc{Flares} simulation \citep{Lovell2021, Vijayan2022, Wilkins2023}. This is perhaps unsurprising because \textsc{Flares} consists of a suite of zoom-in hydrodynamical simulations within larger cosmological volumes to study rare, massive structures at $z > 5$. 
The fact that \textsc{Flares} probes the widest range of environments, from extremely underdense voids, to the most overdense high redshift structures, can explain their higher SMF and the closest agreement to ours. \cite{Lovell2021} show a strong dependence of the SMF on the environment and suggest that the higher SMF compared to other observations and simulations at $M_{\star} \gtrsim 10^{10}\, M_{\odot}$ is due to small volume probed by the latter, which does not probe extreme environments that can have a strong impact on the SMF. 

The \textsc{Astraeus} semi-analytical model \citep{Hutter2021} provides insight on how the IMF assumption impacts the resulting SMF. \cite{Cueto2023} compute the SMF using both a universal IMF and one that evolves towards an increasingly top-heavy IMF at higher redshifts. This results in lower mass-to-light ratios and a slower buildup of stellar mass, as shown by the $\sim 1$ dex lower SMF for the evolving IMF compared to the universal IMF (yellow solid and dashed lines in Fig.~\ref{fig:smf-compare-literature}). In our work, we assume a universal \cite{Chabrier03} IMF, so applying an evolving top-heavy IMF would result in lower stellar masses and number densities. However, the SMF from \textsc{Astraeus} shows steeper slopes and increased number densities at the low-mass at all redshifts.

One aspect that these comparisons with simulations highlight is the overabundance of massive galaxies, especially at $z\gtrsim 7$, measured in our work, but also in recent work from JWST \citep[e.g.,][cf. \S\ref{sec:most-extreme}]{Weibel2024}. The surprising abundances of such systems stretch our current theories of galaxy formation and potentially the $\Lambda$CDM model. As a consequence, they can be powerful probes of our theory. In the following, we discuss possible scenarios that can explain these observations.

\subsubsection{Possible physical mechanisms that can produce an overabundance of massive galaxies}

\textbf{The feedback-free starbursts (FFB) model} by \cite{Dekel2023, Li2023} postulates that at early times, the most massive dark matter halos can have a very high star formation efficiency ($\sim 100\%$) due to starbursts free of feedback. These feedback-free bursts can happen in dense star-forming clouds ($\sim 3\times 10^3\, {\rm cm}^{-3}$), when the free fall time is $\lesssim 1$ Myr, shorter than the time for massive stars to develop winds and supernovae feedback \citep{Torrey2017, Grudic2018}. The bursts can last a few Myr and reoccur during a phase of about 100 Myr. The FFB regime is activated above halo mass threshold, dependent on the redshift $M_{\rm h}^{\rm FFB} \sim 10^{10.8} \, [(1+z)/10]^{-6.2}$ \citep
{Dekel2023}, in which the SFE reaches a maximum $\epsilon_{\rm max}^{\rm FFB}$. In the \cite{Li2023} implementation of the FFB, the mean SFR of a galaxy in the FFB regime is modulated by the maximum SFE:  $\langle {\rm SFR}_{\rm FFB} \rangle = \epsilon_{\rm max}^{\rm FFB} \, f_{\rm b} \, \Dot{M}_{\rm h} $, where $\Dot{M}_{\rm h}$ is the halo growth rate. Then stellar masses (and consequently SMF) are obtained by adding the time-integrated SFR contribution from the FFB mode on top of the nominal SFR based on the \cite{Behroozi2019_UniverseMachine} \textsc{UniverseMachine} empirical model \citep[see][for details]{Li2023}. In Fig.~\ref{fig:smf-compare-literature} we show the predictions from the FFB model for different $\epsilon_{\rm max}^{\rm FFB}$, including the no FFB regime of $\epsilon_{\rm max}^{\rm FFB} = 0$, which is essentially the \textsc{UniverseMachine} model. Out to $z\sim7$ our results agree with the no FFB regime; while at higher redshifts our SMF would be consistent with an evolving $\epsilon_{\rm max}^{\rm FFB}$ of $\sim0.1$ at $z\sim8$ to $\sim0.2-0.5$ at $z>8.5$. Additionally, our measurements also suggest a possible mass dependence of the $\epsilon_{\rm max}^{\rm FFB}$ -- the most massive end of our SMF moves towards higher FFB efficiencies. This is not excluded by the FFB model, since the halo mass is a key ingredient and the $\epsilon_{\rm max}^{\rm FFB}$ can, in principle, depend on mass. Therefore, these observations can serve to properly tune the parameters of the FFB model, something that we do not attempt and leave for future work.
However, given the uncertainties in the SMF (and $M_{\star}$) at these redshifts, it is difficult to provide robust constraints from such photometric surveys alone.

\textbf{Stochastic bursts of star formation} in the early phases of high-$z$ galaxies can produce rapid stellar mass growth. These bursts can happen on a scale of about 5 Myr and can reoccur in a time span of about 100 Myr, leading to an increased integrated SFE ($\epsilon_{\star}$), qualitatively consistent with our observations of the SMF. 
This hypothesis has been studied theoretically \cite{PallottiniFerrara2023}, and is found to explain the observed abundances of bright galaxies at $z>10$ \citep{Shen2023, Sun2023}. The stochastic and bursty SFH is also consistent with the FFB model \citep{Dekel2023,Li2023}. The transition to stochastic star formation around $z=9$ has also been identified observationally by \cite{Ciesla23b}, showing that at $z>9$ about $87\%$ of massive galaxies have evidence for a stochastic star formation in the last 100 Myr.

Bursty star formation histories can produce a scatter in galaxy UV luminosities at a given halo mass, leading to a dispersion in the $M_{\rm UV}-M_{\rm h}$ relation that is found to increase with decreasing halo mass \citep{Sun2023}. According to this model, the $M_{\rm UV}-M_{\rm h}$ scatter can reproduce increased abundances of luminous galaxies without the need of enhanced star-formation efficiencies.
\cite{Gelli2024} implement such a model and find that it can successfully match the observations of the UVLF and $\rho_{\rm SFR}$ up to $z\sim12$, using a constant star-formation efficiency, while it still falls short at $z\sim14$. Their prediction on the $\rho_{\rm SFR}$ is shown in the purple curve in Fig.~\ref{fig:smd-sfrd}. However, this model remains to be extended to stellar masses and compared against SMF. Clustering analysis in COSMOS-Web by Paquereau et al. in prep. measures galaxy bias that tends to disfavor a scenario with high shochasticity and scatter.

\textbf{Positive feedback} from supermassive black hole (SMBH)-driven AGN in the very early Universe can considerably increase the star formation efficiency, as argued by \cite{Silk2024}. In this hypothesis, star-forming galaxies at $z\sim10$ can host an AGN that enhances gas accretion onto both star-forming regions and the central SMBH. This, in turn, causes momentum-conserving AGN outflows and radiatively cooled turbulence, which, coupled with efficient cooling of the shocked gas due to the ultracompact galaxy, leads to a dense and cool phase with increased star formation. 
This positive feedback from AGN causes the first episodes of vigorous star formation at $z\gtrsim10$, which then transitions to negative feedback and quenching due to gas depletion by energy-conserving AGN outflows at $z\lesssim6$. The positive feedback phase could be active in compact galaxies that have their central AGN obscured \citep{Silk2024}. Therefore, this scenario is also consistently linked to the emerging abundant population of compact AGN-dominated sources at $z>5$ with massive ($10^{7}-10^{9} \, M_{\odot}$) black holes \citep{Labbe2023, Greene2024, Matthee2024, Kokorev2024}. 

Qualitatively, our measurements are consistent with the positive feedback scenario. The transition from the Schecter function to the double power law, as well as the monotonic increase with mass of the integrated star formation efficiency close to $100\%$ (discussed in \S\ref{sec:dm-connection-sfe}) at $z\sim6$, all coincide with the transition from negative to positive AGN feedback.
However, adequate data to constrain the parameters of this unified theory of the coevolution of SMBH and galaxies is still lacking, and details on how the positive feedback scenario would reflect on galaxy and AGN abundances (the former being the scope of this paper) are not yet developed. One way to detect the hidden momentum-driven positive AGN feedback according to \cite{Silk2024} is indirectly by searching for very high specific star formation rates. This would require follow-up spectroscopy of the most massive high-$z$ candidates in this work.

\textbf{Modifications of the $\Lambda$CDM cosmology} are another alternative explanation. One proponent is the Early Dark Energy (EDE) model that can produce higher abundance of massive halos \citep{Klypin2021}. This excess number of massive halos can host galaxies assembled with a 
more modest $\epsilon_{\star}$ than the higher one implied by our measurements within the standard cosmology. This can explain the observed abundance of massive galaxies without considerable modification to the galaxy physics that regulates the SFE. \cite{Liu2024} have studied the implications of the EDE model on the SMF, showing that it indeed predicts increased abundances. These predictions are shown in the gray lines and square symbols in Fig.~\ref{fig:smf-compare-literature}. This model largely overpredicts the SMF out to $z\sim9$ but agrees at $z>10$. However, in their application, \cite{Liu2024} assume a stellar-to-halo mass relation (SHMR) calibrated on pre-\JWST{} data \citep[by][obtained from abundance matching]{Stefanon21}, which as we show in the next section can be very uncertain \citep[see also][]{Shuntov2022}. Accurately constraining the SHMR requires the implementation of halo occupation distribution (HOD) models that need to predict both galaxy abundance and clustering. Therefore, coupling clustering measurements to the abundances is necessary to provide more meaningful constraints on the EDE model.

\subsection{Connection to dark matter halos and the inferred star formation efficiency} \label{sec:dm-connection-sfe}

\begin{figure}[t!]
  \centering
    \includegraphics[width=1\columnwidth]{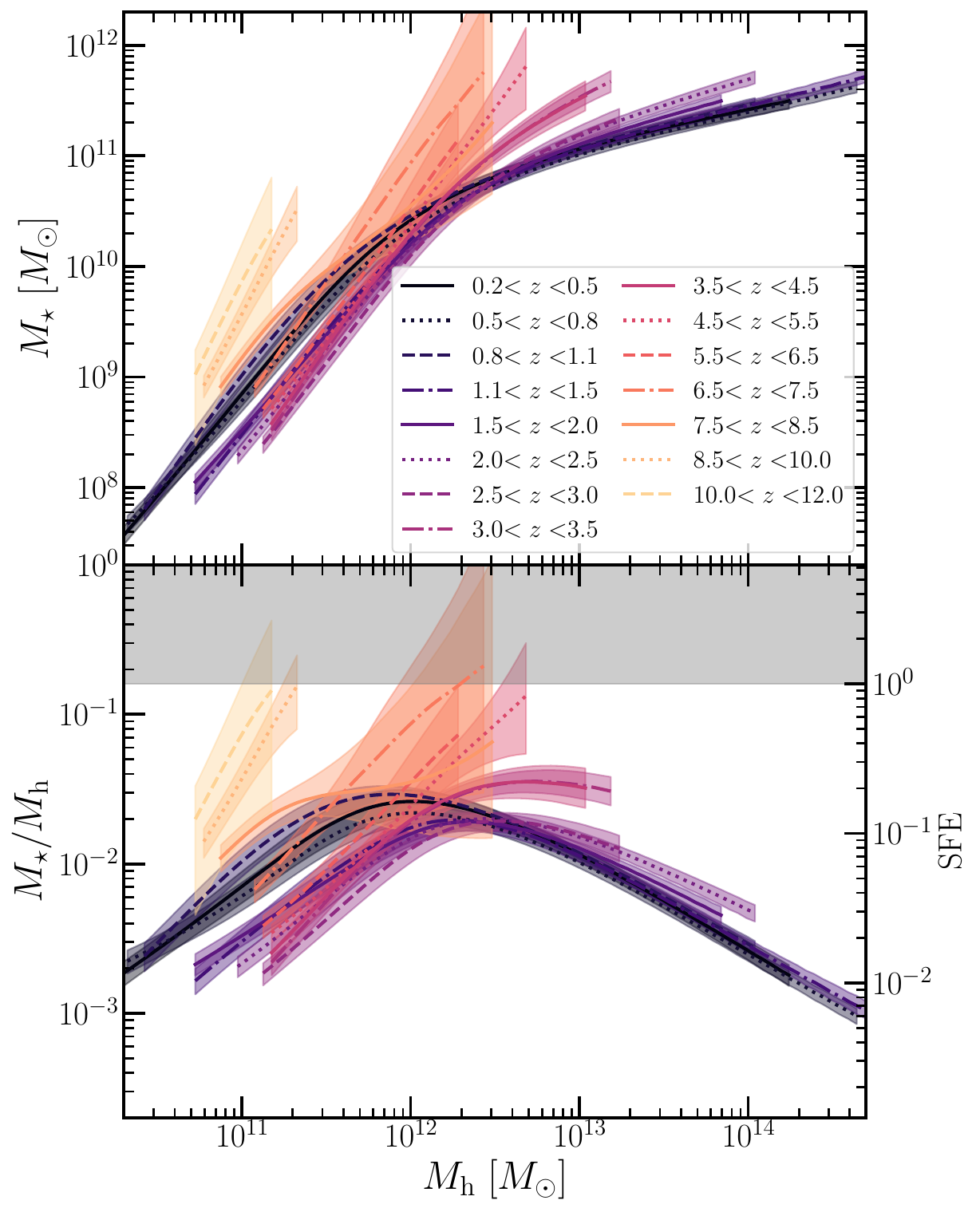}
 \caption{Stellar-to-halo mass relation for each of the redshift bins in this work. This is obtained by doing abundance matching using the best-fit functional forms of the SMF and the HMF by \cite{2013MNRAS.433.1230W}. The top panel shows stellar mass as a function of the inferred halo mass, while the bottom panel shows the ratio between the two as a function of halo mass. The right axis indicates the star formation efficiency (SFE), $\epsilon_{\star} = M_{\star}\,M_{\rm h}^{-1}\,f_b^{-1}$, and the gray shaded region marks $\epsilon_{\star}\geq100\%$. The lines and their confidence intervals are shown only in the $M_{\star}$ range probed by our SMF measurements.
 }
  \label{fig:Ms-Mh_SHMR_allz}
\end{figure}

In this section, we use the SMF measurements to explore the connection between galaxies and dark matter. Galaxies grow in dark matter halos through star formation by converting the available baryonic gas, given by the cosmic baryon fraction $f_{\rm b}\approx0.16$, and through merging. This results in a relation between the stellar and the halo mass $M_{\star}=\epsilon_{\star}\,f_{\rm b}\,M_{\rm h}$, referred to as the {\it stellar-to-halo mass relation} \citep[SHMR, see][for a review]{wechsler_connection_2018}. Correspondingly, this establishes a relation between the cumulative number densities of galaxies and halos $ \Phi_{\star}(M_{\star}, z) = \Phi_{\rm h}(M_{\star}\, f_{\rm b}^{-1}\, \epsilon_{\star}^{-1},z)$ \citep{BehrooziSilk2018}.
$\epsilon_{\star}$ is the star-formation efficiency (SFE) integrated over the lifetime of the halo\footnote{Here, it is important to make the distinction between the integrated SFE $\epsilon_{\star} = M_{\star}/(M_{\rm h}\, f_{\rm b})$ that we use in this work, instantaneous SFE $\epsilon = {\rm SFR}/(f_{\rm b}\, \Dot{M}_{\rm h})$, and the SFE in the context of gas depletion $\epsilon_{\rm gas} = {\rm SFR}/M_{\rm gas}$. Given the fact that ${\rm SHMR} \equiv M_{\star}/M_{\rm h} = \epsilon_{\star}\, f_{\rm b}$, we use the SMHR and $\epsilon_{\star}$ interchangeably to refer to the integrated SFE.}.  In this paper, we provide some of the first systematic quantification of the integrated SFE from \JWST{}, down to the earliest ages of the Universe. Additionally, a study by Paquereau et al. in prep provides complementary constraints from HOD modeling of galaxy clustering in COSMOS-Web.

\begin{figure*}[t!]
  \centering
    \includegraphics[width=1\textwidth]{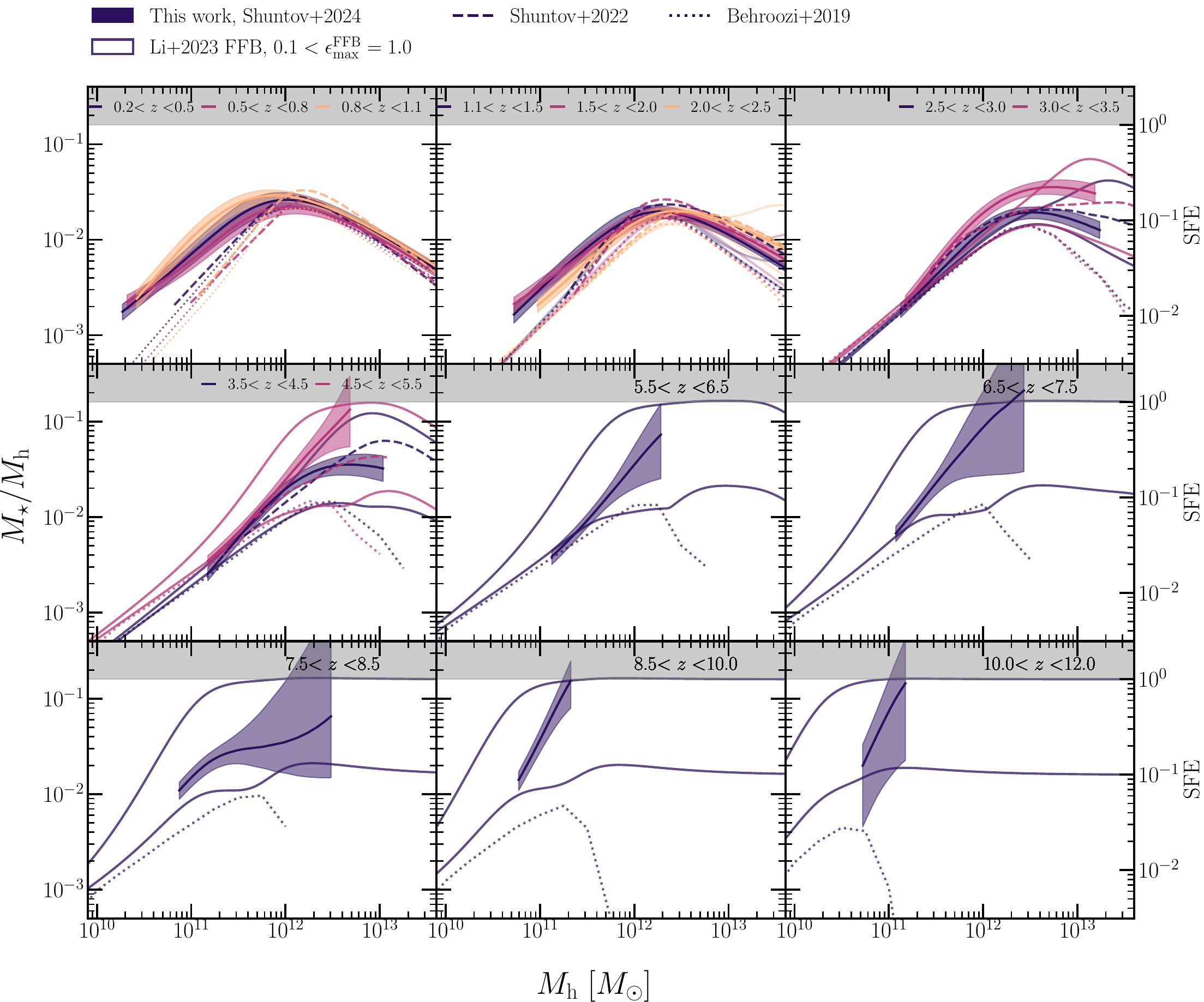}
 \caption{ Stellar-to-halo mass ratio (SHMR) and the implied star formation efficiency in different redshift bins. The right axis indicates the star formation efficiency (SFE), $\epsilon = M_{\star}\,M_{\rm h}^{-1}\,f_b^{-1}$, and the gray shaded region marks $\epsilon=100\%$. Each panel shows the SHMR in several redshift bins from our work (solid lines with $1\,\sigma$ uncertainty envelope), and literature works from \citet[][dashed]{Shuntov2022}, \citet[][dotted]{Behroozi2019_UniverseMachine}, and \citet[][points]{Stefanon21}. The FFB predictions from \cite{Li2023} are shown in the solid line envelope that encloses maximum SFE in the FFB regime in the $0.1 - 1.0$ range.
 }
  \label{fig:shmr}
\end{figure*}

We infer the SHMR by carrying out non-parametric {\it abundance matching} \citep[AM,][]{marinoni_mass--light_2002, kravtsov_dark_2004,ValeOstriker2004,Tasitsiomi2004}. For each $z$-bin, we match the cumulative forms of best-fit SMFs (presented in \S\ref{subsec:intrinsic-SMF}) and the \cite{2013MNRAS.433.1230W} halo mass functions (HMF) to assign $M_{\rm h}$ to each $M_{\star}$. We note that this is a very simplistic and only a first-order implementation of AM using directly the HMF and does not consider any stochasticity in matching galaxies and halos. In principle, other halo properties, such as the maximal halo mass ($M_{\rm h, max}$) over the halo's lifetime, the peak maximum velocity of the particles in the halo across its formation history ($V_{\rm peak}$), or the maximum circular velocity at the time of accretion $v_{\rm acc}$, correlate better with the total baryonic content of halos and better reproduce clustering properties  \cite[e.g.,][]{Reddick2013}. This is because (sub)halos can be subject of significant stripping, while the galaxy inside can retain the stellar mass.
However, this should have an increasing impact on lower redshift, which is not so much the focus of this work, since the SMHR has been robustly constrained at lower $z$ by numerous other works. Moreover, \cite{Stefanon21}, for example, have shown that at high $z$, the inferred halo masses from AM of the SMF to the HMF directly, differ only by $\lesssim 0.04$ dex from those inferred from matching to $v_{\rm acc}$. Therefore, we adopt this simplistic procedure for this paper. A more detailed analysis of the SHMR using clustering and HOD modeling will be presented in Paquereau et al. in prep.

\textbf{The stellar-to-halo mass relation} resulting from our work is shown in Fig~\ref{fig:Ms-Mh_SHMR_allz} for each of the 15 $z$-bins. The top panel shows the relation between $M_{\star}$ and $M_{\rm h}$, while the bottom shows the ratio between the two masses $M_{\star}/M_{\rm h}$ as a function of halo mass.  The right $y$-axis of the bottom panel  shows the integrated SFE, $\epsilon_{\star} = M_{\star}/(M_{\rm h}\, f_{\rm b})$.  We limit the mass range to the minimum and maximum $M_{\star}$ probed by our survey. Fig~\ref{fig:Ms-Mh_SHMR_allz} shows the characteristic strong dependence of the SHMR on $M_{\rm h}$,  with the SFE remaining in the range of $0.02-0.2$ for all masses and out to $z\sim7$.  The peak of the SHMR is reached at $M_{\rm h}^{\rm peak}\approx10^{12}\, M_{\odot}$, $M_{\star}^{\rm peak}\approx2\times 10^{10}\, M_{\odot}$, and $\epsilon^{\rm peak}_{\star}\approx 0.2$ at $z<2$ . The SHMR decreases at low (due to stellar and SNe feedback) and high masses \citep[due to AGN feedback, ][]{silk_current_2012}, with the slopes at both ends remain roughly constant with $z$. The SHMR, including the peak, shifts to lower maximum SFE and higher halo masses out to $z\sim3.5$, after which the limit in maximum stellar mass and sample size prevent us from robustly establishing the peak; this is consistent with previous studies in COSMOS \citep{Legrand2019, Shuntov2022}. After $z\gtrsim 3.5$, there is an upturn, and the SHMR sharply increases and monotonically approaches $\epsilon_{\star} \sim 0.8-1$ at the highest masses and redshifts, albeit with large uncertainties. Within the uncertainties, none of our measurements enter the regime of $\epsilon_{\star}>1$ that signifies more stellar mass than available baryons. Therefore, these results do not suggest any significant tension with $\Lambda$CDM.

\textbf{An evolving SFE and the upturn at $\mathbf{z\sim3.5}$} is an important feature our analysis reveals (see also Paquereau et al. in prep.), because it provides an observational imprint of all the processes involved in regulating galaxy growth. It suggests that the SFE is not constant and that galaxies and halos do not grow at the same rate over cosmic history. A relatively simple explanation can be given in the framework of the gas-regulated model of galaxies, closely coupled with halo evolution \citep{Lilly2013}. The cosmic SFE is tightly linked to the interplay between the halo growth rate (that quantifies the growth through accretion of dark matter) and the star-formation rate (that quantifies the conversion of the gas reservoir into stars). 

Our results indicate that from the earliest times until $z\sim3.5$, the halo growth rate $\Dot{M_{\rm h}}$ outpaces the SFR of the residing galaxies, causing the decrease of SFE with time. The halo growth rate slows down, while the gas reservoirs built over time in the halo can keep the SFR going and at $z\lesssim3.5$, galaxies outpace halo growth, resulting in the increase of the SFE. This is especially the case at $M_{\rm h} \lesssim 10^{12} \, M_{\odot}$, in the regime where halo mass quenching has not occurred.
Indeed, Paquereau et al. in prep analyze the ratio ${\rm SFR}/\Dot{M_{\rm h}}$ in \textsc{UniverseMachine} and find that it decreases from $0<z<3.8$ turns at $z\sim3.8$ and increases at higher redshifts -- consistent with our observations and interpretations.

Figure~\ref{fig:shmr} shows another rendering of our SHMR measurements, where redshift bins are separated in different panels for clarity, and where we compare with \cite{Shuntov2022}, and the \textsc{UniverseMachine} and FFB models. In general, there is very good agreement at lower redshifts, with a notable exception in the first three bins at $z<1.1$. Our SHMR is consistently higher by $\sim 0.4$ dex at the low mass end, and the peak is at lower halo mass compared to \textsc{UniverseMachine} and \cite{Shuntov2022}. This can be a systematic from the simplistic AM using the $M_{\rm h}$, as opposed to other more tightly correlated halo properties. At higher redshifts (out to $z\sim7$) the low mass end remains in good agreement, with the high mass end being increasingly higher compared to \textsc{UniverseMachine}, but in agreement with \cite{Shuntov2022} within the uncertainties. The sharp rise of the SHMR after $z\gtrsim4$ in our work is in striking contrast to the \textsc{UniverseMachine} model that shows a decrease.
The FFB predictions are shown in the transparent envelope that enclose maximum SFE in the range of $0.1 < \epsilon_{\star}^{\rm FFB} < 1.0$. The FFB regime becomes important at the very high masses at lower redshifts and moves towards affecting lower masses at high redshifts, where it can reach maximum SFE at $M_{\rm h} \approx 10^{12}\, M_{\odot}$ at $z\sim 6$ and $M_{\rm h} \approx 10^{11}\, M_{\odot}$ by $z\sim 10$. In all cases, our measurements are enclosed by the $0.1 < \epsilon_{\star}^{\rm FFB} < 1.0$ FFB predictions. However, the increase of our SHMR towards larger values and potentially its steepening with redshift indicates that $\epsilon_{\star}^{\rm FFB}$ should be increasing with both redshift and halo mass.

\begin{figure*}[ht!]
  \centering
  \begin{subfigure}{0.49\textwidth}
    \includegraphics[width=\linewidth]{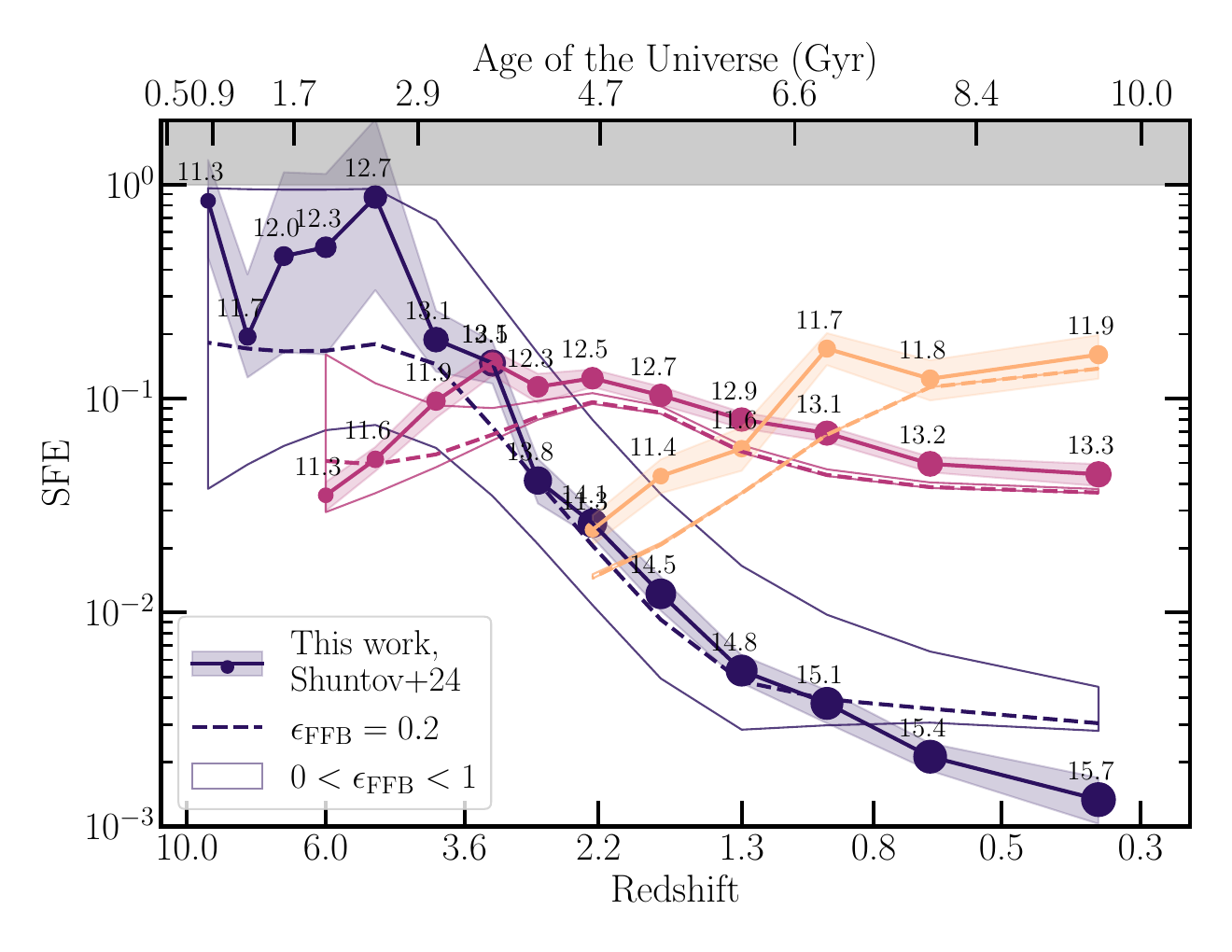}
  \end{subfigure}
  \hfill
  \begin{subfigure}{0.49\textwidth}
    \includegraphics[width=\linewidth]{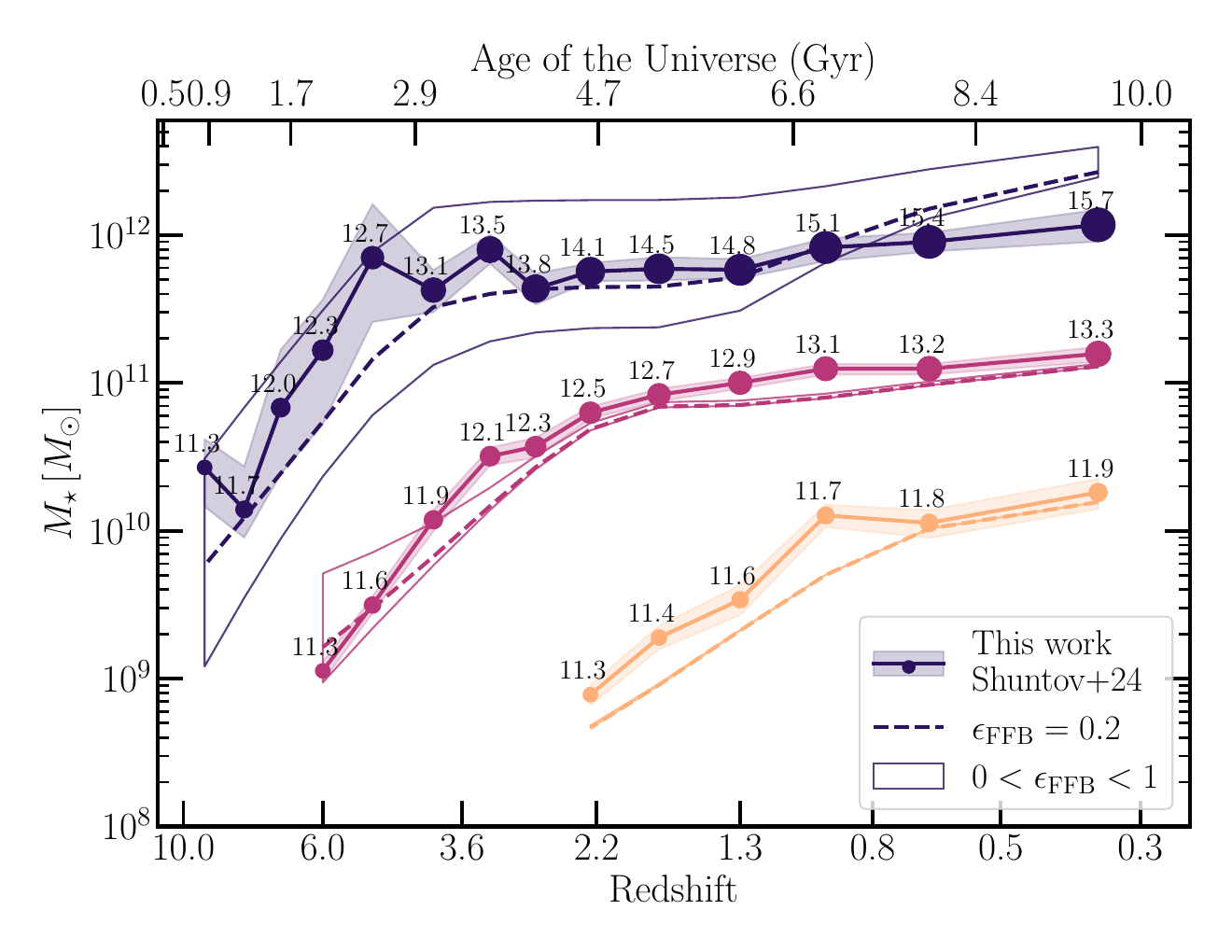}
  \end{subfigure}
  \caption{{\bf Left:} Evolution of the star formation efficiency in dark matter halos, tracing the halo growth with cosmic time. The different color-coded points show three different halos of the same initial mass, $M_{\rm h} = 2\times10^{11}\, M_{\odot}$, starting their evolution at $z=9.3$, $z=6.0$ and $z=2.25$ respectively. The halo mass growth follows \cite{Dekel2013} parametrization, and the SFE is computed from the SHMR at the corresponding redshift and halo mass. The size of the points corresponds to the halo mass, also annotated above each point (in log scale).
  {\bf Right:} Stellar mass growth in the same halos as the left panel, where $M_{\star } = M_{\rm h}\, \epsilon_{\star}\, f_{\rm b}$. The envelope corresponds to the $1\, \sigma$ confidence interval propagated from the uncertainty in the SHMR.
  The dashed lines are computed in the same way, using the resulting SHMR from the FFB model at $\epsilon_{\star}^{\rm FFB}=0.2$, while the transparent envelope encloses between no FFB and maximum efficiency $0.0 < \epsilon_{\star}^{\rm FFB} < 1.0$ \citep{Liu2024}. This shows that halos of $10^{11}\, M_{\odot}$ starting their growth at $z>3.5$ would be more efficient in assembling stellar mass by the time they reach $10^{12}\, M_{\odot}$.
  }
  \label{fig:tracing-halo-growth}
\end{figure*}

\subsection{Tracing the SFE and stellar mass growth throughout the halo history.}
We use our results on the SHMR coupled with evolutionary tracks of the halo growth to study the evolution of the SFE and stellar mass within a given halo throughout cosmic history. For this purpose, we use the \cite{Dekel2013} halo growth parametrization to compute the evolution of $M_{\rm h}$ with redshift. We choose three halos of same initial mass $M_{\rm h} = 2\times10^{11}\, M_{\odot}$ starting their evolution at $z=9.3$, $z=6.0$ and $z=2.25$ respectively. Then at the median of each redshift bin, we compute the $M_{\rm h}(z)$ and take the SFE at that $M_{\rm h}$ from our SHMR measurements. We chose $M_{\rm h} = 2\times10^{11}\, M_{\odot}$ because it is the only mass probed at all redshifts (Fig.~\ref{fig:Ms-Mh_SHMR_allz}).

In Fig.~\ref{fig:tracing-halo-growth} (left) we plot the evolution of the SFE in the three halos as a function of their mass and redshift. The $M_{\rm h} = 2\times10^{11}\, M_{\odot}$ halo starting at $z=9.3$ is massive for its epoch and has a very high SFE $\sim0.8$, meaning is efficiently converting most of the available baryons to stars. Since the halo growth is proportional to its mass, it grows rapidly, reaches $M_{\rm h}\approx 10^{12.5}\, M_{\odot}$ and enters into the hot halo mode already at $z\sim5$. This results in a sharp decline in the SFE down to $0.001$ at $z\sim0.3$ in a supercluster halo of $M_{\rm h} \approx 5 \times 10^{15} \, M_{\odot}$. A halo of the same initial mass starting its growth at $z\sim6$, has a lower starting SFE that remains roughly constant, with only a mild increase of about 0.5 dex until cosmic noon, reaching $\epsilon_{\star}\approx 0.1$, and then decreases by about the same amount since. In both scenarios the downturn in SFE happens after the halo has reached a similar mass-scale of $\approx 10^{12.5} \, M_{\odot}$, consistent with hot halo powered by AGN feedback \citep{gabor_hot_2015}. This is somehow higher than the characteristic $\sim10^{12} \, M_{\odot}$ \citep{Cattaneo06}, but consistent with the cold stream paradigm -- at higher redshifts, cold streams can efficiently penetrate massive hot halos and fuel an increased SFE \citep{Dekel2009}. In the third case, the halo of the same initial mass at $z\sim2.3$ has the lowest starting SFE $\sim 0.02$, which increases steadily to $\epsilon_{\star} \sim 0.2$, a result of the slow halo growth of only about $0.6$ dex in  $\sim5.3$ Gyr.

In Fig.~\ref{fig:tracing-halo-growth} (right) we show the stellar mass growth in the three halos, which we obtain in the same way as in the left panel, using $M_{\star } = M_{\rm h}\, \epsilon_{\star}\, f_{\rm b}$. The high SFE at $z\sim9$ means high stellar content ($\sim 2\times 10^{10} \, M_{\odot}$) that sharply increases by about 1.5 dex in $\sim3$ Gyr after which it stagnates and reaches a maximum of $\sim 10^{12} \, M_{\odot}$. This would mean that in such an early and massive halo, a galaxy as massive as some of the brightest cluster galaxies (BCG) in the local Universe \citep{Bellstedt2016} would be assembled by $z\sim3.5$. However, in these massive halos, new stellar mass likely accumulates in satellite galaxies, which is not reflected in the SHMR from AM that only considers central galaxies. The more moderate but constant SFE of the $z\sim6$ halo shows a longer and steadier stellar mass buildup of almost 2 dex until $z\sim 2$ and a slower growth of about 0.2 dex since, to reach $10^{12}\, M_{\odot}$ at $z\sim0.3$. In this case, the galaxy and its host halo grow at the same rate. Finally, in the halo starting at $z\sim2.2$, the stellar growth outpaces the halo, thanks to the increasing SFE, with 0.6 vs 1.5 dex increase in the halo and stellar mass respectively. These trends are a signature of the downsizing scenario \citep{de_lucia_formation_2006, thomas_environment_2010} -- the massive and early-type galaxies today have stellar populations that are formed earlier in a short period of high SFE in some of the most massive halos for its epoch.

The shape evolutionary tracks we obtain in this analysis are a result of the evolving SFE, the upturn at $z\sim3.5$ and galaxies not closely following halo growth. We can see that the SFE of a $10^{12}\, M_{\odot}$ halo decreases from the $z=10$ to the $z=6$ track, but then increases again for the $z=2$ track. Consequently, a halo starting its growth at $z>3.5$ would assemble the most stellar mass by the time it reaches $M_{\rm h} \approx 10^{12.5}\, M_{\odot}$, compared to similar halos that start their growth earlier. This can be explained within the same gas-regulated model coupled with halo evolution framework that we discussed in the previous subsection, where after $z\sim3.5$ the stagnating halo growth and the available gas reservoir keep the SFE high.

In Fig~\ref{fig:tracing-halo-growth} we also compare with the predictions from the FFB model, obtained in the same way. The transparent envelope encloses between the no FFB regime and maximum FFB efficiency $0.0 < \epsilon_{\star}^{\rm FFB} < 1.0$. The FFB, as expected, and has the most impact in the halo that starts its growth early. Our results for the earliest are in best agreement with $\epsilon_{\star}^{\rm FFB} = 0.2$. For the halos at lower redshifts, the FFB has no impact and the underlying \textsc{UniverseMachine} model is consistently lower by 0.1-0.3 dex. However, as discussed previously, the highest redshift points do suggest an increasing $\epsilon_{\star}^{\rm FFB}$.

We have to note that beyond $M_{\rm h} \sim 10^{14}$, these results are based on extrapolations of the SHMR at high masses. Furthermore, the SHMR from our simple AM application relates only the mass of the (sub)halo to the mass of the {\it central} galaxy. It has been shown that beyond the cluster scale halo mass of $M_{\rm h} \gtrsim 10^{13}\, M_{\odot}$, {\it satellite} galaxies dominate the stellar mass budget of the halo \citep{leauthaud_new_2012, coupon_galaxy-halo_2015, Shuntov2022}. Therefore, the resulting SFE would be higher at the high-mass end if satellites were considered in the analysis.

\section{Conclusions}\label{sec:conclusions}
In this paper, we have presented measurements of the SMF from \JWST{} in $98\%$ of cosmic history ($0.2 < z <12.0$). We leverage the largest contiguous \JWST{} imaging survey over $\sim0.5 \, {\rm deg}^2$, COSMOS-Web, along with the wealth of ground- and space-based data in COSMOS to construct complete samples that probe a very large dynamic range in stellar mass. The large area allows us to study the evolution of some of the most massive and rare galaxies in the Universe. We summarize our main findings as follows.\\

\noindent\textbf{Evolution of the SMF}
\begin{itemize}
    \item The normalization of the GSMF monotonically decreases from $z=0.2$ to $z=12$, with a strong mass-dependent evolution. The density evolution at masses just below the knee (log$M_{\star}/M_{\odot} \sim 10.4$) is faster than the low-mass (log$M_{\star}/M_{\odot} \sim 7$) with $\sim 3$ dex and $\sim 2$ dex change since $z \sim 7$, respectively. Beyond the knee, (log$M_{\star}/M_{\odot} \gtrsim 11$) the number densities are within 1.5 dex since $z\sim5$.
    
    \item Our measurements show an excellent agreement with pre-\JWST{} results at low and intermediate redshift, and good agreement with the latest \JWST{} results from deeper but smaller surveys. Our results corroborate the findings of increased abundances of massive galaxies at high redshifts.

    \item The Double Schechter function provides the best fit out to $z=3.5$, the single Schechter out to $z=5.5$ and the double power law at $z>5.5$, due to a flattening of the high mass end. This flattening indicates that the most massive galaxies have assembled very efficiently in the first few Gyr and did not grow significantly, suggesting an onset of negative feedback in massive galaxies at $z\sim5.5$.
    
\end{itemize}

\noindent\textbf{Evolution of the SMD and the inferred SFRD}

\begin{itemize}
    \item The cosmic SMD shows a steady increase with redshift at $1<z<9$, with no significant change in slope, indicating that stellar mass has assembled at a constant rate as a function of redshift, throughout most of cosmic history. 

    \item We infer the cosmic SFRD by integrating and fitting a parametric form to the SMD measurements. At $z\lesssim 3.5$, we find lower values of the SFRD inferred from the SMD compared to instantaneous SFR indicators, corroborating the tension between the two repeatedly reported in the past. The causes remain poorly understood, the most likely being uncertain SFR calibration factors, effects of dust attenuation and SED modeling systematics on stellar mass and SFR measurements.

    \item At $z>7.5$ we find good agreement with recent \JWST{} UVLF measurements.
    This remarkable consistency solidifies the emerging picture of rapid galaxy formation leading to increased abundances of bright and massive galaxies in the first $\sim 0.7$ Gyrs.

\end{itemize}

\noindent\textbf{The most massive galaxies in COSMOS}

\begin{itemize}
    \item We apply the EVS to search for the most extreme sources for their epoch, and find about 40 galaxies with masses that exceed the limits from halo abundances and local-Universe SFE factors. These are red, extended and likely highly dust-attenuated, with properties difficult to estimate from optical-NIR SED fitting, and as such warrant further investigation. However, we do not report any tension with $\Lambda$CDM within the confidence intervals.
    
\end{itemize}

\noindent\textbf{Stellar-to-halo mass relation and the integrated star formation efficiency}
\begin{itemize}
    \item Using abundance matching, we infer the SHMR in 15 redshift bins from $z=0.2$ to $z\sim12$, finding a non-monotonic evolution suggesting that galaxies and halos do not grow at the same rate throughout cosmic history. The integrated SFE has the characteristic strong dependence with mass and mildly decreasing trend at the low mass end out to $z\sim 3.5$ in the range of $ 0.02 < \epsilon_{\star} < 0.2$.

    \item After $z>3.5$ the SFE increases sharply, going from $\epsilon_{\star} \approx 0.1$ to $\epsilon_{\star} \approx 0.8-1$ at $z\approx10$ for log$(M_{\rm h}/M_{\odot}) \approx 11.5$, albeit with large uncertainties. The upturn at $z\sim3.5$ indicates that halo growth rate dominates over the SFR until $z\sim3.5$ when it slows down, but accumulated gas reservoirs can keep the SFR high until present day, resulting in rising SFE.

    \item The SHMR, coupled with halo growth curves, shows that a halo of initial mass $M_{\rm h} = 10^{11}\, M_{\odot}$ starting its growth at $z<3.5$ would assemble the most stellar mass by the time it reaches $M_{\rm h} = 10^{12.5}\, M_{\odot}$. This suggests that the stagnating halo growth after $z\sim3.5$ and the available gas reservoir that results in efficient SF for longer periods.

\end{itemize}

\noindent\textbf{Implications for galaxy formation models}
\begin{itemize}
    \item The comparison of the SMF with simulation shows a good agreement out to $z\sim6-7$; at higher redshifts our measurements show increasingly higher abundances with both mass and redshift.

    \item We investigate several proposed theoretical models that explain the increased abundances at high redshift. We find that the FFB model can successfully reproduce our observations, but requires increasing maximum FFB efficiency with both redshift (from $z>6$) and mass. Furthermore, qualitatively, our results are also consistent with the positive feedback scenario, as indicated by the Schechter to DPL transition at $z\sim6$ and the monotonic increase of the SFE close to $80-100\%$ with mass.

\end{itemize}

To make our results transparent and facilitate comparison we provide all our measurements in tabulated form at \url{https://github.com/mShuntov/SMF_in_COSMOS-Web_Shuntov2024}.

\noindent\textbf{Future prospects}.
Confirmation of the high redshift candidates from spectroscopy will be crucial to accurately measure abundances of massive galaxies and put robust constraints on the emerging theoretical models. We identify that the way forward is a complete spectroscopic survey of massive galaxies at high redshift in the COSMOS, that will also allow robust clustering measurements deep into EoR. HOD analysis of the clustering and abundance measurements will put robust constrains on the host halo mass and galaxy bias at these redshifts, and unveil the true nature of the increased abundance of massive galaxies. 

\begin{acknowledgements}

    The Cosmic Dawn Center (DAWN) is funded by the Danish National Research Foundation (DNRF140). This work has received funding from the Swiss State Secretariat for Education, Research, and Innovation (SERI) under contract number MB22.00072
    This work was made possible thanks to the CANDIDE cluster at the Institut d’Astrophysique de Paris, which was funded through grants from the PNCG, CNES, DIM-ACAV, and the Cosmic Dawn Center; CANDIDE is maintained by S. Rouberol. The French contingent of the COSMOS team is partly supported by the Centre National d’Etudes Spatiales (CNES). OI acknowledges the funding of the French Agence Nationale de la Recherche for the project iMAGE (grant ANR-22-CE31-0007). 
    Some of the data products presented herein were retrieved from the Dawn JWST Archive (DJA). DJA is an initiative of the Cosmic Dawn Center, which is funded by the Danish National Research Foundation under grant DNRF140.
    Based in part on observations collected at the European Southern Observatory under ESO programmes 179.A-2005 and 198.A-2003 and on data obtained from the ESO Science Archive Facility with DOI https://doi.org/10.18727/archive/52, and on data products produced by CALET and the Cambridge Astronomy Survey Unit on behalf of the UltraVISTA consortium.

\end{acknowledgements}

%
%

\bibliographystyle{aa_url}
\bibliography{biblio.bib}

\begin{thebibliography}{207}
\expandafter\ifx\csname natexlab\endcsname\relax\def\natexlab#1{#1}\fi

\bibitem[{{Aihara} {et~al.}(2022){Aihara}, {AlSayyad}, {Ando}, {Armstrong}, {Bosch}, {Egami}, {Furusawa}, {Furusawa}, {Harasawa}, {Harikane}, {Hsieh}, {Ikeda}, {Ito}, {Iwata}, {Kodama}, {Koike}, {Kokubo}, {Komiyama}, {Li}, {Liang}, {Lin}, {Lupton}, {Lust}, {MacArthur}, {Mawatari}, {Mineo}, {Miyatake}, {Miyazaki}, {More}, {Morishima}, {Murayama}, {Nakajima}, {Nakata}, {Nishizawa}, {Oguri}, {Okabe}, {Okura}, {Ono}, {Osato}, {Ouchi}, {Pan}, {Plazas Malag{\'o}n}, {Price}, {Reed}, {Rykoff}, {Shibuya}, {Simunovic}, {Strauss}, {Sugimori}, {Suto}, {Suzuki}, {Takada}, {Takagi}, {Takata}, {Takita}, {Tanaka}, {Tang}, {Taranu}, {Terai}, {Toba}, {Turner}, {Uchiyama}, {Vijarnwannaluk}, {Waters}, {Yamada}, {Yamamoto}, \& {Yamashita}}]{Aihara21}
{Aihara}, H., {AlSayyad}, Y., {Ando}, M., {et~al.} 2022, \href{http://dx.doi.org/10.1093/pasj/psab122}{\color{magenta}\pasj}, \href{https://ui.adsabs.harvard.edu/abs/2022PASJ...74..247A}{74, 247}

\bibitem[{Aihara {et~al.}(2018)Aihara, Arimoto, Armstrong, Arnouts, Bahcall, Bickerton, Bosch, Bundy, Capak, Chan, Chiba, Coupon, Egami, Enoki, Finet, Fujimori, Fujimoto, Furusawa, Furusawa, Goto, Goulding, Greco, Greene, Gunn, Hamana, Harikane, Hashimoto, Hattori, Hayashi, Hayashi, Hełminiak, Higuchi, Hikage, Ho, Hsieh, Huang, Huang, Ikeda, Imanishi, Inoue, Iwasawa, Iwata, Jaelani, Jian, Kamata, Karoji, Kashikawa, Katayama, Kawanomoto, Kayo, Koda, Koike, Kojima, Komiyama, Konno, Koshida, Koyama, Kusakabe, Leauthaud, Lee, Lin, Lin, Lupton, Mandelbaum, Matsuoka, Medezinski, Mineo, Miyama, Miyatake, Miyazaki, Momose, More, More, Moritani, Moriya, Morokuma, Mukae, Murata, Murayama, Nagao, Nakata, Niida, Niikura, Nishizawa, Obuchi, Oguri, Oishi, Okabe, Okamoto, Okura, Ono, Onodera, Onoue, Osato, Ouchi, Price, Pyo, Sako, Sawicki, Shibuya, Shimasaku, Shimono, Shirasaki, Silverman, Simet, Speagle, Spergel, Strauss, Sugahara, Sugiyama, Suto, Suyu, Suzuki, Tait, Takada, Takata, Tamura, Tanaka, Tanaka, Tanaka, Tanaka,
  Terai, Terashima, Toba, Tominaga, Toshikawa, Turner, Uchida, Uchiyama, Umetsu, Uraguchi, Urata, Usuda, Utsumi, Wang, Wang, Wong, Yabe, Yamada, Yamanoi, Yasuda, Yeh, Yonehara, \& Yuma}]{Aihara18}
Aihara, H., Arimoto, N., Armstrong, R., {et~al.} 2018, \href{http://dx.doi.org/10.1093/pasj/psx066}{\color{magenta}\pasj}, 70, S4

\bibitem[{{Akins} {et~al.}(2023){Akins}, {Casey}, {Allen}, {Bagley}, {Dickinson}, {Finkelstein}, {Franco}, {Harish}, {Arrabal Haro}, {Ilbert}, {Kartaltepe}, {Koekemoer}, {Liu}, {Long}, {McCracken}, {Paquereau}, {Papovich}, {Pirzkal}, {Rhodes}, {Robertson}, {Shuntov}, {Toft}, {Yang}, {Barro}, {Bisigello}, {Buat}, {Champagne}, {Cooper}, {Costantin}, {de La Vega}, {Drakos}, {Faisst}, {Fontana}, {Fujimoto}, {Gillman}, {G{\'o}mez-Guijarro}, {Gozaliasl}, {Hathi}, {Hayward}, {Hirschmann}, {Holwerda}, {Jin}, {Kocevski}, {Kokorev}, {Lambrides}, {Lucas}, {Magdis}, {Magnelli}, {McKinney}, {Mobasher}, {P{\'e}rez-Gonz{\'a}lez}, {Rich}, {Seill{\'e}}, {Talia}, {Urry}, {Valentino}, {Whitaker}, {Yung}, {Zavala}, {Cosmos-Web Team}, \& {Ceers Team}}]{Akins2023}
{Akins}, H.~B., {Casey}, C.~M., {Allen}, N., {et~al.} 2023, \href{http://dx.doi.org/10.3847/1538-4357/acef21}{\color{magenta}\apj}, \href{https://ui.adsabs.harvard.edu/abs/2023ApJ...956...61A}{956, 61}

\bibitem[{{Akins} {et~al.}(2024){Akins}, {Casey}, {Lambrides}, {Allen}, {Andika}, {Brinch}, {Champagne}, {Cooper}, {Ding}, {Drakos}, {Faisst}, {Finkelstein}, {Franco}, {Fujimoto}, {Gentile}, {Gillman}, {Gozaliasl}, {Harish}, {Hayward}, {Hirschmann}, {Ilbert}, {Kartaltepe}, {Kocevski}, {Koekemoer}, {Kokorev}, {Liu}, {Long}, {McCracken}, {McKinney}, {Onoue}, {Paquereau}, {Renzini}, {Rhodes}, {Robertson}, {Shuntov}, {Silverman}, {Tanaka}, {Toft}, {Trakhtenbrot}, {Valentino}, \& {Zavala}}]{Akins2024}
{Akins}, H.~B., {Casey}, C.~M., {Lambrides}, E., {et~al.} 2024, \href{https://ui.adsabs.harvard.edu/abs/2024arXiv240610341A}{\href{http://dx.doi.org/10.48550/arXiv.2406.10341}{\color{magenta}arXiv e-prints}, arXiv:2406.10341}

\bibitem[{{Arango-Toro} {et~al.}(2023){Arango-Toro}, {Ciesla}, {Ilbert}, {Magnelli}, {Jim{\'e}nez-Andrade}, \& {Buat}}]{Arango23}
{Arango-Toro}, R.~C., {Ciesla}, L., {Ilbert}, O., {et~al.} 2023, \href{http://dx.doi.org/10.1051/0004-6361/202345848}{\color{magenta}\aap}, \href{https://ui.adsabs.harvard.edu/abs/2023A&A...675A.126A}{675, A126}

\bibitem[{{Arnouts} {et~al.}(2013){Arnouts}, {Le Floc'h}, {Chevallard}, {Johnson}, {Ilbert}, {Treyer}, {Aussel}, {Capak}, {Sanders}, {Scoville}, {McCracken}, {Milliard}, {Pozzetti}, \& {Salvato}}]{Arnouts2013}
{Arnouts}, S., {Le Floc'h}, E., {Chevallard}, J., {et~al.} 2013, \href{http://dx.doi.org/10.1051/0004-6361/201321768}{\color{magenta}\aap}, 558, A67

\bibitem[{Arnouts {et~al.}(2002)Arnouts, Moscardini, Vanzella, Colombi, Cristiani, Fontana, Giallongo, Matarrese, \& Saracco}]{Arnouts02}
Arnouts, S., Moscardini, L., Vanzella, E., {et~al.} 2002, \href{http://dx.doi.org/10.1046/j.1365-8711.2002.04988.x}{\color{magenta}\mnras}, 329, 355

\bibitem[{{Baldry} {et~al.}(2012){Baldry}, {Driver}, {Loveday}, {Taylor}, {Kelvin}, {Liske}, {Norberg}, {Robotham}, {Brough}, {Hopkins}, {Bamford}, {Peacock}, {Bland-Hawthorn}, {Conselice}, {Croom}, {Jones}, {Parkinson}, {Popescu}, {Prescott}, {Sharp}, \& {Tuffs}}]{Baldry2012}
{Baldry}, I.~K., {Driver}, S.~P., {Loveday}, J., {et~al.} 2012, \href{http://dx.doi.org/10.1111/j.1365-2966.2012.20340.x}{\color{magenta}\mnras}, \href{https://ui.adsabs.harvard.edu/abs/2012MNRAS.421..621B}{421, 621}

\bibitem[{{Barrufet} {et~al.}(2023){Barrufet}, {Oesch}, {Weibel}, {Brammer}, {Bezanson}, {Bouwens}, {Fudamoto}, {Gonzalez}, {Gottumukkala}, {Illingworth}, {Heintz}, {Holden}, {Labbe}, {Magee}, {Naidu}, {Nelson}, {Stefanon}, {Smit}, {van Dokkum}, {Weaver}, \& {Williams}}]{Barrufet2023}
{Barrufet}, L., {Oesch}, P.~A., {Weibel}, A., {et~al.} 2023, \href{http://dx.doi.org/10.1093/mnras/stad947}{\color{magenta}\mnras}, \href{https://ui.adsabs.harvard.edu/abs/2023MNRAS.522..449B}{522, 449}

\bibitem[{{Behroozi} \& {Silk}(2018)}]{BehrooziSilk2018}
{Behroozi}, P. \& {Silk}, J. 2018, \href{http://dx.doi.org/10.1093/mnras/sty945}{\color{magenta}\mnras}, \href{https://ui.adsabs.harvard.edu/abs/2018MNRAS.477.5382B}{477, 5382}

\bibitem[{{Behroozi} {et~al.}(2019){Behroozi}, {Wechsler}, {Hearin}, \& {Conroy}}]{Behroozi2019_UniverseMachine}
{Behroozi}, P., {Wechsler}, R.~H., {Hearin}, A.~P., \& {Conroy}, C. 2019, \href{http://dx.doi.org/10.1093/mnras/stz1182}{\color{magenta}\mnras}, \href{https://ui.adsabs.harvard.edu/abs/2019MNRAS.488.3143B}{488, 3143}

\bibitem[{{Behroozi} {et~al.}(2013){Behroozi}, {Wechsler}, \& {Conroy}}]{Behroozi13}
{Behroozi}, P.~S., {Wechsler}, R.~H., \& {Conroy}, C. 2013, \href{http://dx.doi.org/10.1088/0004-637X/770/1/57}{\color{magenta}\apj}, \href{https://ui.adsabs.harvard.edu/abs/2013ApJ...770...57B}{770, 57}

\bibitem[{{Bell} {et~al.}(2003){Bell}, {McIntosh}, {Katz}, \& {Weinberg}}]{Bell2003}
{Bell}, E.~F., {McIntosh}, D.~H., {Katz}, N., \& {Weinberg}, M.~D. 2003, \href{http://dx.doi.org/10.1086/378847}{\color{magenta}\apjs}, \href{https://ui.adsabs.harvard.edu/abs/2003ApJS..149..289B}{149, 289}

\bibitem[{{Bellstedt} {et~al.}(2016){Bellstedt}, {Lidman}, {Muzzin}, {Franx}, {Guatelli}, {Hill}, {Hoekstra}, {Kurinsky}, {Labbe}, {Marchesini}, {Marsan}, {Safavi-Naeini}, {Sif{\'o}n}, {Stefanon}, {van de Sande}, {van Dokkum}, \& {Weigel}}]{Bellstedt2016}
{Bellstedt}, S., {Lidman}, C., {Muzzin}, A., {et~al.} 2016, \href{http://dx.doi.org/10.1093/mnras/stw1184}{\color{magenta}\mnras}, \href{https://ui.adsabs.harvard.edu/abs/2016MNRAS.460.2862B}{460, 2862}

\bibitem[{{Bertin} {et~al.}(2020){Bertin}, {Schefer}, {Apostolakos}, {{\'A}lvarez-Ayll{\'o}n}, {Dubath}, \& {K{\"u}mmel}}]{bertin20}
{Bertin}, E., {Schefer}, M., {Apostolakos}, N., {et~al.} 2020, in Astronomical Society of the Pacific Conference Series, Vol. 527, Astronomical Data Analysis Software and Systems XXIX, ed. R.~{Pizzo}, E.~R. {Deul}, J.~D. {Mol}, J.~{de Plaa}, \& H.~{Verkouter}, \href{https://ui.adsabs.harvard.edu/abs/2020ASPC..527..461B}{461}

\bibitem[{{Bhatawdekar} {et~al.}(2019){Bhatawdekar}, {Conselice}, {Margalef-Bentabol}, \& {Duncan}}]{Bhatawdekar19}
{Bhatawdekar}, R., {Conselice}, C.~J., {Margalef-Bentabol}, B., \& {Duncan}, K. 2019, \href{http://dx.doi.org/10.1093/mnras/stz866}{\color{magenta}\mnras}, \href{https://ui.adsabs.harvard.edu/abs/2019MNRAS.486.3805B}{486, 3805}

\bibitem[{{Boquien} {et~al.}(2019){Boquien}, {Burgarella}, {Roehlly}, {Buat}, {Ciesla}, {Corre}, {Inoue}, \& {Salas}}]{Boquien19}
{Boquien}, M., {Burgarella}, D., {Roehlly}, Y., {et~al.} 2019, \href{http://dx.doi.org/10.1051/0004-6361/201834156}{\color{magenta}\aap}, \href{https://ui.adsabs.harvard.edu/abs/2019A&A...622A.103B}{622, A103}

\bibitem[{{Bouwens} {et~al.}(2023{\natexlab{a}}){Bouwens}, {Illingworth}, {Oesch}, {Stefanon}, {Naidu}, {van Leeuwen}, \& {Magee}}]{Bouwens2023b}
{Bouwens}, R., {Illingworth}, G., {Oesch}, P., {et~al.} 2023{\natexlab{a}}, \href{http://dx.doi.org/10.1093/mnras/stad1014}{\color{magenta}\mnras}, \href{https://ui.adsabs.harvard.edu/abs/2023MNRAS.523.1009B}{523, 1009}

\bibitem[{{Bouwens} {et~al.}(2012{\natexlab{a}}){Bouwens}, {Illingworth}, {Oesch}, {Franx}, {Labb{\'e}}, {Trenti}, {van Dokkum}, {Carollo}, {Gonz{\'a}lez}, {Smit}, \& {Magee}}]{Bouwens2012a}
{Bouwens}, R.~J., {Illingworth}, G.~D., {Oesch}, P.~A., {et~al.} 2012{\natexlab{a}}, \href{http://dx.doi.org/10.1088/0004-637X/754/2/83}{\color{magenta}\apj}, \href{https://ui.adsabs.harvard.edu/abs/2012ApJ...754...83B}{754, 83}

\bibitem[{{Bouwens} {et~al.}(2012{\natexlab{b}}){Bouwens}, {Illingworth}, {Oesch}, {Trenti}, {Labb{\'e}}, {Franx}, {Stiavelli}, {Carollo}, {van Dokkum}, \& {Magee}}]{Bouwens2012b}
{Bouwens}, R.~J., {Illingworth}, G.~D., {Oesch}, P.~A., {et~al.} 2012{\natexlab{b}}, \href{http://dx.doi.org/10.1088/2041-8205/752/1/L5}{\color{magenta}\apjl}, \href{https://ui.adsabs.harvard.edu/abs/2012ApJ...752L...5B}{752, L5}

\bibitem[{{Bouwens} {et~al.}(2023{\natexlab{b}}){Bouwens}, {Stefanon}, {Brammer}, {Oesch}, {Herard-Demanche}, {Illingworth}, {Matthee}, {Naidu}, {van Dokkum}, \& {van Leeuwen}}]{Bouwens2023a}
{Bouwens}, R.~J., {Stefanon}, M., {Brammer}, G., {et~al.} 2023{\natexlab{b}}, \href{http://dx.doi.org/10.1093/mnras/stad1145}{\color{magenta}\mnras}, \href{https://ui.adsabs.harvard.edu/abs/2023MNRAS.523.1036B}{523, 1036}

\bibitem[{Bowler {et~al.}(2012)Bowler, Dunlop, McLure, McCracken, Milvang-Jensen, Furusawa, Fynbo, {Le F{\`{e}}vre}, Holt, Ideue, Ihara, Rogers, \& Taniguchi}]{Bowler2012}
Bowler, R. A.~A., Dunlop, J.~S., McLure, R.~J., {et~al.} 2012, \href{http://dx.doi.org/10.1111/j.1365-2966.2012.21904.x}{\color{magenta}Monthly Notices of the Royal Astronomical Society}, 426, 2772

\bibitem[{{Bowler} {et~al.}(2014){Bowler}, {Dunlop}, {McLure}, {Rogers}, {McCracken}, {Milvang-Jensen}, {Furusawa}, {Fynbo}, {Taniguchi}, {Afonso}, {Bremer}, \& {Le F{\`e}vre}}]{Bowler2014}
{Bowler}, R.~A.~A., {Dunlop}, J.~S., {McLure}, R.~J., {et~al.} 2014, \href{http://dx.doi.org/10.1093/mnras/stu449}{\color{magenta}\mnras}, \href{https://ui.adsabs.harvard.edu/abs/2014MNRAS.440.2810B}{440, 2810}

\bibitem[{{Bowler} {et~al.}(2020){Bowler}, {Jarvis}, {Dunlop}, {McLure}, {McLeod}, {Adams}, {Milvang-Jensen}, \& {McCracken}}]{Bowler2020}
{Bowler}, R.~A.~A., {Jarvis}, M.~J., {Dunlop}, J.~S., {et~al.} 2020, \href{http://dx.doi.org/10.1093/mnras/staa313}{\color{magenta}\mnras}, \href{https://ui.adsabs.harvard.edu/abs/2020MNRAS.493.2059B}{493, 2059}

\bibitem[{{Boylan-Kolchin}(2023)}]{Boylan-Kolchin2023}
{Boylan-Kolchin}, M. 2023, \href{http://dx.doi.org/10.1038/s41550-023-01937-7}{\color{magenta}Nature Astronomy}, \href{https://ui.adsabs.harvard.edu/abs/2023NatAs...7..731B}{7, 731}

\bibitem[{Bruzual \& Charlot(2003)}]{BC03}
Bruzual, G. \& Charlot, S. 2003, \href{http://dx.doi.org/10.1046/j.1365-8711.2003.06897.x}{\color{magenta}\mnras}, 344, 1000

\bibitem[{{Bruzual} \& {Charlot}(2003)}]{BruzualCharlot03}
{Bruzual}, G. \& {Charlot}, S. 2003, \href{http://dx.doi.org/10.1046/j.1365-8711.2003.06897.x}{\color{magenta}\mnras}, \href{https://ui.adsabs.harvard.edu/abs/2003MNRAS.344.1000B}{344, 1000}

\bibitem[{{Bushouse} {et~al.}(2022){Bushouse}, {Eisenhamer}, {Dencheva}, {Davies}, {Greenfield}, {Morrison}, {Hodge}, {Simon}, {Grumm}, {Droettboom}, {Slavich}, {Sosey}, {Pauly}, {Miller}, {Jedrzejewski}, {Hack}, {Davis}, {Crawford}, {Law}, {Gordon}, {Regan}, {Cara}, {MacDonald}, {Bradley}, {Shanahan}, {Jamieson}, {Teodoro}, \& {Williams}}]{Bushouse2022}
{Bushouse}, H., {Eisenhamer}, J., {Dencheva}, N., {et~al.} 2022, {JWST Calibration Pipeline}, Zenodo

\bibitem[{{Calzetti} {et~al.}(2000){Calzetti}, {Armus}, {Bohlin}, {Kinney}, {Koornneef}, \& {Storchi-Bergmann}}]{Calzetti00}
{Calzetti}, D., {Armus}, L., {Bohlin}, R.~C., {et~al.} 2000, \href{http://dx.doi.org/10.1086/308692}{\color{magenta}\apj}, \href{https://ui.adsabs.harvard.edu/abs/2000ApJ...533..682C}{533, 682}

\bibitem[{{Calzetti} {et~al.}(2007){Calzetti}, {Kennicutt}, {Engelbracht}, {Leitherer}, {Draine}, {Kewley}, {Moustakas}, {Sosey}, {Dale}, {Gordon}, {Helou}, {Hollenbach}, {Armus}, {Bendo}, {Bot}, {Buckalew}, {Jarrett}, {Li}, {Meyer}, {Murphy}, {Prescott}, {Regan}, {Rieke}, {Roussel}, {Sheth}, {Smith}, {Thornley}, \& {Walter}}]{Calzetti2007}
{Calzetti}, D., {Kennicutt}, R.~C., {Engelbracht}, C.~W., {et~al.} 2007, \href{http://dx.doi.org/10.1086/520082}{\color{magenta}\apj}, \href{https://ui.adsabs.harvard.edu/abs/2007ApJ...666..870C}{666, 870}

\bibitem[{{Capak} {et~al.}(2011){Capak}, {Mobasher}, {Scoville}, {McCracken}, {Ilbert}, {Salvato}, {Men{\'e}ndez-Delmestre}, {Aussel}, {Carilli}, {Civano}, {Elvis}, {Giavalisco}, {Jullo}, {Kartaltepe}, {Leauthaud}, {Koekemoer}, {Kneib}, {LeFloch}, {Sanders}, {Schinnerer}, {Shioya}, {Shopbell}, {Tanaguchi}, {Thompson}, \& {Willott}}]{capak11}
{Capak}, P., {Mobasher}, B., {Scoville}, N.~Z., {et~al.} 2011, \href{http://dx.doi.org/10.1088/0004-637X/730/2/68}{\color{magenta}\apj}, \href{https://ui.adsabs.harvard.edu/abs/2011ApJ...730...68C}{730, 68}

\bibitem[{{Carnall} {et~al.}(2023){Carnall}, {McLeod}, {McLure}, {Dunlop}, {Begley}, {Cullen}, {Donnan}, {Hamadouche}, {Jewell}, {Jones}, {Pollock}, \& {Wild}}]{Carnall2023}
{Carnall}, A.~C., {McLeod}, D.~J., {McLure}, R.~J., {et~al.} 2023, \href{http://dx.doi.org/10.1093/mnras/stad369}{\color{magenta}\mnras}, \href{https://ui.adsabs.harvard.edu/abs/2023MNRAS.520.3974C}{520, 3974}

\bibitem[{{Casey} {et~al.}(2024){Casey}, {Akins}, {Shuntov}, {Ilbert}, {Paquereau}, {Franco}, {Hayward}, {Finkelstein}, {Boylan-Kolchin}, {Robertson}, {Allen}, {Brinch}, {Cooper}, {Ding}, {Drakos}, {Faisst}, {Fujimoto}, {Gillman}, {Harish}, {Hirschmann}, {Jin}, {Kartaltepe}, {Koekemoer}, {Kokorev}, {Liu}, {Long}, {Magdis}, {Maraston}, {Martin}, {McCracken}, {McKinney}, {Mobasher}, {Rhodes}, {Rich}, {Sanders}, {Silverman}, {Toft}, {Vijayan}, {Weaver}, {Wilkins}, {Yang}, \& {Zavala}}]{Casey2024}
{Casey}, C.~M., {Akins}, H.~B., {Shuntov}, M., {et~al.} 2024, \href{http://dx.doi.org/10.3847/1538-4357/ad2075}{\color{magenta}\apj}, \href{https://ui.adsabs.harvard.edu/abs/2024ApJ...965...98C}{965, 98}

\bibitem[{{Casey} {et~al.}(2012){Casey}, {Berta}, {B{\'e}thermin}, {Bock}, {Bridge}, {Budynkiewicz}, {Burgarella}, {Chapin}, {Chapman}, {Clements}, {Conley}, {Conselice}, {Cooray}, {Farrah}, {Hatziminaoglou}, {Ivison}, {le Floc'h}, {Lutz}, {Magdis}, {Magnelli}, {Oliver}, {Page}, {Pozzi}, {Rigopoulou}, {Riguccini}, {Roseboom}, {Sanders}, {Scott}, {Seymour}, {Valtchanov}, {Vieira}, {Viero}, \& {Wardlow}}]{casey12}
{Casey}, C.~M., {Berta}, S., {B{\'e}thermin}, M., {et~al.} 2012, \href{http://dx.doi.org/10.1088/0004-637X/761/2/140}{\color{magenta}\apj}, \href{https://ui.adsabs.harvard.edu/abs/2012ApJ...761..140C}{761, 140}

\bibitem[{{Casey} {et~al.}(2023){Casey}, {Kartaltepe}, {Drakos}, {Franco}, {Harish}, {Paquereau}, {Ilbert}, {Rose}, {Cox}, {Nightingale}, {Robertson}, {Silverman}, {Koekemoer}, {Massey}, {McCracken}, {Rhodes}, {Akins}, {Allen}, {Amvrosiadis}, {Arango-Toro}, {Bagley}, {Bongiorno}, {Capak}, {Champagne}, {Chartab}, {Ch{\'a}vez Ortiz}, {Chworowsky}, {Cooke}, {Cooper}, {Darvish}, {Ding}, {Faisst}, {Finkelstein}, {Fujimoto}, {Gentile}, {Gillman}, {Gould}, {Gozaliasl}, {Hayward}, {He}, {Hemmati}, {Hirschmann}, {Jahnke}, {Jin}, {Khostovan}, {Kokorev}, {Lambrides}, {Laigle}, {Larson}, {Leung}, {Liu}, {Liaudat}, {Long}, {Magdis}, {Mahler}, {Mainieri}, {Manning}, {Maraston}, {Martin}, {McCleary}, {McKinney}, {McPartland}, {Mobasher}, {Pattnaik}, {Renzini}, {Rich}, {Sanders}, {Sattari}, {Scognamiglio}, {Scoville}, {Sheth}, {Shuntov}, {Sparre}, {Suzuki}, {Talia}, {Toft}, {Trakhtenbrot}, {Urry}, {Valentino}, {Vanderhoof}, {Vardoulaki}, {Weaver}, {Whitaker}, {Wilkins}, {Yang}, \& {Zavala}}]{Casey2023}
{Casey}, C.~M., {Kartaltepe}, J.~S., {Drakos}, N.~E., {et~al.} 2023, \href{http://dx.doi.org/10.3847/1538-4357/acc2bc}{\color{magenta}\apj}, \href{https://ui.adsabs.harvard.edu/abs/2023ApJ...954...31C}{954, 31}

\bibitem[{{Cattaneo} {et~al.}(2006){Cattaneo}, {Dekel}, {Devriendt}, {Guiderdoni}, \& {Blaizot}}]{Cattaneo06}
{Cattaneo}, A., {Dekel}, A., {Devriendt}, J., {Guiderdoni}, B., \& {Blaizot}, J. 2006, \href{http://dx.doi.org/10.1111/j.1365-2966.2006.10608.x}{\color{magenta}\mnras}, \href{https://ui.adsabs.harvard.edu/abs/2006MNRAS.370.1651C}{370, 1651}

\bibitem[{{Chabrier}(2003)}]{Chabrier03}
{Chabrier}, G. 2003, \href{http://dx.doi.org/10.1086/376392}{\color{magenta}\pasp}, \href{https://ui.adsabs.harvard.edu/abs/2003PASP..115..763C}{115, 763}

\bibitem[{{Chworowsky} {et~al.}(2023){Chworowsky}, {Finkelstein}, {Boylan-Kolchin}, {McGrath}, {Iyer}, {Papovich}, {Dickinson}, {Taylor}, {Yung}, {Arrabal Haro}, {Bagley}, {Backhaus}, {Bhatawdekar}, {Cheng}, {Cleri}, {Cole}, {Cooper}, {Costantin}, {Dekel}, {Franco}, {Fujimoto}, {Hayward}, {Holwerda}, {Huertas-Company}, {Hirschmann}, {Hutchison}, {Koekemoer}, {Larson}, {Li}, {Long}, {Lucas}, {Pirzkal}, {Rodighiero}, {Somerville}, {Vanderhoof}, {de la Vega}, {Wilkins}, {Yang}, \& {Zavala}}]{Chworowsky2023}
{Chworowsky}, K., {Finkelstein}, S.~L., {Boylan-Kolchin}, M., {et~al.} 2023, \href{https://ui.adsabs.harvard.edu/abs/2023arXiv231114804C}{\href{http://dx.doi.org/10.48550/arXiv.2311.14804}{\color{magenta}arXiv e-prints}, arXiv:2311.14804}

\bibitem[{{Ciesla} {et~al.}(2023{\natexlab{a}}){Ciesla}, {Elbaz}, {Ilbert}, {Buat}, {Magnelli}, {Narayanan}, {Daddi}, {G{\'o}mez-Guijarro}, \& {Arango-Toro}}]{Ciesla23b}
{Ciesla}, L., {Elbaz}, D., {Ilbert}, O., {et~al.} 2023{\natexlab{a}}, \href{https://ui.adsabs.harvard.edu/abs/2023arXiv230915720C}{\href{http://dx.doi.org/10.48550/arXiv.2309.15720}{\color{magenta}arXiv e-prints}, arXiv:2309.15720}

\bibitem[{{Ciesla} {et~al.}(2023{\natexlab{b}}){Ciesla}, {G{\'o}mez-Guijarro}, {Buat}, {Elbaz}, {Jin}, {B{\'e}thermin}, {Daddi}, {Franco}, {Inami}, {Magdis}, {Magnelli}, \& {Xiao}}]{Ciesla23}
{Ciesla}, L., {G{\'o}mez-Guijarro}, C., {Buat}, V., {et~al.} 2023{\natexlab{b}}, \href{http://dx.doi.org/10.1051/0004-6361/202245376}{\color{magenta}\aap}, \href{https://ui.adsabs.harvard.edu/abs/2023A&A...672A.191C}{672, A191}

\bibitem[{Civano {et~al.}(2016)Civano, Marchesi, Comastri, Urry, Elvis, Cappelluti, Puccetti, Brusa, Zamorani, Hasinger, Aldcroft, Alexander, Allevato, Brunner, Capak, Finoguenov, Fiore, Fruscione, Gilli, Glotfelty, Griffiths, Hao, Harrison, Jahnke, Kartaltepe, Karim, LaMassa, Lanzuisi, Miyaji, Ranalli, Salvato, Sargent, Scoville, Schawinski, Schinnerer, Silverman, Smolcic, Stern, Toft, Trakhenbrot, Treister, \& Vignali}]{Civano_2016}
Civano, F., Marchesi, S., Comastri, A., {et~al.} 2016, \href{http://dx.doi.org/10.3847/0004-637X/819/1/62}{\color{magenta}The Astrophysical Journal}, 819, 62

\bibitem[{{Cole} {et~al.}(2001){Cole}, {Norberg}, {Baugh}, {Frenk}, {Bland-Hawthorn}, {Bridges}, {Cannon}, {Colless}, {Collins}, {Couch}, {Cross}, {Dalton}, {De Propris}, {Driver}, {Efstathiou}, {Ellis}, {Glazebrook}, {Jackson}, {Lahav}, {Lewis}, {Lumsden}, {Maddox}, {Madgwick}, {Peacock}, {Peterson}, {Sutherland}, \& {Taylor}}]{Cole2001}
{Cole}, S., {Norberg}, P., {Baugh}, C.~M., {et~al.} 2001, \href{http://dx.doi.org/10.1046/j.1365-8711.2001.04591.x}{\color{magenta}\mnras}, \href{https://ui.adsabs.harvard.edu/abs/2001MNRAS.326..255C}{326, 255}

\bibitem[{{Conroy} \& {Wechsler}(2009)}]{ConroyWechsler2009}
{Conroy}, C. \& {Wechsler}, R.~H. 2009, \href{http://dx.doi.org/10.1088/0004-637X/696/1/620}{\color{magenta}\apj}, \href{https://ui.adsabs.harvard.edu/abs/2009ApJ...696..620C}{696, 620}

\bibitem[{Coupon {et~al.}(2015)Coupon, Arnouts, van Waerbeke, Moutard, Ilbert, van Uitert, Erben, Garilli, Guzzo, Heymans, Hildebrandt, Hoekstra, Kilbinger, Kitching, Mellier, Miller, Scodeggio, Bonnett, Branchini, Davidzon, De~Lucia, Fritz, Fu, Hudelot, Hudson, Kuijken, Leauthaud, Fèvre, McCracken, Moscardini, Rowe, Schrabback, Semboloni, \& Velander}]{coupon_galaxy-halo_2015}
Coupon, J., Arnouts, S., van Waerbeke, L., {et~al.} 2015, \href{http://dx.doi.org/10.1093/mnras/stv276}{\color{magenta}\mnras}, 449, 1352

\bibitem[{{Dale} {et~al.}(2014){Dale}, {Helou}, {Magdis}, {Armus}, {D{\'\i}az-Santos}, \& {Shi}}]{Dale14}
{Dale}, D.~A., {Helou}, G., {Magdis}, G.~E., {et~al.} 2014, \href{http://dx.doi.org/10.1088/0004-637X/784/1/83}{\color{magenta}\apj}, \href{https://ui.adsabs.harvard.edu/abs/2014ApJ...784...83D}{784, 83}

\bibitem[{{Davidzon} {et~al.}(2017){Davidzon}, {Ilbert}, {Laigle}, {Coupon}, {McCracken}, {Delvecchio}, {Masters}, {Capak}, {Hsieh}, {Le F{\`e}vre}, {Tresse}, {Bethermin}, {Chang}, {Faisst}, {Le Floc'h}, {Steinhardt}, {Toft}, {Aussel}, {Dubois}, {Hasinger}, {Salvato}, {Sanders}, {Scoville}, \& {Silverman}}]{Davidzon2017}
{Davidzon}, I., {Ilbert}, O., {Laigle}, C., {et~al.} 2017, \href{http://dx.doi.org/10.1051/0004-6361/201730419}{\color{magenta}\aap}, \href{https://ui.adsabs.harvard.edu/abs/2017A&A...605A..70D}{605, A70}

\bibitem[{{De Lucia} {et~al.}(2024){De Lucia}, {Fontanot}, {Xie}, \& {Hirschmann}}]{DeLucia2024}
{De Lucia}, G., {Fontanot}, F., {Xie}, L., \& {Hirschmann}, M. 2024, \href{https://ui.adsabs.harvard.edu/abs/2024arXiv240106211D}{\href{http://dx.doi.org/10.48550/arXiv.2401.06211}{\color{magenta}arXiv e-prints}, arXiv:2401.06211}

\bibitem[{De~Lucia {et~al.}(2006)De~Lucia, Springel, White, Croton, \& Kauffmann}]{de_lucia_formation_2006}
De~Lucia, G., Springel, V., White, S. D.~M., Croton, D., \& Kauffmann, G. 2006, \href{http://dx.doi.org/10.1111/j.1365-2966.2005.09879.x}{\color{magenta}\mnras}, 366, 499

\bibitem[{{Dekel} {et~al.}(2009){Dekel}, {Birnboim}, {Engel}, {Freundlich}, {Goerdt}, {Mumcuoglu}, {Neistein}, {Pichon}, {Teyssier}, \& {Zinger}}]{Dekel2009}
{Dekel}, A., {Birnboim}, Y., {Engel}, G., {et~al.} 2009, \href{http://dx.doi.org/10.1038/nature07648}{\color{magenta}\nat}, \href{https://ui.adsabs.harvard.edu/abs/2009Natur.457..451D}{457, 451}

\bibitem[{{Dekel} {et~al.}(2023){Dekel}, {Sarkar}, {Birnboim}, {Mandelker}, \& {Li}}]{Dekel2023}
{Dekel}, A., {Sarkar}, K.~C., {Birnboim}, Y., {Mandelker}, N., \& {Li}, Z. 2023, \href{http://dx.doi.org/10.1093/mnras/stad1557}{\color{magenta}\mnras}, \href{https://ui.adsabs.harvard.edu/abs/2023MNRAS.523.3201D}{523, 3201}

\bibitem[{{Dekel} {et~al.}(2013){Dekel}, {Zolotov}, {Tweed}, {Cacciato}, {Ceverino}, \& {Primack}}]{Dekel2013}
{Dekel}, A., {Zolotov}, A., {Tweed}, D., {et~al.} 2013, \href{http://dx.doi.org/10.1093/mnras/stt1338}{\color{magenta}\mnras}, \href{https://ui.adsabs.harvard.edu/abs/2013MNRAS.435..999D}{435, 999}

\bibitem[{{Donnan} {et~al.}(2023){Donnan}, {McLeod}, {Dunlop}, {McLure}, {Carnall}, {Begley}, {Cullen}, {Hamadouche}, {Bowler}, {Magee}, {McCracken}, {Milvang-Jensen}, {Moneti}, \& {Targett}}]{Donnan2023}
{Donnan}, C.~T., {McLeod}, D.~J., {Dunlop}, J.~S., {et~al.} 2023, \href{http://dx.doi.org/10.1093/mnras/stac3472}{\color{magenta}\mnras}, \href{https://ui.adsabs.harvard.edu/abs/2023MNRAS.518.6011D}{518, 6011}

\bibitem[{{Donnan} {et~al.}(2024){Donnan}, {McLure}, {Dunlop}, {McLeod}, {Magee}, {Arellano-C{\'o}rdova}, {Barrufet}, {Begley}, {Bowler}, {Carnall}, {Cullen}, {Ellis}, {Fontana}, {Illingworth}, {Grogin}, {Hamadouche}, {Koekemoer}, {Liu}, {Mason}, {Santini}, \& {Stanton}}]{Donnan2024}
{Donnan}, C.~T., {McLure}, R.~J., {Dunlop}, J.~S., {et~al.} 2024, \href{https://ui.adsabs.harvard.edu/abs/2024arXiv240303171D}{\href{http://dx.doi.org/10.48550/arXiv.2403.03171}{\color{magenta}arXiv e-prints}, arXiv:2403.03171}

\bibitem[{{Drakos} {et~al.}(2022){Drakos}, {Villasenor}, {Robertson}, {Hausen}, {Dickinson}, {Ferguson}, {Furlanetto}, {Greene}, {Madau}, {Shapley}, {Stark}, \& {Wechsler}}]{Drakos2022}
{Drakos}, N.~E., {Villasenor}, B., {Robertson}, B.~E., {et~al.} 2022, \href{http://dx.doi.org/10.3847/1538-4357/ac46fb}{\color{magenta}\apj}, \href{https://ui.adsabs.harvard.edu/abs/2022ApJ...926..194D}{926, 194}

\bibitem[{{Drlica-Wagner} {et~al.}(2018){Drlica-Wagner}, {Sevilla-Noarbe}, {Rykoff}, {Gruendl}, {Yanny}, {Tucker}, {Hoyle}, {Carnero Rosell}, {Bernstein}, {Bechtol}, {Becker}, {Benoit-L{\'e}vy}, {Bertin}, {Carrasco Kind}, {Davis}, {de Vicente}, {Diehl}, {Gruen}, {Hartley}, {Leistedt}, {Li}, {Marshall}, {Neilsen}, {Rau}, {Sheldon}, {Smith}, {Troxel}, {Wyatt}, {Zhang}, {Abbott}, {Abdalla}, {Allam}, {Banerji}, {Brooks}, {Buckley-Geer}, {Burke}, {Capozzi}, {Carretero}, {Cunha}, {D'Andrea}, {da Costa}, {DePoy}, {Desai}, {Dietrich}, {Doel}, {Evrard}, {Fausti Neto}, {Flaugher}, {Fosalba}, {Frieman}, {Garc{\'\i}a-Bellido}, {Gerdes}, {Giannantonio}, {Gschwend}, {Gutierrez}, {Honscheid}, {James}, {Jeltema}, {Kuehn}, {Kuhlmann}, {Kuropatkin}, {Lahav}, {Lima}, {Lin}, {Maia}, {Martini}, {McMahon}, {Melchior}, {Menanteau}, {Miquel}, {Nichol}, {Ogando}, {Plazas}, {Romer}, {Roodman}, {Sanchez}, {Scarpine}, {Schindler}, {Schubnell}, {Smith}, {Smith}, {Soares-Santos}, {Sobreira}, {Suchyta}, {Tarle}, {Vikram}, {Walker},
  {Wechsler}, {Zuntz}, \& {DES Collaboration}}]{Drlica-Wagner2018}
{Drlica-Wagner}, A., {Sevilla-Noarbe}, I., {Rykoff}, E.~S., {et~al.} 2018, \href{http://dx.doi.org/10.3847/1538-4365/aab4f5}{\color{magenta}\apjs}, \href{https://ui.adsabs.harvard.edu/abs/2018ApJS..235...33D}{235, 33}

\bibitem[{{Duan} {et~al.}(2024){Duan}, {Conselice}, {Li}, {Austin}, {Harvey}, {Adams}, {Duncan}, {Trussler}, {Ferreira}, {Westcott}, {Harris}, {Windhorst}, {Holwerda}, {Broadhurst}, {Coe}, {Cohen}, {Driver}, {Frye}, {Grogin}, {Hathi}, {Jansen}, {Koekemoer}, {Marshall}, {Nonino}, {Ortiz}, {Pirzkal}, {Robotham}, {Ryan}, {Summers}, {D'Silva}, {Willmer}, \& {Yan}}]{Duan2024}
{Duan}, Q., {Conselice}, C.~J., {Li}, Q., {et~al.} 2024, \href{https://ui.adsabs.harvard.edu/abs/2024arXiv240709472D}{\href{http://dx.doi.org/10.48550/arXiv.2407.09472}{\color{magenta}arXiv e-prints}, arXiv:2407.09472}

\bibitem[{{Dubois} {et~al.}(2024){Dubois}, {Rodr{\'\i}guez Montero}, {Guerra}, {Trebitsch}, {Han}, {Beckmann}, {Yi}, {Lewis}, \& {Jang}}]{Dubois2024}
{Dubois}, Y., {Rodr{\'\i}guez Montero}, F., {Guerra}, C., {et~al.} 2024, \href{https://ui.adsabs.harvard.edu/abs/2024arXiv240218515D}{\href{http://dx.doi.org/10.48550/arXiv.2402.18515}{\color{magenta}arXiv e-prints}, arXiv:2402.18515}

\bibitem[{Dunlop(2016)}]{dunlop_ultravista_2016}
Dunlop, J. 2016, {UltraVISTA} - {Ultra} {Deep} {Survey} with {VISTA}

\bibitem[{{Dunlop} {et~al.}(2021){Dunlop}, {Abraham}, {Ashby}, {Bagley}, {Best}, {Bongiorno}, {Bouwens}, {Bowler}, {Brammer}, {Bremer}, {Calabro'}, {Carnall}, {Castellano}, {Cirasuolo}, {Conselice}, {Cullen}, {Dave}, {Dayal}, {Dekel}, {Dickinson}, {Duncan}, {Elbaz}, {Ellis}, {Ferguson}, {Ferrara}, {Finkelstein}, {Fontana}, {Furlanetto}, {Fynbo}, {Gallerani}, {Gardner}, {Giavalisco}, {Grazian}, {Grogin}, {Harikane}, {Hopkins}, {Ilbert}, {Illingworth}, {Juneau}, {Jung}, {Kartaltepe}, {Kassin}, {Kauffmann}, {Khochfar}, {Kirkpatrick}, {Kocevski}, {Koekemoer}, {Labbe}, {Laporte}, {Larson}, {Lucas}, {Magee}, {Mason}, {McCracken}, {McLeod}, {McLure}, {Merlin}, {Mesinger}, {Milvang-Jensen}, {Newman}, {Oesch}, {Ouchi}, {Pacifici}, {Papovich}, {Peacock}, {Peeples}, {Pentericci}, {Perez-Gonzalez}, {Pirzkal}, {Pope}, {Pye}, {Reddy}, {Robertson}, {Salvato}, {Santini}, {Schaerer}, {Shapley}, {Simons}, {Smit}, {Smith}, {Snyder}, {Somerville}, {Stanway}, {Stefanon}, {Tasca}, {Tikkanen}, {Tresse}, {Trump}, {Whitaker},
  {Wilkins}, {Wright}, {Wyithe}, {van Dokkum}, \& {van der Werf}}]{PrimerDunlop2021}
{Dunlop}, J.~S., {Abraham}, R.~G., {Ashby}, M. L.~N., {et~al.} 2021, {PRIMER: Public Release IMaging for Extragalactic Research}, JWST Proposal. Cycle 1, ID. \#1837

\bibitem[{{Eddington}(1913)}]{Eddington1913}
{Eddington}, A.~S. 1913, \href{http://dx.doi.org/10.1093/mnras/73.5.359}{\color{magenta}\mnras}, \href{https://ui.adsabs.harvard.edu/abs/1913MNRAS..73..359E}{73, 359}

\bibitem[{{Euclid Collaboration} {et~al.}(2024){Euclid Collaboration}, {McPartland}, {Zalesky}, {Weaver}, {Toft}, {Sanders}, {Mobasher}, {Suzuki}, {Szapudi}, {Valdes}, {Murphree}, {Chartab}, {Allen}, {Taamoli}, {Eisenhardt}, {Arnouts}, {Atek}, {Brinchmann}, {Castellano}, {Chary}, {Ch{\'a}vez Ortiz}, {Cuby}, {Finkelstein}, {Goto}, {Gwyn}, {Harikane}, {Inoue}, {McCracken}, {Mohr}, {Oesch}, {Ouchi}, {Oguri}, {Rhodes}, {Rottgering}, {Sawicki}, {Scaramella}, {Scarlata}, {Silverman}, {Stern}, {Teplitz}, {Shuntov}, {Altieri}, {Amara}, {Andreon}, {Auricchio}, {Aussel}, {Baccigalupi}, {Baldi}, {Bardelli}, {Bender}, {Bonino}, {Branchini}, {Brescia}, {Camera}, {Capobianco}, {Carbone}, {Carretero}, {Casas}, {Castander}, {Castignani}, {Cavuoti}, {Cimatti}, {Colodro-Conde}, {Congedo}, {Conselice}, {Conversi}, {Copin}, {Courbin}, {Courtois}, {Da Silva}, {Degaudenzi}, {De Lucia}, {Di Giorgio}, {Dinis}, {Douspis}, {Dubath}, {Dupac}, {Dusini}, {Fabricius}, {Farina}, {Farrens}, {Ferriol}, {Fotopoulou}, {Frailis}, {Franceschi},
  {Fumana}, {Galeotta}, {Garilli}, {George}, {Gillis}, {Giocoli}, {Grazian}, {Grupp}, {Guzzo}, {Hoekstra}, {Holmes}, {Hook}, {Hormuth}, {Hornstrup}, {Hudelot}, {Jahnke}, {Keih{\"a}nen}, {Kermiche}, {Kiessling}, {Kilbinger}, {Kitching}, {Kubik}, {Kunz}, {Kurki-Suonio}, {Lilje}, {Lindholm}, {Lloro}, {Mainetti}, {Maiorano}, {Mansutti}, {Marggraf}, {Markovic}, {Martinelli}, {Martinet}, {Marulli}, {Massey}, {Maurogordato}, {Medinaceli}, {Mei}, {Melchior}, {Mellier}, {Meneghetti}, {Merlin}, {Meylan}, {Moresco}, {Moscardini}, {Munari}, {Nakajima}, {Neissner}, {Niemi}, {Nightingale}, {Padilla}, {Paltani}, {Pasian}, {Pedersen}, {Percival}, {Pettorino}, {Polenta}, {Poncet}, {Popa}, {Pozzetti}, {Raison}, {Rebolo}, {Renzi}, {Riccio}, {Romelli}, {Roncarelli}, {Rossetti}, {Saglia}, {Sakr}, {S{\'a}nchez}, {Sapone}, {Sartoris}, {Schirmer}, {Schneider}, {Schrabback}, {Secroun}, {Seidel}, {Serrano}, {Sirignano}, {Sirri}, {Stanco}, {Steinwagner}, {Surace}, {Tallada-Crespi}, {Tavagnacco}, {Tereno}, {Toledo-Moreo}, {Torradeflot},
  {Tutusaus}, {Valentijn}, {Valenziano}, {Vassallo}, {Veropalumbo}, {Wang}, {Weller}, {Zamorani}, {Zoubian}, {Zucca}, {Biviano}, {Bolzonella}, {Boucaud}, {Bozzo}, {Burigana}, {Di Ferdinando}, {Farinelli}, {Gracia-Carpio}, {Mauri}, {Scottez}, {Tenti}, {Viel}, {Wiesmann}, {Akrami}, {Allevato}, {Anselmi}, {Ballardini}, {Bethermin}, {Borgani}, {Borlaff}, {Bruton}, {Cabanac}, {Calabro}, {Ca{\~n}as-Herrera}, {Cappi}, {Carvalho}, {Castro}, {Chambers}, {Contarini}, {Cooray}, {Coupon}, {Davini}, {de la Torre}, {Desprez}, {D{\'\i}az-S{\'a}nchez}, {Di Domizio}, {Dole}, {Escartin Vigo}, {Escoffier}, {Ferrari}, {Ferreira}, {Ferrero}, {Finelli}, {Fornari}, {Gabarra}, {Ganga}, {Garc{\'\i}a-Bellido}, {Gautard}, {Gaztanaga}, {Giacomini}, {Gozaliasl}, {Gregorio}, {Hall}, {Hartley}, {Hildebrandt}, {Hjorth}, {Huertas-Company}, {Ilbert}, {Kajava}, {Kansal}, {Karagiannis}, {Kirkpatrick}, {Legrand}, {Libet}, {Loureiro}, {Macias-Perez}, {Maggio}, {Magliocchetti}, {Mancini}, {Mannucci}, {Maoli}, {Martins}, {Matthew}, {Maturi},
  {Maurin}, {Metcalf}, {Monaco}, {Moretti}, {Morgante}, {Musi}, {Walton}, {Odier}, {Patrizii}, {P{\"o}ntinen}, {Popa}, {Porciani}, {Potter}, {Reimberg}, {Risso}, {Rocci}, {Sahl{\'e}n}, {Schneider}, {Sereno}, {Simon}, {Spurio Mancini}, {Stanford}, {Tao}, {Testera}, {Teyssier}, {Tosi}, {Troja}, {Tucci}, {Valieri}, {Valiviita}, {Vergani}, {Verza}, \& {Shankar}}]{CDS2024}
{Euclid Collaboration}, {McPartland}, C.~J.~R., {Zalesky}, L., {et~al.} 2024, \href{https://ui.adsabs.harvard.edu/abs/2024arXiv240805275E}{\href{http://dx.doi.org/10.48550/arXiv.2408.05275}{\color{magenta}arXiv e-prints}, arXiv:2408.05275}

\bibitem[{{Feng} {et~al.}(2016){Feng}, {Di-Matteo}, {Croft}, {Bird}, {Battaglia}, \& {Wilkins}}]{Feng2016}
{Feng}, Y., {Di-Matteo}, T., {Croft}, R.~A., {et~al.} 2016, \href{http://dx.doi.org/10.1093/mnras/stv2484}{\color{magenta}\mnras}, \href{https://ui.adsabs.harvard.edu/abs/2016MNRAS.455.2778F}{455, 2778}

\bibitem[{{Finkelstein} \& {Bagley}(2022)}]{Finkelstein2022}
{Finkelstein}, S.~L. \& {Bagley}, M.~B. 2022, \href{http://dx.doi.org/10.3847/1538-4357/ac89eb}{\color{magenta}\apj}, \href{https://ui.adsabs.harvard.edu/abs/2022ApJ...938...25F}{938, 25}

\bibitem[{{Finkelstein} {et~al.}(2023){Finkelstein}, {Leung}, {Bagley}, {Dickinson}, {Ferguson}, {Papovich}, {Akins}, {Arrabal Haro}, {Dave}, {Dekel}, {Kartaltepe}, {Kocevski}, {Koekemoer}, {Pirzkal}, {Somerville}, {Yung}, {Amorin}, {Backhaus}, {Behroozi}, {Bisigello}, {Bromm}, {Casey}, {Chavez Ortiz}, {Cheng}, {Chworowsky}, {Cleri}, {Cooper}, {Davis}, {de la Vega}, {Elbaz}, {Franco}, {Fontana}, {Fujimoto}, {Giavalisco}, {Grogin}, {Holwerda}, {Huertas-Company}, {Hirschmann}, {Iyer}, {Jogee}, {Jung}, {Larson}, {Lucas}, {Mobasher}, {Morales}, {Morley}, {Mukherjee}, {Perez-Gonzalez}, {Ravindranath}, {Rodighiero}, {Rowland}, {Tacchella}, {Taylor}, {Trump}, \& {Wilkins}}]{Finkelstein2023}
{Finkelstein}, S.~L., {Leung}, G. C.~K., {Bagley}, M.~B., {et~al.} 2023, \href{https://ui.adsabs.harvard.edu/abs/2023arXiv231104279F}{\href{http://dx.doi.org/10.48550/arXiv.2311.04279}{\color{magenta}arXiv e-prints}, arXiv:2311.04279}

\bibitem[{{Fontana} {et~al.}(2006){Fontana}, {Salimbeni}, {Grazian}, {Giallongo}, {Pentericci}, {Nonino}, {Fontanot}, {Menci}, {Monaco}, {Cristiani}, {Vanzella}, {de Santis}, \& {Gallozzi}}]{Fontana06}
{Fontana}, A., {Salimbeni}, S., {Grazian}, A., {et~al.} 2006, \href{http://dx.doi.org/10.1051/0004-6361:20065475}{\color{magenta}\aap}, \href{https://ui.adsabs.harvard.edu/abs/2006A&A...459..745F}{459, 745}

\bibitem[{Foreman-Mackey {et~al.}(2013)Foreman-Mackey, Hogg, Lang, \& Goodman}]{foreman-mackey_emcee_2013}
Foreman-Mackey, D., Hogg, D.~W., Lang, D., \& Goodman, J. 2013, \href{http://dx.doi.org/10.1086/670067}{\color{magenta}\pasp}, 125, 306

\bibitem[{{Franco} {et~al.}(2024){Franco}, {Akins}, {Casey}, {Finkelstein}, {Shuntov}, {Chworowsky}, {Faisst}, {Fujimoto}, {Ilbert}, {Koekemoer}, {Liu}, {Lovell}, {Maraston}, {McCracken}, {McKinney}, {Robertson}, {Bagley}, {Champagne}, {Cooper}, {Ding}, {Drakos}, {Enia}, {Gillman}, {Gozaliasl}, {Harish}, {Hayward}, {Hirschmann}, {Jin}, {Kartaltepe}, {Kokorev}, {Laigle}, {Long}, {Magdis}, {Mahler}, {Martin}, {Massey}, {Mobasher}, {Paquereau}, {Renzini}, {Rhodes}, {Rich}, {Sheth}, {Silverman}, {Sparre}, {Talia}, {Trakhtenbrot}, {Valentino}, {Vijayan}, {Wilkins}, {Yang}, \& {Zavala}}]{Franco2024}
{Franco}, M., {Akins}, H.~B., {Casey}, C.~M., {et~al.} 2024, \href{http://dx.doi.org/10.3847/1538-4357/ad5e6a}{\color{magenta}\apj}, \href{https://ui.adsabs.harvard.edu/abs/2024ApJ...973...23F}{973, 23}

\bibitem[{{Fujimoto} {et~al.}(2023){Fujimoto}, {Arrabal Haro}, {Dickinson}, {Finkelstein}, {Kartaltepe}, {Larson}, {Burgarella}, {Bagley}, {Behroozi}, {Chworowsky}, {Hirschmann}, {Trump}, {Wilkins}, {Yung}, {Koekemoer}, {Papovich}, {Pirzkal}, {Ferguson}, {Fontana}, {Grogin}, {Grazian}, {Kewley}, {Kocevski}, {Lotz}, {Pentericci}, {Ravindranath}, {Somerville}, {Wilkins}, {Amor{\'\i}n}, {Backhaus}, {Calabr{\`o}}, {Casey}, {Cooper}, {Fern{\'a}ndez}, {Franco}, {Giavalisco}, {Hathi}, {Harish}, {Hutchison}, {Iyer}, {Jung}, {Lucas}, \& {Zavala}}]{Fujimoto2023}
{Fujimoto}, S., {Arrabal Haro}, P., {Dickinson}, M., {et~al.} 2023, \href{http://dx.doi.org/10.3847/2041-8213/acd2d9}{\color{magenta}\apjl}, \href{https://ui.adsabs.harvard.edu/abs/2023ApJ...949L..25F}{949, L25}

\bibitem[{{Furtak} {et~al.}(2023){Furtak}, {Shuntov}, {Atek}, {Zitrin}, {Richard}, {Lehnert}, \& {Chevallard}}]{Furtak20230}
{Furtak}, L.~J., {Shuntov}, M., {Atek}, H., {et~al.} 2023, \href{http://dx.doi.org/10.1093/mnras/stac3717}{\color{magenta}\mnras}, \href{https://ui.adsabs.harvard.edu/abs/2023MNRAS.519.3064F}{519, 3064}

\bibitem[{Gabor \& Davé(2015)}]{gabor_hot_2015}
Gabor, J.~M. \& Davé, R. 2015, \href{http://dx.doi.org/10.1093/mnras/stu2399}{\color{magenta}\mnras}, 447, 374

\bibitem[{{Gehrels}(1986)}]{Gehrels1986}
{Gehrels}, N. 1986, \href{http://dx.doi.org/10.1086/164079}{\color{magenta}\apj}, \href{https://ui.adsabs.harvard.edu/abs/1986ApJ...303..336G}{303, 336}

\bibitem[{{Gelli} {et~al.}(2024){Gelli}, {Mason}, \& {Hayward}}]{Gelli2024}
{Gelli}, V., {Mason}, C., \& {Hayward}, C.~C. 2024, \href{https://ui.adsabs.harvard.edu/abs/2024arXiv240513108G}{\href{http://dx.doi.org/10.48550/arXiv.2405.13108}{\color{magenta}arXiv e-prints}, arXiv:2405.13108}

\bibitem[{{Gentile} {et~al.}(2024){Gentile}, {Casey}, {Akins}, {Franco}, {McKinney}, {Berman}, {Cooper}, {Drakos}, {Hirschmann}, {Long}, {Magdis}, {Koekemoer}, {Kokorev}, {Shuntov}, {Talia}, {Allen}, {Harish}, {Ilbert}, {McCracken}, {Kartaltepe}, {Liu}, {Paquereau}, {Rhodes}, {Rich}, {Robertson}, {Toft}, \& {Gozaliasl}}]{Gentile2024}
{Gentile}, F., {Casey}, C.~M., {Akins}, H.~B., {et~al.} 2024, \href{https://ui.adsabs.harvard.edu/abs/2024arXiv240810305G}{\href{http://dx.doi.org/10.48550/arXiv.2408.10305}{\color{magenta}arXiv e-prints}, arXiv:2408.10305}

\bibitem[{{Gim{\'e}nez-Arteaga} {et~al.}(2023){Gim{\'e}nez-Arteaga}, {Oesch}, {Brammer}, {Valentino}, {Mason}, {Weibel}, {Barrufet}, {Fujimoto}, {Heintz}, {Nelson}, {Strait}, {Suess}, \& {Gibson}}]{Gimenez-Arteaga2023}
{Gim{\'e}nez-Arteaga}, C., {Oesch}, P.~A., {Brammer}, G.~B., {et~al.} 2023, \href{http://dx.doi.org/10.3847/1538-4357/acc5ea}{\color{magenta}\apj}, \href{https://ui.adsabs.harvard.edu/abs/2023ApJ...948..126G}{948, 126}

\bibitem[{{Gottumukkala} {et~al.}(2023){Gottumukkala}, {Barrufet}, {Oesch}, {Weibel}, {Allen}, {Alcalde Pampliega}, {Nelson}, {Williams}, {Brammer}, {Fudamoto}, {Gonz{\'a}lez}, {Heintz}, {Illingworth}, {Magee}, {Naidu}, {Shuntov}, {Stefanon}, {Toft}, {Valentino}, \& {Xiao}}]{Gottumukkala2023}
{Gottumukkala}, R., {Barrufet}, L., {Oesch}, P.~A., {et~al.} 2023, \href{https://ui.adsabs.harvard.edu/abs/2023arXiv231003787G}{\href{http://dx.doi.org/10.48550/arXiv.2310.03787}{\color{magenta}arXiv e-prints}, arXiv:2310.03787}

\bibitem[{{Grazian} {et~al.}(2015){Grazian}, {Fontana}, {Santini}, {Dunlop}, {Ferguson}, {Castellano}, {Amorin}, {Ashby}, {Barro}, {Behroozi}, {Boutsia}, {Caputi}, {Chary}, {Dekel}, {Dickinson}, {Faber}, {Fazio}, {Finkelstein}, {Galametz}, {Giallongo}, {Giavalisco}, {Grogin}, {Guo}, {Kocevski}, {Koekemoer}, {Koo}, {Lee}, {Lu}, {Merlin}, {Mobasher}, {Nonino}, {Papovich}, {Paris}, {Pentericci}, {Reddy}, {Renzini}, {Salmon}, {Salvato}, {Sommariva}, {Song}, \& {Vanzella}}]{Grazian2015}
{Grazian}, A., {Fontana}, A., {Santini}, P., {et~al.} 2015, \href{http://dx.doi.org/10.1051/0004-6361/201424750}{\color{magenta}\aap}, \href{https://ui.adsabs.harvard.edu/abs/2015A&A...575A..96G}{575, A96}

\bibitem[{{Greene} {et~al.}(2024){Greene}, {Labbe}, {Goulding}, {Furtak}, {Chemerynska}, {Kokorev}, {Dayal}, {Volonteri}, {Williams}, {Wang}, {Setton}, {Burgasser}, {Bezanson}, {Atek}, {Brammer}, {Cutler}, {Feldmann}, {Fujimoto}, {Glazebrook}, {de Graaff}, {Khullar}, {Leja}, {Marchesini}, {Maseda}, {Matthee}, {Miller}, {Naidu}, {Nanayakkara}, {Oesch}, {Pan}, {Papovich}, {Price}, {van Dokkum}, {Weaver}, {Whitaker}, \& {Zitrin}}]{Greene2024}
{Greene}, J.~E., {Labbe}, I., {Goulding}, A.~D., {et~al.} 2024, \href{http://dx.doi.org/10.3847/1538-4357/ad1e5f}{\color{magenta}\apj}, \href{https://ui.adsabs.harvard.edu/abs/2024ApJ...964...39G}{964, 39}

\bibitem[{{Grudi{\'c}} {et~al.}(2018){Grudi{\'c}}, {Hopkins}, {Faucher-Gigu{\`e}re}, {Quataert}, {Murray}, \& {Kere{\v{s}}}}]{Grudic2018}
{Grudi{\'c}}, M.~Y., {Hopkins}, P.~F., {Faucher-Gigu{\`e}re}, C.-A., {et~al.} 2018, \href{http://dx.doi.org/10.1093/mnras/sty035}{\color{magenta}\mnras}, \href{https://ui.adsabs.harvard.edu/abs/2018MNRAS.475.3511G}{475, 3511}

\bibitem[{{Hainline} {et~al.}(2011){Hainline}, {Blain}, {Smail}, {Alexander}, {Armus}, {Chapman}, \& {Ivison}}]{Hainline2011}
{Hainline}, L.~J., {Blain}, A.~W., {Smail}, I., {et~al.} 2011, \href{http://dx.doi.org/10.1088/0004-637X/740/2/96}{\color{magenta}\apj}, \href{https://ui.adsabs.harvard.edu/abs/2011ApJ...740...96H}{740, 96}

\bibitem[{{Harikane} {et~al.}(2022){Harikane}, {Ono}, {Ouchi}, {Liu}, {Sawicki}, {Shibuya}, {Behroozi}, {He}, {Shimasaku}, {Arnouts}, {Coupon}, {Fujimoto}, {Gwyn}, {Huang}, {Inoue}, {Kashikawa}, {Komiyama}, {Matsuoka}, \& {Willott}}]{Harikane2022}
{Harikane}, Y., {Ono}, Y., {Ouchi}, M., {et~al.} 2022, \href{http://dx.doi.org/10.3847/1538-4365/ac3dfc}{\color{magenta}\apjs}, \href{https://ui.adsabs.harvard.edu/abs/2022ApJS..259...20H}{259, 20}

\bibitem[{{Harikane} {et~al.}(2023{\natexlab{a}}){Harikane}, {Ouchi}, {Oguri}, {Ono}, {Nakajima}, {Isobe}, {Umeda}, {Mawatari}, \& {Zhang}}]{Harikane2023}
{Harikane}, Y., {Ouchi}, M., {Oguri}, M., {et~al.} 2023{\natexlab{a}}, \href{http://dx.doi.org/10.3847/1538-4365/acaaa9}{\color{magenta}\apjs}, \href{https://ui.adsabs.harvard.edu/abs/2023ApJS..265....5H}{265, 5}

\bibitem[{{Harikane} {et~al.}(2023{\natexlab{b}}){Harikane}, {Zhang}, {Nakajima}, {Ouchi}, {Isobe}, {Ono}, {Hatano}, {Xu}, \& {Umeda}}]{Harikane-AGN2023}
{Harikane}, Y., {Zhang}, Y., {Nakajima}, K., {et~al.} 2023{\natexlab{b}}, \href{http://dx.doi.org/10.3847/1538-4357/ad029e}{\color{magenta}\apj}, \href{https://ui.adsabs.harvard.edu/abs/2023ApJ...959...39H}{959, 39}

\bibitem[{{Harvey} {et~al.}(2024){Harvey}, {Conselice}, {Adams}, {Austin}, {Juodzbalis}, {Trussler}, {Li}, {Ormerod}, {Ferreira}, {Duan}, {Westcott}, {Harris}, {Bhatawdekar}, {Coe}, {Cohen}, {Caruana}, {Cheng}, {Driver}, {Frye}, {Furtak}, {Grogin}, {Hathi}, {Holwerda}, {Jansen}, {Koekemoer}, {Lovell}, {Marshall}, {Nonino}, {Smail}, {Vijayan}, {Wilkins}, {Windhorst}, {Willmer}, {Yan}, \& {Zitrin}}]{Harvey2024}
{Harvey}, T., {Conselice}, C., {Adams}, N.~J., {et~al.} 2024, \href{https://ui.adsabs.harvard.edu/abs/2024arXiv240303908H}{\href{http://dx.doi.org/10.48550/arXiv.2403.03908}{\color{magenta}arXiv e-prints}, arXiv:2403.03908}

\bibitem[{{Hasinger} {et~al.}(2018){Hasinger}, {Capak}, {Salvato}, {Barger}, {Cowie}, {Faisst}, {Hemmati}, {Kakazu}, {Kartaltepe}, {Masters}, {Mobasher}, {Nayyeri}, {Sanders}, {Scoville}, {Suh}, {Steinhardt}, \& {Yang}}]{hasinger18}
{Hasinger}, G., {Capak}, P., {Salvato}, M., {et~al.} 2018, \href{http://dx.doi.org/10.3847/1538-4357/aabacf}{\color{magenta}\apj}, \href{https://ui.adsabs.harvard.edu/abs/2018ApJ...858...77H}{858, 77}

\bibitem[{{Haskell} {et~al.}(2024){Haskell}, {Das}, {Smith}, {Cochrane}, {Hayward}, \& {Angl{\'e}s-Alc{\'a}zar}}]{Haskell2024}
{Haskell}, P., {Das}, S., {Smith}, D.~J.~B., {et~al.} 2024, \href{http://dx.doi.org/10.1093/mnrasl/slae019}{\color{magenta}\mnras}, \href{https://ui.adsabs.harvard.edu/abs/2024MNRAS.530L...7H}{530, L7}

\bibitem[{{Haskell} {et~al.}(2023){Haskell}, {Smith}, {Cochrane}, {Hayward}, \& {Angl{\'e}s-Alc{\'a}zar}}]{Haskell2023}
{Haskell}, P., {Smith}, D.~J.~B., {Cochrane}, R.~K., {Hayward}, C.~C., \& {Angl{\'e}s-Alc{\'a}zar}, D. 2023, \href{http://dx.doi.org/10.1093/mnras/stad2315}{\color{magenta}\mnras}, \href{https://ui.adsabs.harvard.edu/abs/2023MNRAS.525.1535H}{525, 1535}

\bibitem[{{Hayward} \& {Smith}(2015)}]{Hayward2015}
{Hayward}, C.~C. \& {Smith}, D. J.~B. 2015, \href{http://dx.doi.org/10.1093/mnras/stu2195}{\color{magenta}\mnras}, \href{https://ui.adsabs.harvard.edu/abs/2015MNRAS.446.1512H}{446, 1512}

\bibitem[{{Helton} {et~al.}(2023){Helton}, {Sun}, {Woodrum}, {Hainline}, {Willmer}, {Rieke}, {Rieke}, {Alberts}, {Eisenstein}, {Tacchella}, {Robertson}, {Johnson}, {Baker}, {Bhatawdekar}, {Bunker}, {Chen}, {Egami}, {Ji}, {Maiolino}, {Willott}, \& {Witstok}}]{Helton2023}
{Helton}, J.~M., {Sun}, F., {Woodrum}, C., {et~al.} 2023, \href{https://ui.adsabs.harvard.edu/abs/2023arXiv231104270H}{\href{http://dx.doi.org/10.48550/arXiv.2311.04270}{\color{magenta}arXiv e-prints}, arXiv:2311.04270}

\bibitem[{{Hopkins} \& {Beacom}(2006)}]{Hopkins2006}
{Hopkins}, A.~M. \& {Beacom}, J.~F. 2006, \href{http://dx.doi.org/10.1086/506610}{\color{magenta}\apj}, \href{https://ui.adsabs.harvard.edu/abs/2006ApJ...651..142H}{651, 142}

\bibitem[{{Hutter} {et~al.}(2021){Hutter}, {Dayal}, {Yepes}, {Gottl{\"o}ber}, {Legrand}, \& {Ucci}}]{Hutter2021}
{Hutter}, A., {Dayal}, P., {Yepes}, G., {et~al.} 2021, \href{http://dx.doi.org/10.1093/mnras/stab602}{\color{magenta}\mnras}, \href{https://ui.adsabs.harvard.edu/abs/2021MNRAS.503.3698H}{503, 3698}

\bibitem[{{Ilbert} {et~al.}(2015){Ilbert}, {Arnouts}, {Le Floc'h}, {Aussel}, {Bethermin}, {Capak}, {Hsieh}, {Kajisawa}, {Karim}, {Le F{\`e}vre}, {Lee}, {Lilly}, {McCracken}, {Michel-Dansac}, {Moutard}, {Renzini}, {Salvato}, {Sanders}, {Scoville}, {Sheth}, {Silverman}, {Smol{\v{c}}i{\'c}}, {Taniguchi}, \& {Tresse}}]{Ilbert15}
{Ilbert}, O., {Arnouts}, S., {Le Floc'h}, E., {et~al.} 2015, \href{http://dx.doi.org/10.1051/0004-6361/201425176}{\color{magenta}\aap}, \href{https://ui.adsabs.harvard.edu/abs/2015A&A...579A...2I}{579, A2}

\bibitem[{Ilbert {et~al.}(2006)Ilbert, Arnouts, McCracken, Bolzonella, Bertin, Le~Fèvre, Mellier, Zamorani, Pellò, Iovino, Tresse, Le~Brun, Bottini, Garilli, Maccagni, Picat, Scaramella, Scodeggio, Vettolani, Zanichelli, Adami, Bardelli, Cappi, Charlot, Ciliegi, Contini, Cucciati, Foucaud, Franzetti, Gavignaud, Guzzo, Marano, Marinoni, Mazure, Meneux, Merighi, Paltani, Pollo, Pozzetti, Radovich, Zucca, Bondi, Bongiorno, Busarello, de~La~Torre, Gregorini, Lamareille, Mathez, Merluzzi, Ripepi, Rizzo, \& Vergani}]{Ilbert06}
Ilbert, O., Arnouts, S., McCracken, H.~J., {et~al.} 2006, \href{http://dx.doi.org/10.1051/0004-6361:20065138}{\color{magenta}A\&A}, 457, 841

\bibitem[{{Ilbert} {et~al.}(2013){Ilbert}, {McCracken}, {Le F{\`e}vre}, {Capak}, {Dunlop}, {Karim}, {Renzini}, {Caputi}, {Boissier}, {Arnouts}, {Aussel}, {Comparat}, {Guo}, {Hudelot}, {Kartaltepe}, {Kneib}, {Krogager}, {Le Floc'h}, {Lilly}, {Mellier}, {Milvang-Jensen}, {Moutard}, {Onodera}, {Richard}, {Salvato}, {Sanders}, {Scoville}, {Silverman}, {Taniguchi}, {Tasca}, {Thomas}, {Toft}, {Tresse}, {Vergani}, {Wolk}, \& {Zirm}}]{Ilbert13}
{Ilbert}, O., {McCracken}, H.~J., {Le F{\`e}vre}, O., {et~al.} 2013, \href{http://dx.doi.org/10.1051/0004-6361/201321100}{\color{magenta}\aap}, \href{https://ui.adsabs.harvard.edu/abs/2013A&A...556A..55I}{556, A55}

\bibitem[{{Jespersen} {et~al.}(2022){Jespersen}, {Cranmer}, {Melchior}, {Ho}, {Somerville}, \& {Gabrielpillai}}]{Jespersen2022_mangrove}
{Jespersen}, C.~K., {Cranmer}, M., {Melchior}, P., {et~al.} 2022, \href{http://dx.doi.org/10.3847/1538-4357/ac9b18}{\color{magenta}\apj}, \href{https://ui.adsabs.harvard.edu/abs/2022ApJ...941....7J}{941, 7}

\bibitem[{{Jespersen} {et~al.}(2024){Jespersen}, {Steinhardt}, {Somerville}, \& {Lovell}}]{Jespersen2024_mostmassive}
{Jespersen}, C.~K., {Steinhardt}, C.~L., {Somerville}, R.~S., \& {Lovell}, C.~C. 2024, \href{https://ui.adsabs.harvard.edu/abs/2024arXiv240300050K}{\href{http://dx.doi.org/10.48550/arXiv.2403.00050}{\color{magenta}arXiv e-prints}, arXiv:2403.00050}

\bibitem[{{Jin} {et~al.}(2019){Jin}, {Daddi}, {Magdis}, {Liu}, {Schinnerer}, {Papadopoulos}, {Gu}, {Gao}, \& {Calabr{\`o}}}]{ShuowenJin2019}
{Jin}, S., {Daddi}, E., {Magdis}, G.~E., {et~al.} 2019, \href{http://dx.doi.org/10.3847/1538-4357/ab55d6}{\color{magenta}\apj}, \href{https://ui.adsabs.harvard.edu/abs/2019ApJ...887..144J}{887, 144}

\bibitem[{{Johnston}(2011)}]{Johnston2011}
{Johnston}, R. 2011, \href{http://dx.doi.org/10.1007/s00159-011-0041-9}{\color{magenta}\aapr}, \href{https://ui.adsabs.harvard.edu/abs/2011A&ARv..19...41J}{19, 41}

\bibitem[{{Kartaltepe} {et~al.}(2010){Kartaltepe}, {Sanders}, {Le Floc'h}, {Frayer}, {Aussel}, {Arnouts}, {Ilbert}, {Salvato}, {Scoville}, {Surace}, {Yan}, {Capak}, {Caputi}, {Carollo}, {Cassata}, {Civano}, {Hasinger}, {Koekemoer}, {Le F{\`e}vre}, {Lilly}, {Liu}, {McCracken}, {Schinnerer}, {Smol{\v{c}}i{\'c}}, {Taniguchi}, {Thompson}, {Trump}, {Baldassare}, \& {Fiorenza}}]{kartaltepe10}
{Kartaltepe}, J.~S., {Sanders}, D.~B., {Le Floc'h}, E., {et~al.} 2010, \href{http://dx.doi.org/10.1088/0004-637X/721/1/98}{\color{magenta}\apj}, \href{https://ui.adsabs.harvard.edu/abs/2010ApJ...721...98K}{721, 98}

\bibitem[{{Kartaltepe} {et~al.}(2015){Kartaltepe}, {Sanders}, {Silverman}, {Kashino}, {Chu}, {Zahid}, {Hasinger}, {Kewley}, {Matsuoka}, {Nagao}, {Riguccini}, {Salvato}, {Schawinski}, {Taniguchi}, {Treister}, {Capak}, {Daddi}, \& {Ohta}}]{kartaltepe15b}
{Kartaltepe}, J.~S., {Sanders}, D.~B., {Silverman}, J.~D., {et~al.} 2015, \href{http://dx.doi.org/10.1088/2041-8205/806/2/L35}{\color{magenta}\apjl}, \href{https://ui.adsabs.harvard.edu/abs/2015ApJ...806L..35K}{806, L35}

\bibitem[{{Kashino} {et~al.}(2019){Kashino}, {Silverman}, {Sanders}, {Kartaltepe}, {Daddi}, {Renzini}, {Rodighiero}, {Puglisi}, {Valentino}, {Juneau}, {Arimoto}, {Nagao}, {Ilbert}, {Le F{\`e}vre}, \& {Koekemoer}}]{kashino19}
{Kashino}, D., {Silverman}, J.~D., {Sanders}, D., {et~al.} 2019, \href{http://dx.doi.org/10.3847/1538-4365/ab06c4}{\color{magenta}\apjs}, \href{https://ui.adsabs.harvard.edu/abs/2019ApJS..241...10K}{241, 10}

\bibitem[{{Kennicutt}(1998)}]{Kennicut1998}
{Kennicutt}, Robert~C., J. 1998, \href{http://dx.doi.org/10.1086/305588}{\color{magenta}\apj}, \href{https://ui.adsabs.harvard.edu/abs/1998ApJ...498..541K}{498, 541}

\bibitem[{{Klypin} {et~al.}(2021){Klypin}, {Poulin}, {Prada}, {Primack}, {Kamionkowski}, {Avila-Reese}, {Rodriguez-Puebla}, {Behroozi}, {Hellinger}, \& {Smith}}]{Klypin2021}
{Klypin}, A., {Poulin}, V., {Prada}, F., {et~al.} 2021, \href{http://dx.doi.org/10.1093/mnras/stab769}{\color{magenta}\mnras}, \href{https://ui.adsabs.harvard.edu/abs/2021MNRAS.504..769K}{504, 769}

\bibitem[{{Kocevski} {et~al.}(2023){Kocevski}, {Onoue}, {Inayoshi}, {Trump}, {Arrabal Haro}, {Grazian}, {Dickinson}, {Finkelstein}, {Kartaltepe}, {Hirschmann}, {Aird}, {Holwerda}, {Fujimoto}, {Juneau}, {Amor{\'\i}n}, {Backhaus}, {Bagley}, {Barro}, {Bell}, {Bisigello}, {Calabr{\`o}}, {Cleri}, {Cooper}, {Ding}, {Grogin}, {Ho}, {Hutchison}, {Inoue}, {Jiang}, {Jones}, {Koekemoer}, {Li}, {Li}, {McGrath}, {Molina}, {Papovich}, {P{\'e}rez-Gonz{\'a}lez}, {Pirzkal}, {Wilkins}, {Yang}, \& {Yung}}]{Kocevski2023}
{Kocevski}, D.~D., {Onoue}, M., {Inayoshi}, K., {et~al.} 2023, \href{http://dx.doi.org/10.3847/2041-8213/ace5a0}{\color{magenta}\apjl}, \href{https://ui.adsabs.harvard.edu/abs/2023ApJ...954L...4K}{954, L4}

\bibitem[{{Koekemoer} {et~al.}(2007){Koekemoer}, {Aussel}, {Calzetti}, {Capak}, {Giavalisco}, {Kneib}, {Leauthaud}, {Le F{\`e}vre}, {McCracken}, {Massey}, {Mobasher}, {Rhodes}, {Scoville}, \& {Shopbell}}]{Koekemoer07}
{Koekemoer}, A.~M., {Aussel}, H., {Calzetti}, D., {et~al.} 2007, \href{http://dx.doi.org/10.1086/520086}{\color{magenta}\apjs}, \href{https://ui.adsabs.harvard.edu/abs/2007ApJS..172..196K}{172, 196}

\bibitem[{{Kokorev} {et~al.}(2024){Kokorev}, {Caputi}, {Greene}, {Dayal}, {Trebitsch}, {Cutler}, {Fujimoto}, {Labb{\'e}}, {Miller}, {Iani}, {Navarro-Carrera}, \& {Rinaldi}}]{Kokorev2024}
{Kokorev}, V., {Caputi}, K.~I., {Greene}, J.~E., {et~al.} 2024, \href{https://ui.adsabs.harvard.edu/abs/2024arXiv240109981K}{\href{http://dx.doi.org/10.48550/arXiv.2401.09981}{\color{magenta}arXiv e-prints}, arXiv:2401.09981}

\bibitem[{Kravtsov {et~al.}(2004)Kravtsov, Berlind, Wechsler, Klypin, Gottlöber, Allgood, \& Primack}]{kravtsov_dark_2004}
Kravtsov, A.~V., Berlind, A.~A., Wechsler, R.~H., {et~al.} 2004, \href{http://dx.doi.org/10.1086/420959}{\color{magenta}\apj}, 609, 35

\bibitem[{{Kriek} {et~al.}(2015){Kriek}, {Shapley}, {Reddy}, {Siana}, {Coil}, {Mobasher}, {Freeman}, {de Groot}, {Price}, {Sanders}, {Shivaei}, {Brammer}, {Momcheva}, {Skelton}, {van Dokkum}, {Whitaker}, {Aird}, {Azadi}, {Kassis}, {Bullock}, {Conroy}, {Dav{\'e}}, {Kere{\v{s}}}, \& {Krumholz}}]{kriek15}
{Kriek}, M., {Shapley}, A.~E., {Reddy}, N.~A., {et~al.} 2015, \href{http://dx.doi.org/10.1088/0067-0049/218/2/15}{\color{magenta}\apjs}, \href{https://ui.adsabs.harvard.edu/abs/2015ApJS..218...15K}{218, 15}

\bibitem[{{K{\"u}mmel} {et~al.}(2020){K{\"u}mmel}, {Bertin}, {Schefer}, {Apostolakos}, {{\'A}lvarez-Ayll{\'o}n}, \& {Dubath}}]{kummel20}
{K{\"u}mmel}, M., {Bertin}, E., {Schefer}, M., {et~al.} 2020, in Astronomical Society of the Pacific Conference Series, Vol. 527, Astronomical Data Analysis Software and Systems XXIX, ed. R.~{Pizzo}, E.~R. {Deul}, J.~D. {Mol}, J.~{de Plaa}, \& H.~{Verkouter}, \href{https://ui.adsabs.harvard.edu/abs/2020ASPC..527...29K}{29}

\bibitem[{{Labb{\'e}} {et~al.}(2023{\natexlab{a}}){Labb{\'e}}, {Greene}, {Bezanson}, {Fujimoto}, {Furtak}, {Goulding}, {Matthee}, {Naidu}, {Oesch}, {Atek}, {Brammer}, {Chemerynska}, {Coe}, {Cutler}, {Dayal}, {Feldmann}, {Franx}, {Glazebrook}, {Leja}, {Marchesini}, {Maseda}, {Nanayakkara}, {Nelson}, {Pan}, {Papovich}, {Price}, {Suess}, {Wang}, {Whitaker}, {Williams}, \& {Zitrin}}]{Labbe2023}
{Labb{\'e}}, I., {Greene}, J.~E., {Bezanson}, R., {et~al.} 2023{\natexlab{a}}, \href{https://ui.adsabs.harvard.edu/abs/2023arXiv230607320L}{\href{http://dx.doi.org/10.48550/arXiv.2306.07320}{\color{magenta}arXiv e-prints}, arXiv:2306.07320}

\bibitem[{{Labb{\'e}} {et~al.}(2023{\natexlab{b}}){Labb{\'e}}, {van Dokkum}, {Nelson}, {Bezanson}, {Suess}, {Leja}, {Brammer}, {Whitaker}, {Mathews}, {Stefanon}, \& {Wang}}]{Labbe2023Natur}
{Labb{\'e}}, I., {van Dokkum}, P., {Nelson}, E., {et~al.} 2023{\natexlab{b}}, \href{http://dx.doi.org/10.1038/s41586-023-05786-2}{\color{magenta}\nat}, \href{https://ui.adsabs.harvard.edu/abs/2023Natur.616..266L}{616, 266}

\bibitem[{{Le F{\`e}vre} {et~al.}(2015){Le F{\`e}vre}, {Tasca}, {Cassata}, {Garilli}, {Le Brun}, {Maccagni}, {Pentericci}, {Thomas}, {Vanzella}, {Zamorani}, {Zucca}, {Amorin}, {Bardelli}, {Capak}, {Cassar{\`a}}, {Castellano}, {Cimatti}, {Cuby}, {Cucciati}, {de la Torre}, {Durkalec}, {Fontana}, {Giavalisco}, {Grazian}, {Hathi}, {Ilbert}, {Lemaux}, {Moreau}, {Paltani}, {Ribeiro}, {Salvato}, {Schaerer}, {Scodeggio}, {Sommariva}, {Talia}, {Taniguchi}, {Tresse}, {Vergani}, {Wang}, {Charlot}, {Contini}, {Fotopoulou}, {L{\'o}pez-Sanjuan}, {Mellier}, \& {Scoville}}]{lefevre15}
{Le F{\`e}vre}, O., {Tasca}, L.~A.~M., {Cassata}, P., {et~al.} 2015, \href{http://dx.doi.org/10.1051/0004-6361/201423829}{\color{magenta}\aap}, \href{https://ui.adsabs.harvard.edu/abs/2015A&A...576A..79L}{576, A79}

\bibitem[{Leauthaud {et~al.}(2012)Leauthaud, Tinker, Bundy, Behroozi, Massey, Rhodes, George, Kneib, Benson, Wechsler, Busha, Capak, Cortes, Ilbert, Koekemoer, Fevre, Lilly, McCracken, Salvato, Schrabback, Scoville, Smith, \& Taylor}]{leauthaud_new_2012}
Leauthaud, A., Tinker, J., Bundy, K., {et~al.} 2012, \href{http://dx.doi.org/10.1088/0004-637X/744/2/159}{\color{magenta}\apj}, 744, 159

\bibitem[{{Legrand} {et~al.}(2019){Legrand}, {McCracken}, {Davidzon}, {Ilbert}, {Coupon}, {Aghanim}, {Douspis}, {Capak}, {Le F{\`e}vre}, \& {Milvang-Jensen}}]{Legrand2019}
{Legrand}, L., {McCracken}, H.~J., {Davidzon}, I., {et~al.} 2019, \href{http://dx.doi.org/10.1093/mnras/stz1198}{\color{magenta}\mnras}, \href{https://ui.adsabs.harvard.edu/abs/2019MNRAS.486.5468L}{486, 5468}

\bibitem[{{Leja} {et~al.}(2015){Leja}, {van Dokkum}, {Franx}, \& {Whitaker}}]{Leja2015}
{Leja}, J., {van Dokkum}, P.~G., {Franx}, M., \& {Whitaker}, K.~E. 2015, \href{http://dx.doi.org/10.1088/0004-637X/798/2/115}{\color{magenta}\apj}, \href{https://ui.adsabs.harvard.edu/abs/2015ApJ...798..115L}{798, 115}

\bibitem[{{Li} {et~al.}(2023){Li}, {Dekel}, {Sarkar}, {Aung}, {Giavalisco}, {Mandelker}, \& {Tacchella}}]{Li2023}
{Li}, Z., {Dekel}, A., {Sarkar}, K.~C., {et~al.} 2023, \href{https://ui.adsabs.harvard.edu/abs/2023arXiv231114662L}{\href{http://dx.doi.org/10.48550/arXiv.2311.14662}{\color{magenta}arXiv e-prints}, arXiv:2311.14662}

\bibitem[{{Lilly} {et~al.}(2013){Lilly}, {Carollo}, {Pipino}, {Renzini}, \& {Peng}}]{Lilly2013}
{Lilly}, S.~J., {Carollo}, C.~M., {Pipino}, A., {Renzini}, A., \& {Peng}, Y. 2013, \href{http://dx.doi.org/10.1088/0004-637X/772/2/119}{\color{magenta}\apj}, \href{https://ui.adsabs.harvard.edu/abs/2013ApJ...772..119L}{772, 119}

\bibitem[{{Lilly} {et~al.}(2009){Lilly}, {Le Brun}, {Maier}, {Mainieri}, {Mignoli}, {Scodeggio}, {Zamorani}, {Carollo}, {Contini}, {Kneib}, {Le F{\`e}vre}, {Renzini}, {Bardelli}, {Bolzonella}, {Bongiorno}, {Caputi}, {Coppa}, {Cucciati}, {de la Torre}, {de Ravel}, {Franzetti}, {Garilli}, {Iovino}, {Kampczyk}, {Kovac}, {Knobel}, {Lamareille}, {Le Borgne}, {Pello}, {Peng}, {P{\'e}rez-Montero}, {Ricciardelli}, {Silverman}, {Tanaka}, {Tasca}, {Tresse}, {Vergani}, {Zucca}, {Ilbert}, {Salvato}, {Oesch}, {Abbas}, {Bottini}, {Capak}, {Cappi}, {Cassata}, {Cimatti}, {Elvis}, {Fumana}, {Guzzo}, {Hasinger}, {Koekemoer}, {Leauthaud}, {Maccagni}, {Marinoni}, {McCracken}, {Memeo}, {Meneux}, {Porciani}, {Pozzetti}, {Sanders}, {Scaramella}, {Scarlata}, {Scoville}, {Shopbell}, \& {Taniguchi}}]{lilly09}
{Lilly}, S.~J., {Le Brun}, V., {Maier}, C., {et~al.} 2009, \href{http://dx.doi.org/10.1088/0067-0049/184/2/218}{\color{magenta}\apjs}, \href{https://ui.adsabs.harvard.edu/abs/2009ApJS..184..218L}{184, 218}

\bibitem[{{Liu} {et~al.}(2019){Liu}, {Schinnerer}, {Groves}, {Magnelli}, {Lang}, {Leslie}, {Jim{\'e}nez-Andrade}, {Riechers}, {Popping}, {Magdis}, {Daddi}, {Sargent}, {Gao}, {Fudamoto}, {Oesch}, \& {Bertoldi}}]{LiuD2019b}
{Liu}, D., {Schinnerer}, E., {Groves}, B., {et~al.} 2019, \href{http://dx.doi.org/10.3847/1538-4357/ab578d}{\color{magenta}\apj}, \href{https://ui.adsabs.harvard.edu/abs/2019ApJ...887..235L}{887, 235}

\bibitem[{{Liu} {et~al.}(2024){Liu}, {Zhan}, {Gong}, \& {Wang}}]{Liu2024}
{Liu}, W., {Zhan}, H., {Gong}, Y., \& {Wang}, X. 2024, \href{https://ui.adsabs.harvard.edu/abs/2024arXiv240214339L}{\href{http://dx.doi.org/10.48550/arXiv.2402.14339}{\color{magenta}arXiv e-prints}, arXiv:2402.14339}

\bibitem[{{Lovell} {et~al.}(2023){Lovell}, {Harrison}, {Harikane}, {Tacchella}, \& {Wilkins}}]{Lovell2023}
{Lovell}, C.~C., {Harrison}, I., {Harikane}, Y., {Tacchella}, S., \& {Wilkins}, S.~M. 2023, \href{http://dx.doi.org/10.1093/mnras/stac3224}{\color{magenta}\mnras}, \href{https://ui.adsabs.harvard.edu/abs/2023MNRAS.518.2511L}{518, 2511}

\bibitem[{{Lovell} {et~al.}(2021){Lovell}, {Vijayan}, {Thomas}, {Wilkins}, {Barnes}, {Irodotou}, \& {Roper}}]{Lovell2021}
{Lovell}, C.~C., {Vijayan}, A.~P., {Thomas}, P.~A., {et~al.} 2021, \href{http://dx.doi.org/10.1093/mnras/staa3360}{\color{magenta}\mnras}, \href{https://ui.adsabs.harvard.edu/abs/2021MNRAS.500.2127L}{500, 2127}

\bibitem[{{Madau}(1995)}]{Madau95}
{Madau}, P. 1995, \href{http://dx.doi.org/10.1086/175332}{\color{magenta}\apj}, \href{https://ui.adsabs.harvard.edu/abs/1995ApJ...441...18M}{441, 18}

\bibitem[{{Madau} \& {Dickinson}(2014)}]{MadauDickinson2014}
{Madau}, P. \& {Dickinson}, M. 2014, \href{http://dx.doi.org/10.1146/annurev-astro-081811-125615}{\color{magenta}\araa}, \href{https://ui.adsabs.harvard.edu/abs/2014ARA&A..52..415M}{52, 415}

\bibitem[{{Madau} {et~al.}(1998){Madau}, {Pozzetti}, \& {Dickinson}}]{Madau1998}
{Madau}, P., {Pozzetti}, L., \& {Dickinson}, M. 1998, \href{http://dx.doi.org/10.1086/305523}{\color{magenta}\apj}, \href{https://ui.adsabs.harvard.edu/abs/1998ApJ...498..106M}{498, 106}

\bibitem[{{Maiolino} {et~al.}(2023){Maiolino}, {Scholtz}, {Curtis-Lake}, {Carniani}, {Baker}, {de Graaff}, {Tacchella}, {{\"U}bler}, {D'Eugenio}, {Witstok}, {Curti}, {Arribas}, {Bunker}, {Charlot}, {Chevallard}, {Eisenstein}, {Egami}, {Ji}, {Jones}, {Lyu}, {Rawle}, {Robertson}, {Rujopakarn}, {Perna}, {Sun}, {Venturi}, {Williams}, \& {Willott}}]{Maiolino2023}
{Maiolino}, R., {Scholtz}, J., {Curtis-Lake}, E., {et~al.} 2023, \href{https://ui.adsabs.harvard.edu/abs/2023arXiv230801230M}{\href{http://dx.doi.org/10.48550/arXiv.2308.01230}{\color{magenta}arXiv e-prints}, arXiv:2308.01230}

\bibitem[{{Malmquist}(1922)}]{Malmquist1922}
{Malmquist}, K.~G. 1922, Meddelanden fran Lunds Astronomiska Observatorium Serie I, \href{https://ui.adsabs.harvard.edu/abs/1922MeLuF.100....1M}{100, 1}

\bibitem[{Marinacci {et~al.}(2018)Marinacci, Vogelsberger, Pakmor, Torrey, Springel, Hernquist, Nelson, Weinberger, Pillepich, Naiman, \& Genel}]{marinacci_first_2018}
Marinacci, F., Vogelsberger, M., Pakmor, R., {et~al.} 2018, \href{http://dx.doi.org/10.1093/mnras/sty2206}{\color{magenta}\mnras}, 480, 5113

\bibitem[{Marinoni \& Hudson(2002)}]{marinoni_mass--light_2002}
Marinoni, C. \& Hudson, M.~J. 2002, \href{http://dx.doi.org/10.1086/339319}{\color{magenta}\apj}, 569, 101

\bibitem[{{Matthee} {et~al.}(2024){Matthee}, {Naidu}, {Brammer}, {Chisholm}, {Eilers}, {Goulding}, {Greene}, {Kashino}, {Labbe}, {Lilly}, {Mackenzie}, {Oesch}, {Weibel}, {Wuyts}, {Xiao}, {Bordoloi}, {Bouwens}, {van Dokkum}, {Illingworth}, {Kramarenko}, {Maseda}, {Mason}, {Meyer}, {Nelson}, {Reddy}, {Shivaei}, {Simcoe}, \& {Yue}}]{Matthee2024}
{Matthee}, J., {Naidu}, R.~P., {Brammer}, G., {et~al.} 2024, \href{http://dx.doi.org/10.3847/1538-4357/ad2345}{\color{magenta}\apj}, \href{https://ui.adsabs.harvard.edu/abs/2024ApJ...963..129M}{963, 129}

\bibitem[{{McCracken} {et~al.}(2012){McCracken}, {Milvang-Jensen}, {Dunlop}, {Franx}, {Fynbo}, {Le F{\`e}vre}, {Holt}, {Caputi}, {Goranova}, {Buitrago}, {Emerson}, {Freudling}, {Hudelot}, {L{\'o}pez-Sanjuan}, {Magnard}, {Mellier}, {M{\o}ller}, {Nilsson}, {Sutherland}, {Tasca}, \& {Zabl}}]{McCracken12}
{McCracken}, H.~J., {Milvang-Jensen}, B., {Dunlop}, J., {et~al.} 2012, \href{http://dx.doi.org/10.1051/0004-6361/201219507}{\color{magenta}\aap}, \href{https://ui.adsabs.harvard.edu/abs/2012A&A...544A.156M}{544, A156}

\bibitem[{{McKinney} {et~al.}(2024){McKinney}, {Casey}, {Long}, {Cooper}, {Manning}, {Franco}, {Akin}, {Lambrides}, {Gammon}, {Silva}, {Gentile}, {Zavala}, {Amvrosiadis}, {Andika}, {Brinch}, {Champagne}, {Chartab}, {Drakos}, {Faisst}, {Fujimoto}, {Gillman}, {Gozaliasl}, {Greve}, {Harish}, {Hayward}, {Hirschmann}, {Ilbert}, {Kalita}, {Kartaltepe}, {Koekemoer}, {Kokorev}, {Liu}, {Magdis}, {McCracken}, {Rhodes}, {Robertson}, {Talia}, {Valentino}, \& {Vijayan}}]{McKinney2024}
{McKinney}, J., {Casey}, C.~M., {Long}, A.~S., {et~al.} 2024, \href{https://ui.adsabs.harvard.edu/abs/2024arXiv240808346M}{\href{http://dx.doi.org/10.48550/arXiv.2408.08346}{\color{magenta}arXiv e-prints}, arXiv:2408.08346}

\bibitem[{{McLeod} {et~al.}(2021){McLeod}, {McLure}, {Dunlop}, {Cullen}, {Carnall}, \& {Duncan}}]{McLeod2021}
{McLeod}, D.~J., {McLure}, R.~J., {Dunlop}, J.~S., {et~al.} 2021, \href{http://dx.doi.org/10.1093/mnras/stab731}{\color{magenta}\mnras}, \href{https://ui.adsabs.harvard.edu/abs/2021MNRAS.503.4413M}{503, 4413}

\bibitem[{{Micha{\l}owski} {et~al.}(2012){Micha{\l}owski}, {Dunlop}, {Cirasuolo}, {Hjorth}, {Hayward}, \& {Watson}}]{Michalowski2012}
{Micha{\l}owski}, M.~J., {Dunlop}, J.~S., {Cirasuolo}, M., {et~al.} 2012, \href{http://dx.doi.org/10.1051/0004-6361/201016308}{\color{magenta}\aap}, \href{https://ui.adsabs.harvard.edu/abs/2012A&A...541A..85M}{541, A85}

\bibitem[{{Mitchell} {et~al.}(2013){Mitchell}, {Lacey}, {Baugh}, \& {Cole}}]{Mitchell2013}
{Mitchell}, P.~D., {Lacey}, C.~G., {Baugh}, C.~M., \& {Cole}, S. 2013, \href{http://dx.doi.org/10.1093/mnras/stt1280}{\color{magenta}\mnras}, \href{https://ui.adsabs.harvard.edu/abs/2013MNRAS.435...87M}{435, 87}

\bibitem[{{Moster} {et~al.}(2011){Moster}, {Somerville}, {Newman}, \& {Rix}}]{Moster2011_cosmicvariancecookbook}
{Moster}, B.~P., {Somerville}, R.~S., {Newman}, J.~A., \& {Rix}, H.-W. 2011, \href{http://dx.doi.org/10.1088/0004-637X/731/2/113}{\color{magenta}\apj}, \href{https://ui.adsabs.harvard.edu/abs/2011ApJ...731..113M}{731, 113}

\bibitem[{{Moutard} {et~al.}(2016){Moutard}, {Arnouts}, {Ilbert}, {Coupon}, {Davidzon}, {Guzzo}, {Hudelot}, {McCracken}, {Van Waerbeke}, {Morrison}, {Le F{\`e}vre}, {Comte}, {Bolzonella}, {Fritz}, {Garilli}, \& {Scodeggio}}]{Moutard2016}
{Moutard}, T., {Arnouts}, S., {Ilbert}, O., {et~al.} 2016, \href{http://dx.doi.org/10.1051/0004-6361/201527294}{\color{magenta}\aap}, \href{https://ui.adsabs.harvard.edu/abs/2016A&A...590A.103M}{590, A103}

\bibitem[{{Muzzin} {et~al.}(2013){Muzzin}, {Marchesini}, {Stefanon}, {Franx}, {McCracken}, {Milvang-Jensen}, {Dunlop}, {Fynbo}, {Brammer}, {Labb{\'e}}, \& {van Dokkum}}]{Muzzin2013}
{Muzzin}, A., {Marchesini}, D., {Stefanon}, M., {et~al.} 2013, \href{http://dx.doi.org/10.1088/0004-637X/777/1/18}{\color{magenta}\apj}, \href{https://ui.adsabs.harvard.edu/abs/2013ApJ...777...18M}{777, 18}

\bibitem[{Naiman {et~al.}(2018)Naiman, Pillepich, Springel, Ramirez-Ruiz, Torrey, Vogelsberger, Pakmor, Nelson, Marinacci, Hernquist, Weinberger, \& Genel}]{naiman_first_2018}
Naiman, J.~P., Pillepich, A., Springel, V., {et~al.} 2018, \href{http://dx.doi.org/10.1093/mnras/sty618}{\color{magenta}\mnras}, 477, 1206

\bibitem[{{Narayanan} {et~al.}(2024){Narayanan}, {Lower}, {Torrey}, {Brammer}, {Cui}, {Dav{\'e}}, {Iyer}, {Li}, {Lovell}, {Sales}, {Stark}, {Marinacci}, \& {Vogelsberger}}]{Narayanan2024}
{Narayanan}, D., {Lower}, S., {Torrey}, P., {et~al.} 2024, \href{http://dx.doi.org/10.3847/1538-4357/ad0966}{\color{magenta}\apj}, \href{https://ui.adsabs.harvard.edu/abs/2024ApJ...961...73N}{961, 73}

\bibitem[{{Navarro-Carrera} {et~al.}(2023){Navarro-Carrera}, {Rinaldi}, {Caputi}, {Iani}, {Kokorev}, \& {van Mierlo}}]{Navarro-Carrera2023}
{Navarro-Carrera}, R., {Rinaldi}, P., {Caputi}, K.~I., {et~al.} 2023, \href{https://ui.adsabs.harvard.edu/abs/2023arXiv230516141N}{\href{http://dx.doi.org/10.48550/arXiv.2305.16141}{\color{magenta}arXiv e-prints}, arXiv:2305.16141}

\bibitem[{Nelson {et~al.}(2018)Nelson, Pillepich, Springel, Weinberger, Hernquist, Pakmor, Genel, Torrey, Vogelsberger, Kauffmann, Marinacci, \& Naiman}]{nelson_first_2018}
Nelson, D., Pillepich, A., Springel, V., {et~al.} 2018, \href{http://dx.doi.org/10.1093/mnras/stx3040}{\color{magenta}\mnras}, 475, 624

\bibitem[{Oke \& Gunn(1983)}]{1983ApJ...266..713O}
Oke, J.~B. \& Gunn, J.~E. 1983, \href{http://dx.doi.org/10.1086/160817}{\color{magenta}\apj}, 266, 713

\bibitem[{{Pacifici} {et~al.}(2023){Pacifici}, {Iyer}, {Mobasher}, {da Cunha}, {Acquaviva}, {Burgarella}, {Calistro Rivera}, {Carnall}, {Chang}, {Chartab}, {Cooke}, {Fairhurst}, {Kartaltepe}, {Leja}, {Ma{\l}ek}, {Salmon}, {Torelli}, {Vidal-Garc{\'\i}a}, {Boquien}, {Brammer}, {Brown}, {Capak}, {Chevallard}, {Circosta}, {Croton}, {Davidzon}, {Dickinson}, {Duncan}, {Faber}, {Ferguson}, {Fontana}, {Guo}, {Haeussler}, {Hemmati}, {Jafariyazani}, {Kassin}, {Larson}, {Lee}, {Mantha}, {Marchi}, {Nayyeri}, {Newman}, {Pandya}, {Pforr}, {Reddy}, {Sanders}, {Shah}, {Shahidi}, {Stevans}, {Triani}, {Tyler}, {Vanderhoof}, {de la Vega}, {Wang}, \& {Weston}}]{Pacifici2023}
{Pacifici}, C., {Iyer}, K.~G., {Mobasher}, B., {et~al.} 2023, \href{http://dx.doi.org/10.3847/1538-4357/acacff}{\color{magenta}\apj}, \href{https://ui.adsabs.harvard.edu/abs/2023ApJ...944..141P}{944, 141}

\bibitem[{{Pallottini} \& {Ferrara}(2023)}]{PallottiniFerrara2023}
{Pallottini}, A. \& {Ferrara}, A. 2023, \href{http://dx.doi.org/10.1051/0004-6361/202347384}{\color{magenta}\aap}, \href{https://ui.adsabs.harvard.edu/abs/2023A&A...677L...4P}{677, L4}

\bibitem[{{Peng} {et~al.}(2010){Peng}, {Lilly}, {Kova{\v{c}}}, {Bolzonella}, {Pozzetti}, {Renzini}, {Zamorani}, {Ilbert}, {Knobel}, {Iovino}, {Maier}, {Cucciati}, {Tasca}, {Carollo}, {Silverman}, {Kampczyk}, {de Ravel}, {Sanders}, {Scoville}, {Contini}, {Mainieri}, {Scodeggio}, {Kneib}, {Le F{\`e}vre}, {Bardelli}, {Bongiorno}, {Caputi}, {Coppa}, {de la Torre}, {Franzetti}, {Garilli}, {Lamareille}, {Le Borgne}, {Le Brun}, {Mignoli}, {Perez Montero}, {Pello}, {Ricciardelli}, {Tanaka}, {Tresse}, {Vergani}, {Welikala}, {Zucca}, {Oesch}, {Abbas}, {Barnes}, {Bordoloi}, {Bottini}, {Cappi}, {Cassata}, {Cimatti}, {Fumana}, {Hasinger}, {Koekemoer}, {Leauthaud}, {Maccagni}, {Marinoni}, {McCracken}, {Memeo}, {Meneux}, {Nair}, {Porciani}, {Presotto}, \& {Scaramella}}]{Peng2010}
{Peng}, Y.-j., {Lilly}, S.~J., {Kova{\v{c}}}, K., {et~al.} 2010, \href{http://dx.doi.org/10.1088/0004-637X/721/1/193}{\color{magenta}\apj}, \href{https://ui.adsabs.harvard.edu/abs/2010ApJ...721..193P}{721, 193}

\bibitem[{{P{\'e}rez-Gonz{\'a}lez} {et~al.}(2023){P{\'e}rez-Gonz{\'a}lez}, {Costantin}, {Langeroodi}, {Rinaldi}, {Annunziatella}, {Ilbert}, {Colina}, {N{\o}rgaard-Nielsen}, {Greve}, {{\"O}stlin}, {Wright}, {Alonso-Herrero}, {{\'A}lvarez-M{\'a}rquez}, {Caputi}, {Eckart}, {Le F{\`e}vre}, {Labiano}, {Garc{\'\i}a-Mar{\'\i}n}, {Hjorth}, {Kendrew}, {Pye}, {Tikkanen}, {van der Werf}, {Walter}, {Ward}, {Bik}, {Boogaard}, {Bosman}, {G{\'o}mez}, {Gillman}, {Iani}, {Jermann}, {Melinder}, {Meyer}, {Moutard}, {van Dishoek}, {Henning}, {Lagage}, {Guedel}, {Peissker}, {Ray}, {Vandenbussche}, {Garc{\'\i}a-Argum{\'a}nez}, \& {Mar{\'\i}a M{\'e}rida}}]{Perez-Gonzale2023}
{P{\'e}rez-Gonz{\'a}lez}, P.~G., {Costantin}, L., {Langeroodi}, D., {et~al.} 2023, \href{http://dx.doi.org/10.3847/2041-8213/acd9d0}{\color{magenta}\apjl}, \href{https://ui.adsabs.harvard.edu/abs/2023ApJ...951L...1P}{951, L1}

\bibitem[{{Picouet} {et~al.}(2023){Picouet}, {Arnouts}, {Le Floc'h}, {Moutard}, {Kraljic}, {Ilbert}, {Sawicki}, {Desprez}, {Laigle}, {Schiminovich}, {de la Torre}, {Gwyn}, {McCracken}, {Dubois}, {Dav{\'e}}, {Toft}, {Weaver}, {Shuntov}, \& {Kauffmann}}]{Picouet2023}
{Picouet}, V., {Arnouts}, S., {Le Floc'h}, E., {et~al.} 2023, \href{http://dx.doi.org/10.1051/0004-6361/202245756}{\color{magenta}\aap}, \href{https://ui.adsabs.harvard.edu/abs/2023A&A...675A.164P}{675, A164}

\bibitem[{Pillepich {et~al.}(2018)Pillepich, Nelson, Hernquist, Springel, Pakmor, Torrey, Weinberger, Genel, Naiman, Marinacci, \& Vogelsberger}]{pillepich_first_2018}
Pillepich, A., Nelson, D., Hernquist, L., {et~al.} 2018, \href{http://dx.doi.org/10.1093/mnras/stx3112}{\color{magenta}\mnras}, 475, 648

\bibitem[{{Popesso} {et~al.}(2023){Popesso}, {Concas}, {Cresci}, {Belli}, {Rodighiero}, {Inami}, {Dickinson}, {Ilbert}, {Pannella}, \& {Elbaz}}]{Popesso2023}
{Popesso}, P., {Concas}, A., {Cresci}, G., {et~al.} 2023, \href{http://dx.doi.org/10.1093/mnras/stac3214}{\color{magenta}\mnras}, \href{https://ui.adsabs.harvard.edu/abs/2023MNRAS.519.1526P}{519, 1526}

\bibitem[{{Pozzetti} {et~al.}(2007){Pozzetti}, {Bolzonella}, {Lamareille}, {Zamorani}, {Franzetti}, {Le F{\`e}vre}, {Iovino}, {Temporin}, {Ilbert}, {Arnouts}, {Charlot}, {Brinchmann}, {Zucca}, {Tresse}, {Scodeggio}, {Guzzo}, {Bottini}, {Garilli}, {Le Brun}, {Maccagni}, {Picat}, {Scaramella}, {Vettolani}, {Zanichelli}, {Adami}, {Bardelli}, {Cappi}, {Ciliegi}, {Contini}, {Foucaud}, {Gavignaud}, {McCracken}, {Marano}, {Marinoni}, {Mazure}, {Meneux}, {Merighi}, {Paltani}, {Pell{\`o}}, {Pollo}, {Radovich}, {Bondi}, {Bongiorno}, {Cucciati}, {de la Torre}, {Gregorini}, {Mellier}, {Merluzzi}, {Vergani}, \& {Walcher}}]{Pozetti2007}
{Pozzetti}, L., {Bolzonella}, M., {Lamareille}, F., {et~al.} 2007, \href{http://dx.doi.org/10.1051/0004-6361:20077609}{\color{magenta}\aap}, \href{https://ui.adsabs.harvard.edu/abs/2007A&A...474..443P}{474, 443}

\bibitem[{{Pozzetti} {et~al.}(2010){Pozzetti}, {Bolzonella}, {Zucca}, {Zamorani}, {Lilly}, {Renzini}, {Moresco}, {Mignoli}, {Cassata}, {Tasca}, {Lamareille}, {Maier}, {Meneux}, {Halliday}, {Oesch}, {Vergani}, {Caputi}, {Kova{\v{c}}}, {Cimatti}, {Cucciati}, {Iovino}, {Peng}, {Carollo}, {Contini}, {Kneib}, {Le F{\'e}vre}, {Mainieri}, {Scodeggio}, {Bardelli}, {Bongiorno}, {Coppa}, {de la Torre}, {de Ravel}, {Franzetti}, {Garilli}, {Kampczyk}, {Knobel}, {Le Borgne}, {Le Brun}, {Pell{\`o}}, {Perez Montero}, {Ricciardelli}, {Silverman}, {Tanaka}, {Tresse}, {Abbas}, {Bottini}, {Cappi}, {Guzzo}, {Koekemoer}, {Leauthaud}, {Maccagni}, {Marinoni}, {McCracken}, {Memeo}, {Porciani}, {Scaramella}, {Scarlata}, \& {Scoville}}]{pozzetti2010}
{Pozzetti}, L., {Bolzonella}, M., {Zucca}, E., {et~al.} 2010, \href{http://dx.doi.org/10.1051/0004-6361/200913020}{\color{magenta}\aap}, \href{https://ui.adsabs.harvard.edu/abs/2010A&A...523A..13P}{523, A13}

\bibitem[{{Rasmussen Cueto} {et~al.}(2023){Rasmussen Cueto}, {Hutter}, {Dayal}, {Gottl{\"o}ber}, {Heintz}, {Mason}, {Trebitsch}, \& {Yepes}}]{Cueto2023}
{Rasmussen Cueto}, E., {Hutter}, A., {Dayal}, P., {et~al.} 2023, \href{https://ui.adsabs.harvard.edu/abs/2023arXiv231212109R}{\href{http://dx.doi.org/10.48550/arXiv.2312.12109}{\color{magenta}arXiv e-prints}, arXiv:2312.12109}

\bibitem[{{Reddick} {et~al.}(2013){Reddick}, {Wechsler}, {Tinker}, \& {Behroozi}}]{Reddick2013}
{Reddick}, R.~M., {Wechsler}, R.~H., {Tinker}, J.~L., \& {Behroozi}, P.~S. 2013, \href{http://dx.doi.org/10.1088/0004-637X/771/1/30}{\color{magenta}\apj}, \href{https://ui.adsabs.harvard.edu/abs/2013ApJ...771...30R}{771, 30}

\bibitem[{{Renzini} \& {Buzzoni}(1986)}]{RenziniBuzzoni1986}
{Renzini}, A. \& {Buzzoni}, A. 1986, in Astrophysics and Space Science Library, Vol. 122, Spectral Evolution of Galaxies, ed. C.~{Chiosi} \& A.~{Renzini}, \href{https://ui.adsabs.harvard.edu/abs/1986ASSL..122..195R}{195--231}

\bibitem[{{Renzini} \& {Voli}(1981)}]{RenziniVoli1981}
{Renzini}, A. \& {Voli}, M. 1981, \aap, \href{https://ui.adsabs.harvard.edu/abs/1981A&A....94..175R}{94, 175}

\bibitem[{{Robertson}(2010)}]{Robertson2010}
{Robertson}, B.~E. 2010, \href{http://dx.doi.org/10.1088/0004-637X/713/2/1266}{\color{magenta}\apj}, \href{https://ui.adsabs.harvard.edu/abs/2010ApJ...713.1266R}{713, 1266}

\bibitem[{{Saito} {et~al.}(2020){Saito}, {de la Torre}, {Ilbert}, {Dubois}, {Yabe}, \& {Coupon}}]{Saito20}
{Saito}, S., {de la Torre}, S., {Ilbert}, O., {et~al.} 2020, \href{http://dx.doi.org/10.1093/mnras/staa727}{\color{magenta}\mnras}, \href{https://ui.adsabs.harvard.edu/abs/2020MNRAS.494..199S}{494, 199}

\bibitem[{{Salim} {et~al.}(2018){Salim}, {Boquien}, \& {Lee}}]{Salim18}
{Salim}, S., {Boquien}, M., \& {Lee}, J.~C. 2018, \href{http://dx.doi.org/10.3847/1538-4357/aabf3c}{\color{magenta}\apj}, \href{https://ui.adsabs.harvard.edu/abs/2018ApJ...859...11S}{859, 11}

\bibitem[{Sawicki {et~al.}(2019)Sawicki, Arnouts, Huang, Coupon, Golob, Gwyn, Foucaud, Moutard, Iwata, Liu, Chen, Desprez, Harikane, Ono, Strauss, Tanaka, Thibert, Balogh, Bundy, Chapman, Gunn, Hsieh, Ilbert, Jing, LeFèvre, Li, Matsuda, Miyazaki, Nagao, Nishizawa, Ouchi, Shimasaku, Silverman, de~la Torre, Tresse, Wang, Willott, Yamada, Yang, \& Yee}]{Sawicki19}
Sawicki, M., Arnouts, S., Huang, J., {et~al.} 2019, \href{http://dx.doi.org/10.1093/mnras/stz2522}{\color{magenta}\mnras}, 489, 5202

\bibitem[{{Schaerer} \& {de Barros}(2009)}]{Schaerer09}
{Schaerer}, D. \& {de Barros}, S. 2009, \href{http://dx.doi.org/10.1051/0004-6361/200911781}{\color{magenta}\aap}, \href{http://adsabs.harvard.edu/abs/2009A%26A...502..423S}{502, 423}

\bibitem[{{Schechter}(1976)}]{Schechter76}
{Schechter}, P. 1976, \href{http://dx.doi.org/10.1086/154079}{\color{magenta}\apj}, \href{https://ui.adsabs.harvard.edu/abs/1976ApJ...203..297S}{203, 297}

\bibitem[{Schmidt(1968)}]{schmidt_space_1968}
Schmidt, M. 1968, \href{http://dx.doi.org/10.1086/149446}{\color{magenta}\apj}, 151, 393

\bibitem[{{Schreiber} {et~al.}(2015){Schreiber}, {Pannella}, {Elbaz}, {B{\'e}thermin}, {Inami}, {Dickinson}, {Magnelli}, {Wang}, {Aussel}, {Daddi}, {Juneau}, {Shu}, {Sargent}, {Buat}, {Faber}, {Ferguson}, {Giavalisco}, {Koekemoer}, {Magdis}, {Morrison}, {Papovich}, {Santini}, \& {Scott}}]{Schreiber2015}
{Schreiber}, C., {Pannella}, M., {Elbaz}, D., {et~al.} 2015, \href{http://dx.doi.org/10.1051/0004-6361/201425017}{\color{magenta}\aap}, \href{https://ui.adsabs.harvard.edu/abs/2015A&A...575A..74S}{575, A74}

\bibitem[{Schwarz(1978)}]{Schwartz-BIC}
Schwarz, G. 1978, \href{http://dx.doi.org/10.1214/aos/1176344136}{\color{magenta}The Annals of Statistics}, 6, 461

\bibitem[{{Scoville} {et~al.}(2007){Scoville}, {Abraham}, {Aussel}, {Barnes}, {Benson}, {Blain}, {Calzetti}, {Comastri}, {Capak}, {Carilli}, {Carlstrom}, {Carollo}, {Colbert}, {Daddi}, {Ellis}, {Elvis}, {Ewald}, {Fall}, {Franceschini}, {Giavalisco}, {Green}, {Griffiths}, {Guzzo}, {Hasinger}, {Impey}, {Kneib}, {Koda}, {Koekemoer}, {Lefevre}, {Lilly}, {Liu}, {McCracken}, {Massey}, {Mellier}, {Miyazaki}, {Mobasher}, {Mould}, {Norman}, {Refregier}, {Renzini}, {Rhodes}, {Rich}, {Sanders}, {Schiminovich}, {Schinnerer}, {Scodeggio}, {Sheth}, {Shopbell}, {Taniguchi}, {Tyson}, {Urry}, {Van Waerbeke}, {Vettolani}, {White}, \& {Yan}}]{Scoville2007}
{Scoville}, N., {Abraham}, R.~G., {Aussel}, H., {et~al.} 2007, \href{http://dx.doi.org/10.1086/516580}{\color{magenta}\apjs}, \href{https://ui.adsabs.harvard.edu/abs/2007ApJS..172...38S}{172, 38}

\bibitem[{{Shen} {et~al.}(2023){Shen}, {Vogelsberger}, {Boylan-Kolchin}, {Tacchella}, \& {Kannan}}]{Shen2023}
{Shen}, X., {Vogelsberger}, M., {Boylan-Kolchin}, M., {Tacchella}, S., \& {Kannan}, R. 2023, \href{http://dx.doi.org/10.1093/mnras/stad2508}{\color{magenta}\mnras}, \href{https://ui.adsabs.harvard.edu/abs/2023MNRAS.525.3254S}{525, 3254}

\bibitem[{{Shuntov} {et~al.}(2022){Shuntov}, {McCracken}, {Gavazzi}, {Laigle}, {Weaver}, {Davidzon}, {Ilbert}, {Kauffmann}, {Faisst}, {Dubois}, {Koekemoer}, {Moneti}, {Milvang-Jensen}, {Mobasher}, {Sanders}, \& {Toft}}]{Shuntov2022}
{Shuntov}, M., {McCracken}, H.~J., {Gavazzi}, R., {et~al.} 2022, \href{http://dx.doi.org/10.1051/0004-6361/202243136}{\color{magenta}\aap}, \href{https://ui.adsabs.harvard.edu/abs/2022A&A...664A..61S}{664, A61}

\bibitem[{{Silk} {et~al.}(2024){Silk}, {Begelman}, {Norman}, {Nusser}, \& {Wyse}}]{Silk2024}
{Silk}, J., {Begelman}, M.~C., {Norman}, C., {Nusser}, A., \& {Wyse}, R. F.~G. 2024, \href{http://dx.doi.org/10.3847/2041-8213/ad1bf0}{\color{magenta}\apjl}, \href{https://ui.adsabs.harvard.edu/abs/2024ApJ...961L..39S}{961, L39}

\bibitem[{Silk \& Mamon(2012)}]{silk_current_2012}
Silk, J. \& Mamon, G.~A. 2012, \href{http://dx.doi.org/10.1088/1674-4527/12/8/004}{\color{magenta}Research in Astronomy and Astrophysics}, 12, 917

\bibitem[{{Silverman} {et~al.}(2015){Silverman}, {Kashino}, {Sanders}, {Kartaltepe}, {Arimoto}, {Renzini}, {Rodighiero}, {Daddi}, {Zahid}, {Nagao}, {Kewley}, {Lilly}, {Sugiyama}, {Baronchelli}, {Capak}, {Carollo}, {Chu}, {Hasinger}, {Ilbert}, {Juneau}, {Kajisawa}, {Koekemoer}, {Kovac}, {Le F{\`e}vre}, {Masters}, {McCracken}, {Onodera}, {Schulze}, {Scoville}, {Strazzullo}, \& {Taniguchi}}]{silverman15}
{Silverman}, J.~D., {Kashino}, D., {Sanders}, D., {et~al.} 2015, \href{http://dx.doi.org/10.1088/0067-0049/220/1/12}{\color{magenta}\apjs}, \href{https://ui.adsabs.harvard.edu/abs/2015ApJS..220...12S}{220, 12}

\bibitem[{{Song} {et~al.}(2016){Song}, {Finkelstein}, {Ashby}, {Grazian}, {Lu}, {Papovich}, {Salmon}, {Somerville}, {Dickinson}, {Duncan}, {Faber}, {Fazio}, {Ferguson}, {Fontana}, {Guo}, {Hathi}, {Lee}, {Merlin}, \& {Willner}}]{Song2016}
{Song}, M., {Finkelstein}, S.~L., {Ashby}, M. L.~N., {et~al.} 2016, \href{http://dx.doi.org/10.3847/0004-637X/825/1/5}{\color{magenta}\apj}, \href{https://ui.adsabs.harvard.edu/abs/2016ApJ...825....5S}{825, 5}

\bibitem[{Springel {et~al.}(2018)Springel, Pakmor, Pillepich, Weinberger, Nelson, Hernquist, Vogelsberger, Genel, Torrey, Marinacci, \& Naiman}]{springel_first_2018}
Springel, V., Pakmor, R., Pillepich, A., {et~al.} 2018, \href{http://dx.doi.org/10.1093/mnras/stx3304}{\color{magenta}\mnras}, 475, 676

\bibitem[{{Stefanon} {et~al.}(2021){Stefanon}, {Bouwens}, {Labb{\'e}}, {Illingworth}, {Gonzalez}, \& {Oesch}}]{Stefanon21}
{Stefanon}, M., {Bouwens}, R.~J., {Labb{\'e}}, I., {et~al.} 2021, \href{http://dx.doi.org/10.3847/1538-4357/ac1bb6}{\color{magenta}\apj}, \href{https://ui.adsabs.harvard.edu/abs/2021ApJ...922...29S}{922, 29}

\bibitem[{{Steinhardt} {et~al.}(2016){Steinhardt}, {Capak}, {Masters}, \& {Speagle}}]{Steinhardt2016}
{Steinhardt}, C.~L., {Capak}, P., {Masters}, D., \& {Speagle}, J.~S. 2016, \href{http://dx.doi.org/10.3847/0004-637X/824/1/21}{\color{magenta}\apj}, \href{https://ui.adsabs.harvard.edu/abs/2016ApJ...824...21S}{824, 21}

\bibitem[{{Steinhardt} {et~al.}(2021){Steinhardt}, {Jespersen}, \& {Linzer}}]{Steinhardt2021_cosmicvariance}
{Steinhardt}, C.~L., {Jespersen}, C.~K., \& {Linzer}, N.~B. 2021, \href{http://dx.doi.org/10.3847/1538-4357/ac2a2f}{\color{magenta}\apj}, \href{https://ui.adsabs.harvard.edu/abs/2021ApJ...923....8S}{923, 8}

\bibitem[{{Steinhardt} {et~al.}(2023){Steinhardt}, {Kokorev}, {Rusakov}, {Garcia}, \& {Sneppen}}]{Steinhardt2023}
{Steinhardt}, C.~L., {Kokorev}, V., {Rusakov}, V., {Garcia}, E., \& {Sneppen}, A. 2023, \href{http://dx.doi.org/10.3847/2041-8213/acdef6}{\color{magenta}\apjl}, \href{https://ui.adsabs.harvard.edu/abs/2023ApJ...951L..40S}{951, L40}

\bibitem[{{Steinhardt} {et~al.}(2022){Steinhardt}, {Sneppen}, {Mostafa}, {Hensley}, {Jermyn}, {Lopez}, {Weaver}, {Brammer}, {Clark}, {Davidzon}, {Diaconu}, {Mobasher}, {Rusakov}, \& {Toft}}]{Steinhardt2022}
{Steinhardt}, C.~L., {Sneppen}, A., {Mostafa}, B., {et~al.} 2022, \href{http://dx.doi.org/10.3847/1538-4357/ac62d6}{\color{magenta}\apj}, \href{https://ui.adsabs.harvard.edu/abs/2022ApJ...931...58S}{931, 58}

\bibitem[{{Sun} {et~al.}(2023){Sun}, {Faucher-Gigu{\`e}re}, {Hayward}, {Shen}, {Wetzel}, \& {Cochrane}}]{Sun2023}
{Sun}, G., {Faucher-Gigu{\`e}re}, C.-A., {Hayward}, C.~C., {et~al.} 2023, \href{http://dx.doi.org/10.3847/2041-8213/acf85a}{\color{magenta}\apjl}, \href{https://ui.adsabs.harvard.edu/abs/2023ApJ...955L..35S}{955, L35}

\bibitem[{{Szalay} {et~al.}(1999){Szalay}, {Connolly}, \& {Szokoly}}]{Szalay1999}
{Szalay}, A.~S., {Connolly}, A.~J., \& {Szokoly}, G.~P. 1999, \href{http://dx.doi.org/10.1086/300689}{\color{magenta}\aj}, \href{https://ui.adsabs.harvard.edu/abs/1999AJ....117...68S}{117, 68}

\bibitem[{Taniguchi {et~al.}(2015)Taniguchi, Kajisawa, Kobayashi, Shioya, Nagao, Capak, Aussel, Ichikawa, Murayama, Scoville, Ilbert, Salvato, Sanders, Mobasher, Miyazaki, Komiyama, Le~Fèvre, Tasca, Lilly, Carollo, Renzini, Rich, Schinnerer, Kaifu, Karoji, Arimoto, Okamura, Ohta, Shimasaku, \& Hayashino}]{Taniguchi15}
Taniguchi, Y., Kajisawa, M., Kobayashi, M. A.~R., {et~al.} 2015, \href{http://dx.doi.org/10.1093/pasj/psv106}{\color{magenta}\pasj}, 67, 104

\bibitem[{Taniguchi {et~al.}(2007)Taniguchi, Scoville, Murayama, Sanders, Mobasher, Aussel, Capak, Ajiki, Miyazaki, Komiyama, Shioya, Nagao, Sasaki, Koda, Carilli, Giavalisco, Guzzo, Hasinger, Impey, LeFevre, Lilly, Renzini, Rich, Schinnerer, Shopbell, Kaifu, Karoji, Arimoto, Okamura, \& Ohta}]{Taniguchi07}
Taniguchi, Y., Scoville, N., Murayama, T., {et~al.} 2007, \href{http://dx.doi.org/10.1086/516596}{\color{magenta}The Astrophysical Journal Supplement Series}, 172, 9

\bibitem[{Tasitsiomi {et~al.}(2004)Tasitsiomi, Kravtsov, Wechsler, \& Primack}]{Tasitsiomi2004}
Tasitsiomi, A., Kravtsov, A.~V., Wechsler, R.~H., \& Primack, J.~R. 2004, \href{http://dx.doi.org/10.1086/423784}{\color{magenta}\apj}, 614, 533

\bibitem[{Thomas {et~al.}(2010)Thomas, Maraston, Schawinski, Sarzi, \& Silk}]{thomas_environment_2010}
Thomas, D., Maraston, C., Schawinski, K., Sarzi, M., \& Silk, J. 2010, \href{http://dx.doi.org/10.1111/j.1365-2966.2010.16427.x}{\color{magenta}\mnras}

\bibitem[{{Torrey} {et~al.}(2017){Torrey}, {Hopkins}, {Faucher-Gigu{\`e}re}, {Vogelsberger}, {Quataert}, {Kere{\v{s}}}, \& {Murray}}]{Torrey2017}
{Torrey}, P., {Hopkins}, P.~F., {Faucher-Gigu{\`e}re}, C.-A., {et~al.} 2017, \href{http://dx.doi.org/10.1093/mnras/stx254}{\color{magenta}\mnras}, \href{https://ui.adsabs.harvard.edu/abs/2017MNRAS.467.2301T}{467, 2301}

\bibitem[{Vale \& Ostriker(2004)}]{ValeOstriker2004}
Vale, A. \& Ostriker, J.~P. 2004, \href{http://dx.doi.org/10.1111/j.1365-2966.2004.08059.x}{\color{magenta}\mnras}, 353, 189

\bibitem[{{Valentino} {et~al.}(2023){Valentino}, {Brammer}, {Gould}, {Kokorev}, {Fujimoto}, {Jespersen}, {Vijayan}, {Weaver}, {Ito}, {Tanaka}, {Ilbert}, {Magdis}, {Whitaker}, {Faisst}, {Gallazzi}, {Gillman}, {Gim{\'e}nez-Arteaga}, {G{\'o}mez-Guijarro}, {Kubo}, {Heintz}, {Hirschmann}, {Oesch}, {Onodera}, {Rizzo}, {Lee}, {Strait}, \& {Toft}}]{Valentino2023}
{Valentino}, F., {Brammer}, G., {Gould}, K. M.~L., {et~al.} 2023, \href{http://dx.doi.org/10.3847/1538-4357/acbefa}{\color{magenta}\apj}, \href{https://ui.adsabs.harvard.edu/abs/2023ApJ...947...20V}{947, 20}

\bibitem[{{Vijayan} {et~al.}(2022){Vijayan}, {Wilkins}, {Lovell}, {Thomas}, {Camps}, {Baes}, {Trayford}, {Kuusisto}, \& {Roper}}]{Vijayan2022}
{Vijayan}, A.~P., {Wilkins}, S.~M., {Lovell}, C.~C., {et~al.} 2022, \href{http://dx.doi.org/10.1093/mnras/stac338}{\color{magenta}\mnras}, \href{https://ui.adsabs.harvard.edu/abs/2022MNRAS.511.4999V}{511, 4999}

\bibitem[{{Vujeva} {et~al.}(2023){Vujeva}, {Steinhardt}, {Kragh Jespersen}, {Frye}, {Koekemoer}, {Natarajan}, {Faisst}, {Hibon}, {Furtak}, {Atek}, {Cen}, \& {Sneppen}}]{Vujeva2024_jwst_cosmic_variance}
{Vujeva}, L., {Steinhardt}, C.~L., {Kragh Jespersen}, C., {et~al.} 2023, \href{https://ui.adsabs.harvard.edu/abs/2023arXiv231015284V}{\href{http://dx.doi.org/10.48550/arXiv.2310.15284}{\color{magenta}arXiv e-prints}, arXiv:2310.15284}

\bibitem[{{Wang} {et~al.}(2024{\natexlab{a}}){Wang}, {Leja}, {de Graaff}, {Brammer}, {Weibel}, {van Dokkum}, {Baggen}, {Suess}, {Greene}, {Bezanson}, {Cleri}, {Hirschmann}, {Labbe}, {Matthee}, {McConachie}, {Naidu}, {Nelson}, {Oesch}, {Setton}, \& {Williams}}]{Wang2024-rub}
{Wang}, B., {Leja}, J., {de Graaff}, A., {et~al.} 2024{\natexlab{a}}, \href{https://ui.adsabs.harvard.edu/abs/2024arXiv240501473W}{\href{http://dx.doi.org/10.48550/arXiv.2405.01473}{\color{magenta}arXiv e-prints}, arXiv:2405.01473}

\bibitem[{{Wang} {et~al.}(2024{\natexlab{b}}){Wang}, {Sun}, {Zhou}, {Xu}, {Cheng}, {Li}, {Chen}, {Mo}, {Dekel}, {Zheng}, {Cai}, {Yang}, {Dai}, {Elbaz}, \& {Huang}}]{TWang2024}
{Wang}, T., {Sun}, H., {Zhou}, L., {et~al.} 2024{\natexlab{b}}, \href{https://ui.adsabs.harvard.edu/abs/2024arXiv240302399W}{\href{http://dx.doi.org/10.48550/arXiv.2403.02399}{\color{magenta}arXiv e-prints}, arXiv:2403.02399}

\bibitem[{{Watson} {et~al.}(2013){Watson}, {Iliev}, {D'Aloisio}, {Knebe}, {Shapiro}, \& {Yepes}}]{2013MNRAS.433.1230W}
{Watson}, W.~A., {Iliev}, I.~T., {D'Aloisio}, A., {et~al.} 2013, \href{http://dx.doi.org/10.1093/mnras/stt791}{\color{magenta}\mnras}, \href{https://ui.adsabs.harvard.edu/abs/2013MNRAS.433.1230W}{433, 1230}

\bibitem[{{Weaver} {et~al.}(2023){Weaver}, {Davidzon}, {Toft}, {Ilbert}, {McCracken}, {Gould}, {Jespersen}, {Steinhardt}, {Lagos}, {Capak}, {Casey}, {Chartab}, {Faisst}, {Hayward}, {Kartaltepe}, {Kauffmann}, {Koekemoer}, {Kokorev}, {Laigle}, {Liu}, {Long}, {Magdis}, {McPartland}, {Milvang-Jensen}, {Mobasher}, {Moneti}, {Peng}, {Sanders}, {Shuntov}, {Sneppen}, {Valentino}, {Zalesky}, \& {Zamorani}}]{Weaver2023}
{Weaver}, J.~R., {Davidzon}, I., {Toft}, S., {et~al.} 2023, \href{http://dx.doi.org/10.1051/0004-6361/202245581}{\color{magenta}\aap}, \href{https://ui.adsabs.harvard.edu/abs/2023A&A...677A.184W}{677, A184}

\bibitem[{{Weaver} {et~al.}(2022{\natexlab{a}}){Weaver}, {Kauffmann}, {Ilbert}, {McCracken}, {Moneti}, {Toft}, {Brammer}, {Shuntov}, {Davidzon}, {Hsieh}, {Laigle}, {Anastasiou}, {Jespersen}, {Vinther}, {Capak}, {Casey}, {McPartland}, {Milvang-Jensen}, {Mobasher}, {Sanders}, {Zalesky}, {Arnouts}, {Aussel}, {Dunlop}, {Faisst}, {Franx}, {Furtak}, {Fynbo}, {Gould}, {Greve}, {Gwyn}, {Kartaltepe}, {Kashino}, {Koekemoer}, {Kokorev}, {Le F{\`e}vre}, {Lilly}, {Masters}, {Magdis}, {Mehta}, {Peng}, {Riechers}, {Salvato}, {Sawicki}, {Scarlata}, {Scoville}, {Shirley}, {Silverman}, {Sneppen}, {Smolc̆i{\'c}}, {Steinhardt}, {Stern}, {Tanaka}, {Taniguchi}, {Teplitz}, {Vaccari}, {Wang}, \& {Zamorani}}]{Weaver2022cosmos2020}
{Weaver}, J.~R., {Kauffmann}, O.~B., {Ilbert}, O., {et~al.} 2022{\natexlab{a}}, \href{http://dx.doi.org/10.3847/1538-4365/ac3078}{\color{magenta}\apjs}, \href{https://ui.adsabs.harvard.edu/abs/2022ApJS..258...11W}{258, 11}

\bibitem[{{Weaver} {et~al.}(2022{\natexlab{b}}){Weaver}, {Kauffmann}, {Ilbert}, {McCracken}, {Moneti}, {Toft}, {Brammer}, {Shuntov}, {Davidzon}, {Hsieh}, {Laigle}, {Anastasiou}, {Jespersen}, {Vinther}, {Capak}, {Casey}, {McPartland}, {Milvang-Jensen}, {Mobasher}, {Sanders}, {Zalesky}, {Arnouts}, {Aussel}, {Dunlop}, {Faisst}, {Franx}, {Furtak}, {Fynbo}, {Gould}, {Greve}, {Gwyn}, {Kartaltepe}, {Kashino}, {Koekemoer}, {Kokorev}, {Le F{\`e}vre}, {Lilly}, {Masters}, {Magdis}, {Mehta}, {Peng}, {Riechers}, {Salvato}, {Sawicki}, {Scarlata}, {Scoville}, {Shirley}, {Silverman}, {Sneppen}, {Smolc̆i{\'c}}, {Steinhardt}, {Stern}, {Tanaka}, {Taniguchi}, {Teplitz}, {Vaccari}, {Wang}, \& {Zamorani}}]{Weaver2022}
{Weaver}, J.~R., {Kauffmann}, O.~B., {Ilbert}, O., {et~al.} 2022{\natexlab{b}}, \href{http://dx.doi.org/10.3847/1538-4365/ac3078}{\color{magenta}\apjs}, \href{https://ui.adsabs.harvard.edu/abs/2022ApJS..258...11W}{258, 11}

\bibitem[{Wechsler \& Tinker(2018)}]{wechsler_connection_2018}
Wechsler, R.~H. \& Tinker, J.~L. 2018, \href{http://dx.doi.org/10.1146/annurev-astro-081817-051756}{\color{magenta}\araa}, 56, 435

\bibitem[{{Weibel} {et~al.}(2024){Weibel}, {Oesch}, {Barrufet}, {Gottumukkala}, {Ellis}, {Santini}, {Weaver}, {Allen}, {Bouwens}, {Bowler}, {Brammer}, {Carnall}, {Cullen}, {Dayal}, {Donnan}, {Dunlop}, {Giavalisco}, {Grogin}, {Illingworth}, {Koekemoer}, {Labbe}, {Marchesini}, {McLeod}, {McLure}, {Naidu}, {Shuntov}, {Stefanon}, {Toft}, \& {Xiao}}]{Weibel2024}
{Weibel}, A., {Oesch}, P.~A., {Barrufet}, L., {et~al.} 2024, \href{https://ui.adsabs.harvard.edu/abs/2024arXiv240308872W}{\href{http://dx.doi.org/10.48550/arXiv.2403.08872}{\color{magenta}arXiv e-prints}, arXiv:2403.08872}

\bibitem[{{Weigel} {et~al.}(2016){Weigel}, {Schawinski}, \& {Bruderer}}]{weigel2016}
{Weigel}, A.~K., {Schawinski}, K., \& {Bruderer}, C. 2016, \href{http://dx.doi.org/10.1093/mnras/stw756}{\color{magenta}\mnras}, \href{https://ui.adsabs.harvard.edu/abs/2016MNRAS.459.2150W}{459, 2150}

\bibitem[{White \& Rees(1978)}]{WhiteRees1978}
White, S. D.~M. \& Rees, M.~J. 1978, \href{http://dx.doi.org/10.1093/mnras/183.3.341}{\color{magenta}\mnras}, 183, 341

\bibitem[{{Wilkins} {et~al.}(2017){Wilkins}, {Feng}, {Di Matteo}, {Croft}, {Lovell}, \& {Waters}}]{Wilkins2017}
{Wilkins}, S.~M., {Feng}, Y., {Di Matteo}, T., {et~al.} 2017, \href{http://dx.doi.org/10.1093/mnras/stx841}{\color{magenta}\mnras}, \href{https://ui.adsabs.harvard.edu/abs/2017MNRAS.469.2517W}{469, 2517}

\bibitem[{{Wilkins} {et~al.}(2019){Wilkins}, {Lovell}, \& {Stanway}}]{Wilkins2019}
{Wilkins}, S.~M., {Lovell}, C.~C., \& {Stanway}, E.~R. 2019, \href{http://dx.doi.org/10.1093/mnras/stz2894}{\color{magenta}\mnras}, \href{https://ui.adsabs.harvard.edu/abs/2019MNRAS.490.5359W}{490, 5359}

\bibitem[{{Wilkins} {et~al.}(2008){Wilkins}, {Trentham}, \& {Hopkins}}]{Wilkins2008}
{Wilkins}, S.~M., {Trentham}, N., \& {Hopkins}, A.~M. 2008, \href{http://dx.doi.org/10.1111/j.1365-2966.2008.12885.x}{\color{magenta}\mnras}, \href{https://ui.adsabs.harvard.edu/abs/2008MNRAS.385..687W}{385, 687}

\bibitem[{{Wilkins} {et~al.}(2023){Wilkins}, {Vijayan}, {Lovell}, {Roper}, {Irodotou}, {Caruana}, {Seeyave}, {Kuusisto}, {Thomas}, \& {Parris}}]{Wilkins2023}
{Wilkins}, S.~M., {Vijayan}, A.~P., {Lovell}, C.~C., {et~al.} 2023, \href{http://dx.doi.org/10.1093/mnras/stac3280}{\color{magenta}\mnras}, \href{https://ui.adsabs.harvard.edu/abs/2023MNRAS.519.3118W}{519, 3118}

\bibitem[{{Willott} {et~al.}(2024){Willott}, {Desprez}, {Asada}, {Sarrouh}, {Abraham}, {Brada{\v{c}}}, {Brammer}, {Estrada-Carpenter}, {Iyer}, {Martis}, {Matharu}, {Mowla}, {Muzzin}, {Noirot}, {Sawicki}, {Strait}, {Rihtar{\v{s}}i{\v{c}}}, \& {Withers}}]{Willott2024}
{Willott}, C.~J., {Desprez}, G., {Asada}, Y., {et~al.} 2024, \href{http://dx.doi.org/10.3847/1538-4357/ad35bc}{\color{magenta}\apj}, \href{https://ui.adsabs.harvard.edu/abs/2024ApJ...966...74W}{966, 74}

\bibitem[{{Wright} {et~al.}(2018){Wright}, {Driver}, \& {Robotham}}]{Wright2018}
{Wright}, A.~H., {Driver}, S.~P., \& {Robotham}, A.~S.~G. 2018, \href{http://dx.doi.org/10.1093/mnras/sty2136}{\color{magenta}\mnras}, \href{https://ui.adsabs.harvard.edu/abs/2018MNRAS.480.3491W}{480, 3491}

\bibitem[{{Xiao} {et~al.}(2023){Xiao}, {Oesch}, {Elbaz}, {Bing}, {Nelson}, {Weibel}, {Naidu}, {Daddi}, {Bouwens}, {Matthee}, {Wuyts}, {Chisholm}, {Brammer}, {Dickinson}, {Magnelli}, {Leroy}, {van Dokkum}, {Schaerer}, {Herard-Demanche}, {Barrufet}, {Endsley}, {Fudamoto}, {G{\'o}mez-Guijarro}, {Gottumukkala}, {Illingworth}, {Labbe}, {Magee}, {Marchesini}, {Maseda}, {Qin}, {Reddy}, {Shapley}, {Shivaei}, {Shuntov}, {Stefanon}, {Whitaker}, \& {Wyithe}}]{Xiao2023}
{Xiao}, M., {Oesch}, P., {Elbaz}, D., {et~al.} 2023, \href{https://ui.adsabs.harvard.edu/abs/2023arXiv230902492X}{\href{http://dx.doi.org/10.48550/arXiv.2309.02492}{\color{magenta}arXiv e-prints}, arXiv:2309.02492}

\bibitem[{{Yung} {et~al.}(2019){Yung}, {Somerville}, {Popping}, {Finkelstein}, {Ferguson}, \& {Dav{\'e}}}]{Yung2019}
{Yung}, L.~Y.~A., {Somerville}, R.~S., {Popping}, G., {et~al.} 2019, \href{http://dx.doi.org/10.1093/mnras/stz2755}{\color{magenta}\mnras}, \href{https://ui.adsabs.harvard.edu/abs/2019MNRAS.490.2855Y}{490, 2855}

\bibitem[{{Zavala} {et~al.}(2021){Zavala}, {Casey}, {Manning}, {Aravena}, {Bethermin}, {Caputi}, {Clements}, {Cunha}, {Drew}, {Finkelstein}, {Fujimoto}, {Hayward}, {Hodge}, {Kartaltepe}, {Knudsen}, {Koekemoer}, {Long}, {Magdis}, {Man}, {Popping}, {Sanders}, {Scoville}, {Sheth}, {Staguhn}, {Toft}, {Treister}, {Vieira}, \& {Yun}}]{Zavala2021}
{Zavala}, J.~A., {Casey}, C.~M., {Manning}, S.~M., {et~al.} 2021, \href{http://dx.doi.org/10.3847/1538-4357/abdb27}{\color{magenta}\apj}, \href{https://ui.adsabs.harvard.edu/abs/2021ApJ...909..165Z}{909, 165}

\end{thebibliography}

\appendix

\section{Mass uncertainty kernels}\label{appdx:eddington-bias-kernels}

\begin{figure}[t]
  \centering
    \includegraphics[width=0.8\columnwidth]{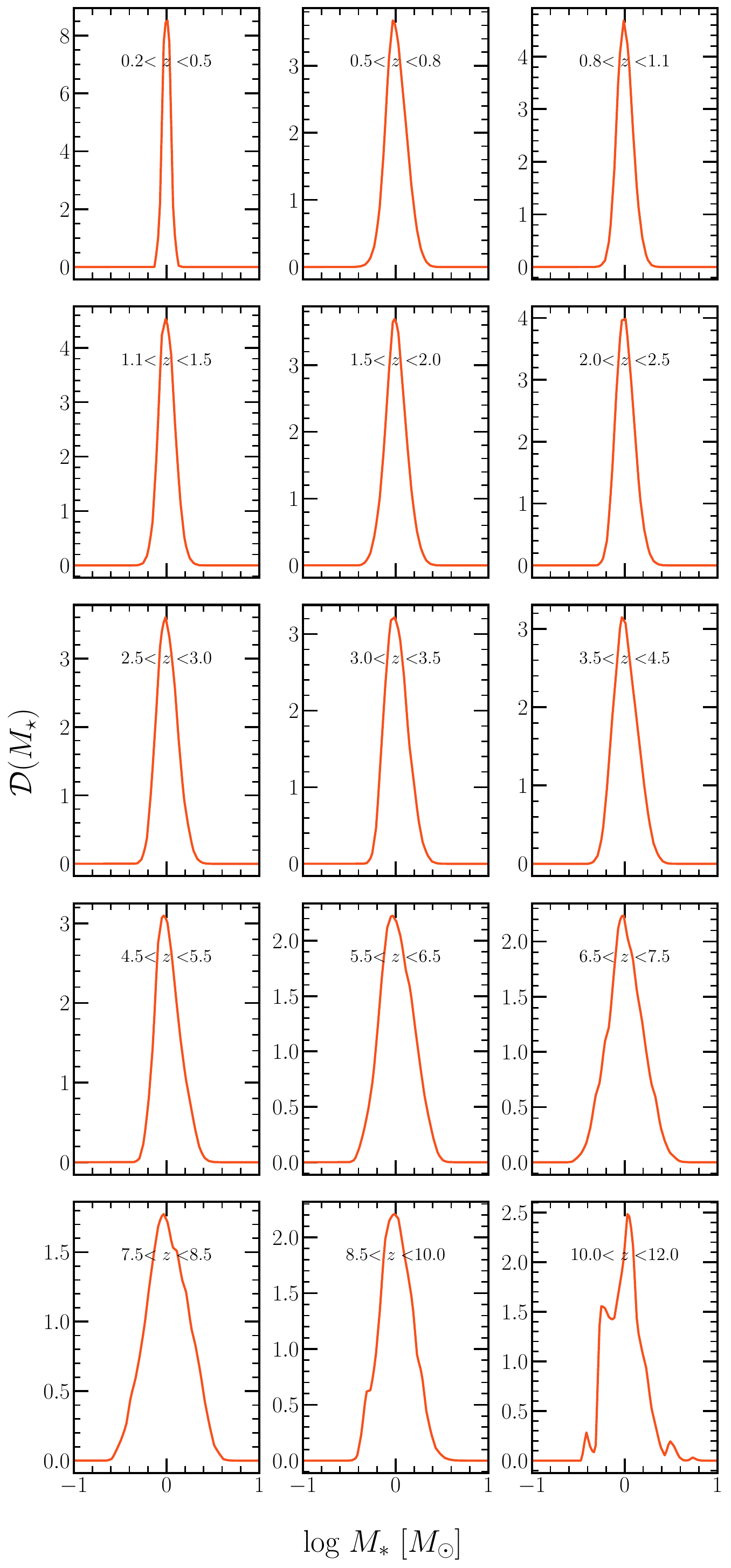}
 \caption{Kernels used in accounting for the Eddington bias for each redshift bin. These are constructed by stacking the PDF($M_{\star}$) of all sources in the redshift bin.
 }
  \label{fig:Mass-uncertainty-kernels}
\end{figure}

Figure~\ref{fig:Mass-uncertainty-kernels} shows the distributions of stellar mass uncertainties obtained by stacking the PDF$(M_{\star})$, centered at the median of the distribution, for each redshift bin.
The distributions are therefore shown as a function of $M_{\star}-\langle M_{\star} \rangle$, where $\langle M_{\star} \rangle$ is the median of PDF$(M_{\star})$. The width of the distribution increases with redshift, quantifying the increasing stellar mass uncertainties with redshift.

\section{Properties of removed AGN/LRD} \label{appdx:lrd-props}
\begin{figure}[t]
  \centering
    \includegraphics[width=0.98\columnwidth]{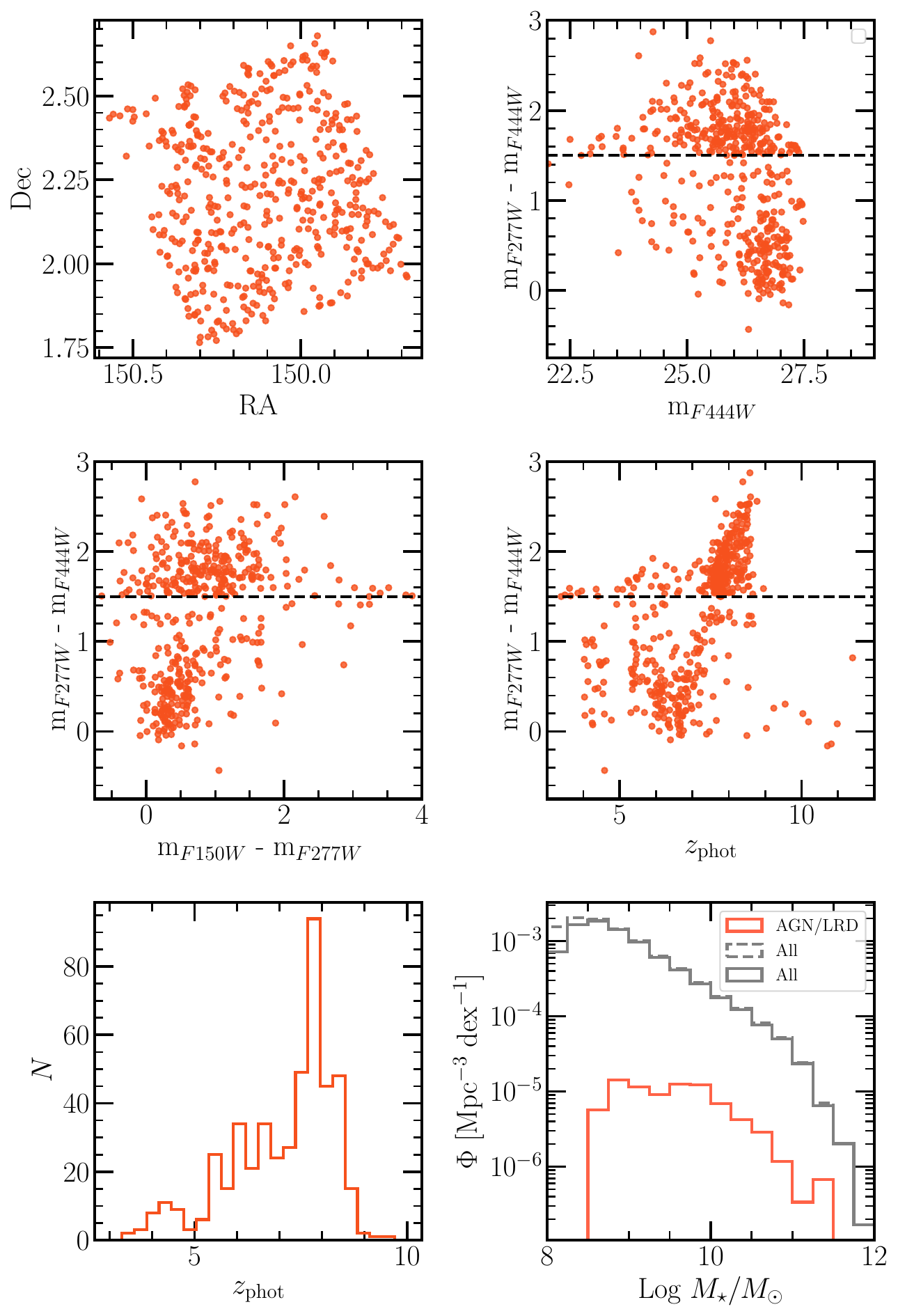}
 \caption{Distribution of properties of the AGN/LRD removed from the primary scientific analysis.
 }
  \label{fig:lrd-agn_dist}
\end{figure}

Figure \ref{fig:lrd-agn_dist} shows the distribution of several properties of the AGN/LRD population that we exclude from our analysis (cf. \S\ref{sec:sample-selection}). The $m_{\rm F277W}-m_{\rm F444W}$ color distribution shows a bimodality because of our double criterion to identify the AGN/LRD population: \textit{Red} $\land$ \textit{Compact} and \textit{AGN-SED} $\land$ \textit{Compact}. With respect to the $z_{\rm phot}$ solution, they are distributed within the $4\lesssim z \lesssim 9$ range, with a mode at $z\sim8$. The middle right panel shows that it is mostly the \textit{Red} $\land$ \textit{Compact} criterion that selects at $z\sim8$. The stellar mass distribution of the AGN/LRD population, weighted by the volume in $3.5<z<12$ and compared to the total SMF, is shown in the bottom right panel of Fig.~\ref{fig:lrd-agn_dist}. The relative contribution of the AGN/LRD to the SMF becomes more important at the high mass end, therefore not removing this population would lead to higher number densities at $z>4$ and log$(M_{\star}/M_{\odot}) > 10.5$. 

\section{The SMF in the MIRI-covered area} \label{appdx:smf-in-miri}
\begin{figure}[t]
  \centering
    \includegraphics[width=0.98\columnwidth]{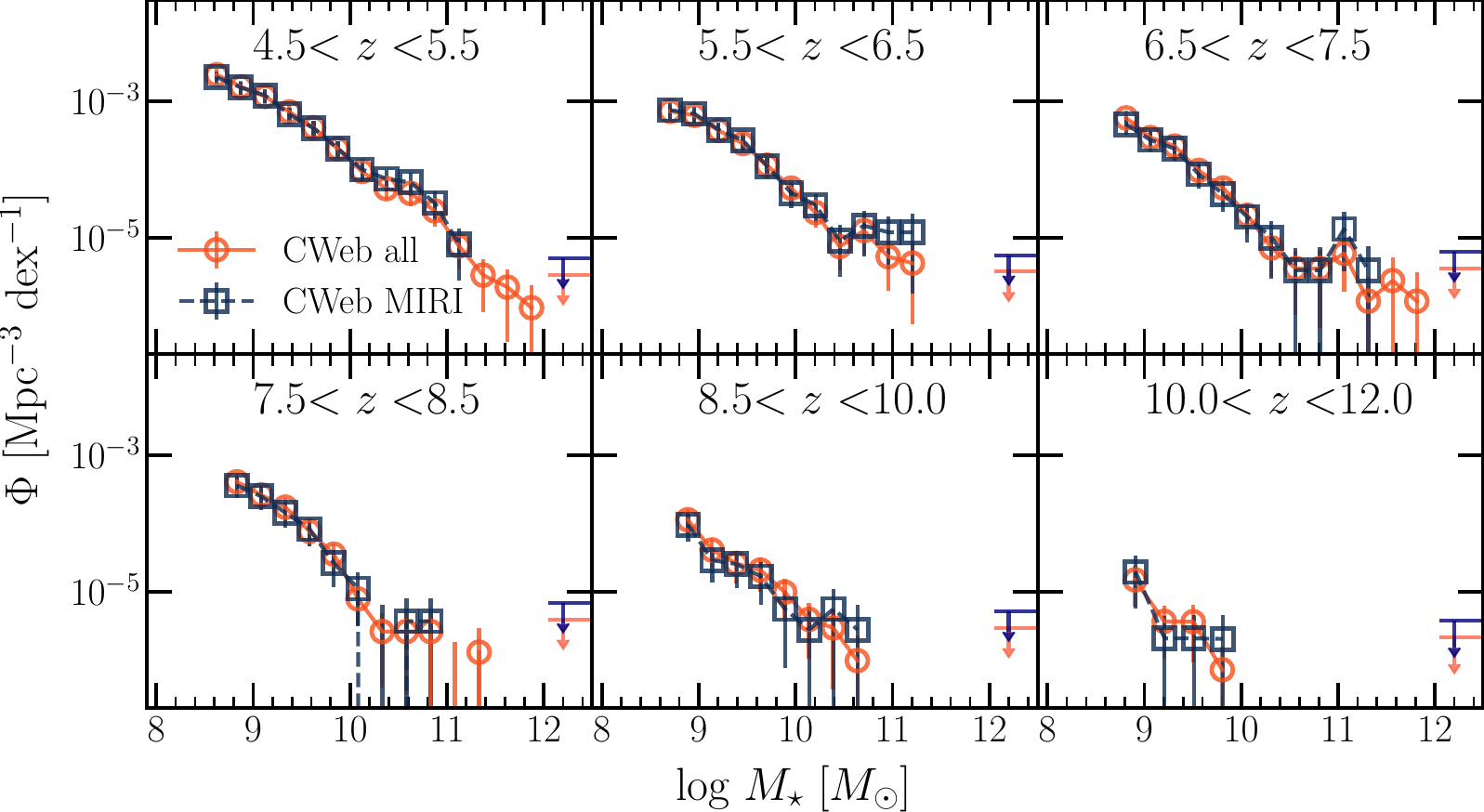}
 \caption{The SMF in the full COSMOS-Web (orange) and in the MIRI-covered area (blue) The downward pointing arrows show the upper limits for empty bins. The agreement between the two shows that the masses are consistent when including MIRI photometry.
 }
  \label{fig:smf-miri}
\end{figure}

Figure~\ref{fig:smf-miri} shows the SMF measured in the area covered by MIRI F770W. After excluding the HSC star masks, this results in $0.15\, {\rm deg}^2$. Not all sources that enter our analysis in the MIRI coverage would have a F770W counterpart above the $5\, \sigma$ depth of AB mag $\sim 25.7$. Nonetheless, we consider sources with ${\rm S/N}<5$ as even non detections can provide useful constrains in the SED fitting.

\section{Examples of the most extreme objects in our sample and their color properties} \label{appdx:extreme-gals}

\begin{figure}[t]
  \centering
    \includegraphics[width=0.98\columnwidth]{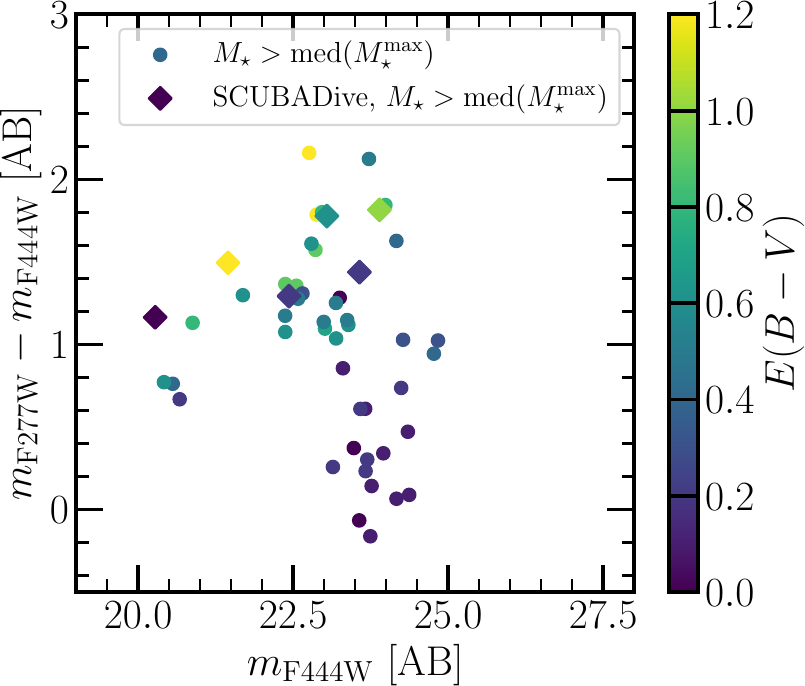}
 \caption{Color vs. magnitude distribution of the extreme sample, defined as $M_{\star} > {\rm med}(M_{\star}^{\rm max})$ (derived from the EVS formalism), color coded by $E(B-V)$. Diamonds show the SMGs from the SCUBADive sample.
 }
  \label{fig:extreme-properties}
\end{figure}

\begin{landscape}
\begin{figure}[p]
  \centering
    \includegraphics[width=\linewidth]{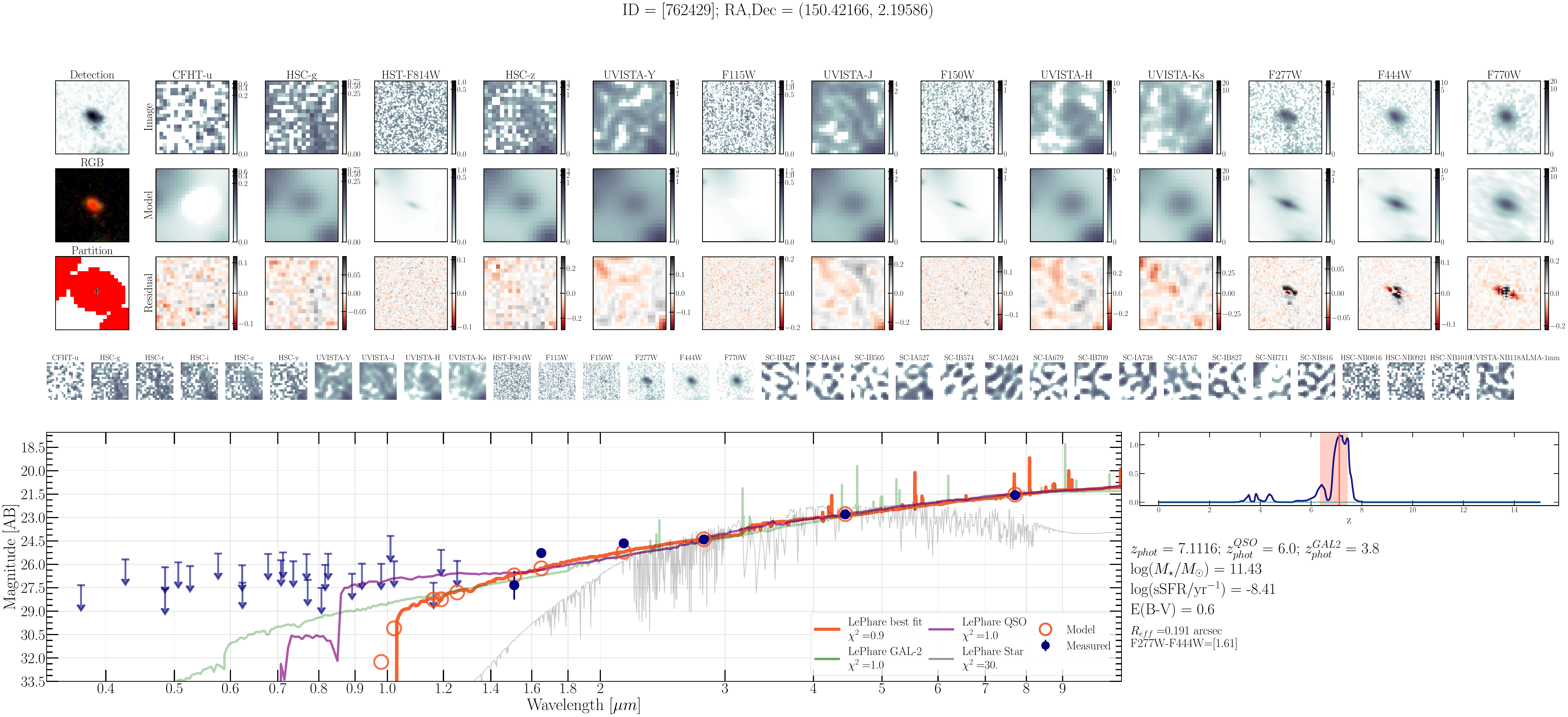}
 \caption{Example of the most extreme source in terms of stellar mass and redshift. In each of the top three rows we show the science, model, and residual images of size $4\,\arcsec$, in the middle row the science images of all the bands used in the analysis. In the bottom left panel, we show the photometry (blue points) and the \LePHARE{} SEDs of the best fit galaxy template (orange), star template (gray), and QSO template (purple). The open orange circles indicate the flux of the best-fit SED integrated in the corresponding filter. The top right panel shows the resulting photo$-z$ distribution.
 }
  \label{fig:Stamps-of-extremes}
\end{figure}
\end{landscape}

Figure~\ref{fig:extreme-properties} shows the F277W $-$ F444W color vs magnitude distribution for the sources in our extreme sample, defined as $M_{\star} > {\rm med}(M_{\star}^{\rm max})$, color coded by $E(B-V)$. Diamonds mark the SCUBADive \citep{McKinney2024} detected SMGs. A significant fraction ($>50\%$) of our extreme sample is very red ($m_{\rm F277W} = m_{\rm F444W} > 1$) and dust attenuated ($E(B-V)>0.5$.
Fig. ~\ref{fig:Stamps-of-extremes} shows examples of two of the most extreme sources in our sample according to the EVS analysis (cf. \ref{sec:most-extreme}). In each of the top three rows we show the science, model and residual images of 4 arcsec size in several selected broad bands. The middle row shows the science images of bands used in the SED fitting.  In the bottom left panel, we show the photometry (blue points) and the \LePHARE{} SEDs of the best fit galaxy template (orange), star template (gray), and QSO template (purple). The open orange circles indicate the flux of the best-fit SED integrated in the corresponding filter. The top right panel shows the resulting photo$-z$ distribution.

\section{Galaxy candidates at $10<z<12$} \label{appdx:zgt10}
\begin{figure*}[t]
  \centering
    \includegraphics[width=0.98\textwidth]{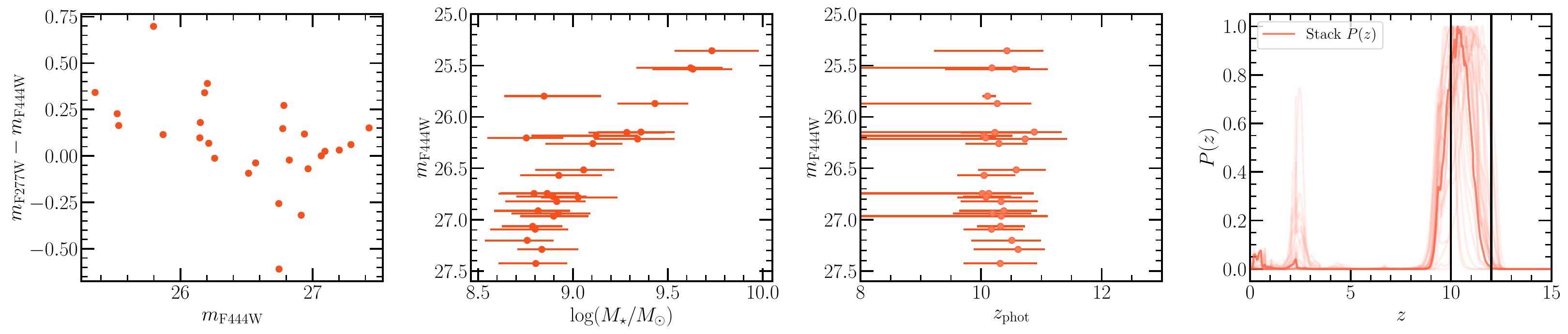}
 \caption{Distribution of some of the properties of the $10<z<12$ galaxy candidates in our work. The right-most panel shows the $P(z)$ of all $10<z<12$ candidates in transparent orange, while the solid orange shows the median stack of all.
 }
  \label{fig:zgt10-properties}
\end{figure*}

In Fig.~\ref{fig:zgt10-properties} we show the distribution of some of the properties of the $10<z<12$ galaxy candidates in our work, and discussed in more detail in \S\ref{sec:the_10z12_smf} and \S\ref{sec:zgt10_candidates}.

\section{Comparison with \textsc{CIGALE}} \label{appdx:cigale-comparison}

\begin{figure}[t]
  \centering
    \includegraphics[width=0.99\columnwidth]{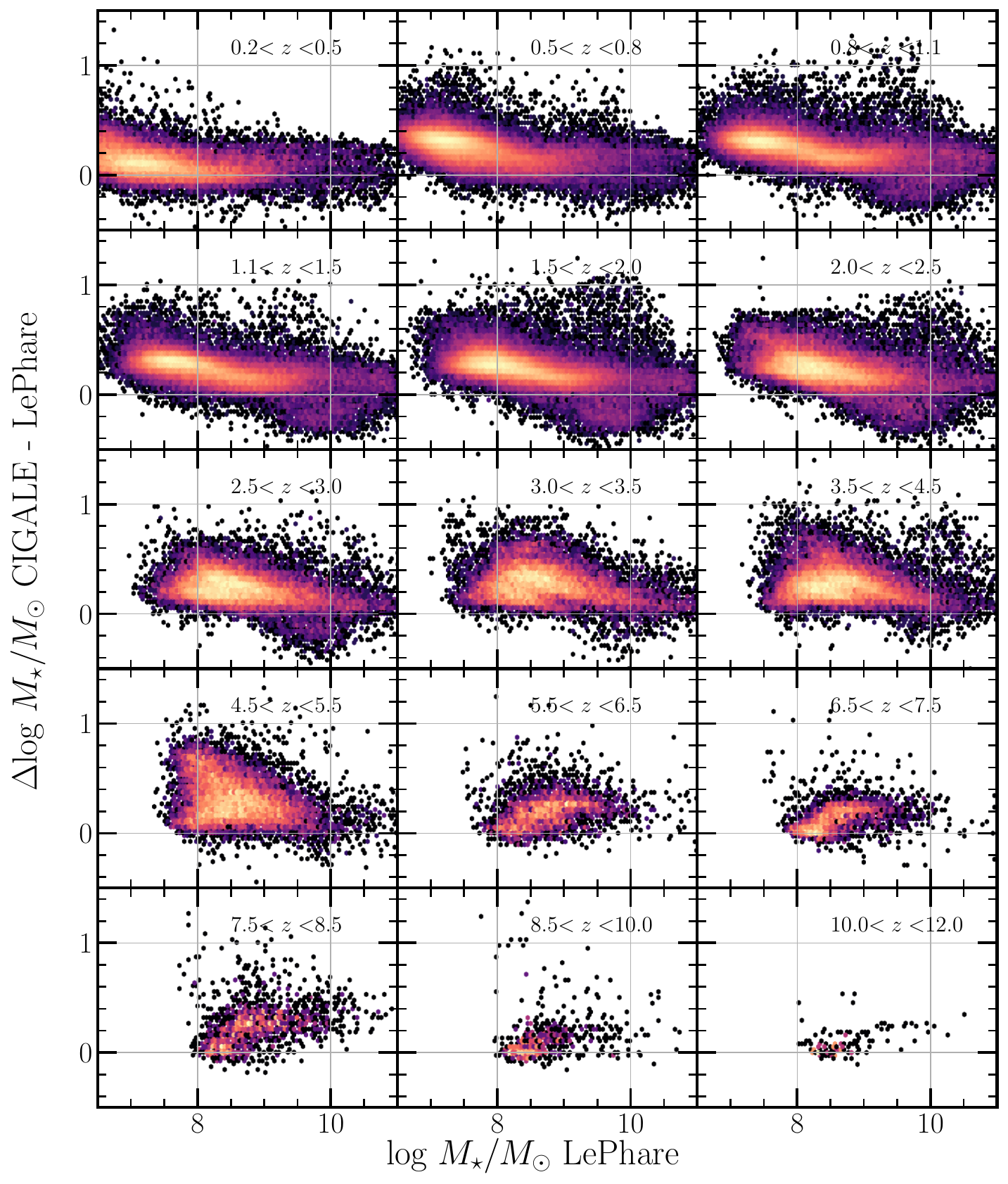}
 \caption{Comparison of the stellar masses from the different SED modelling by \LePHARE{} and \Cigale{}. The figure shows the difference in the log($M_{\star}/M_{\odot}$) between \Cigale{} and \LePHARE{} as a function of the \LePHARE{} mass. Each panel corresponds to the 15 $z$-bins in which we measure the SMF. This shows a constant offset of $0.1-0.3$ dex towards larger \Cigale{} masses that decreases with mass out to $z\sim5.5$ but increases with mass at $z>5.5$.
 }
  \label{fig:cigale-compar-masses}
\end{figure}

\begin{figure}[t]
  \centering
    \includegraphics[width=0.99\columnwidth]{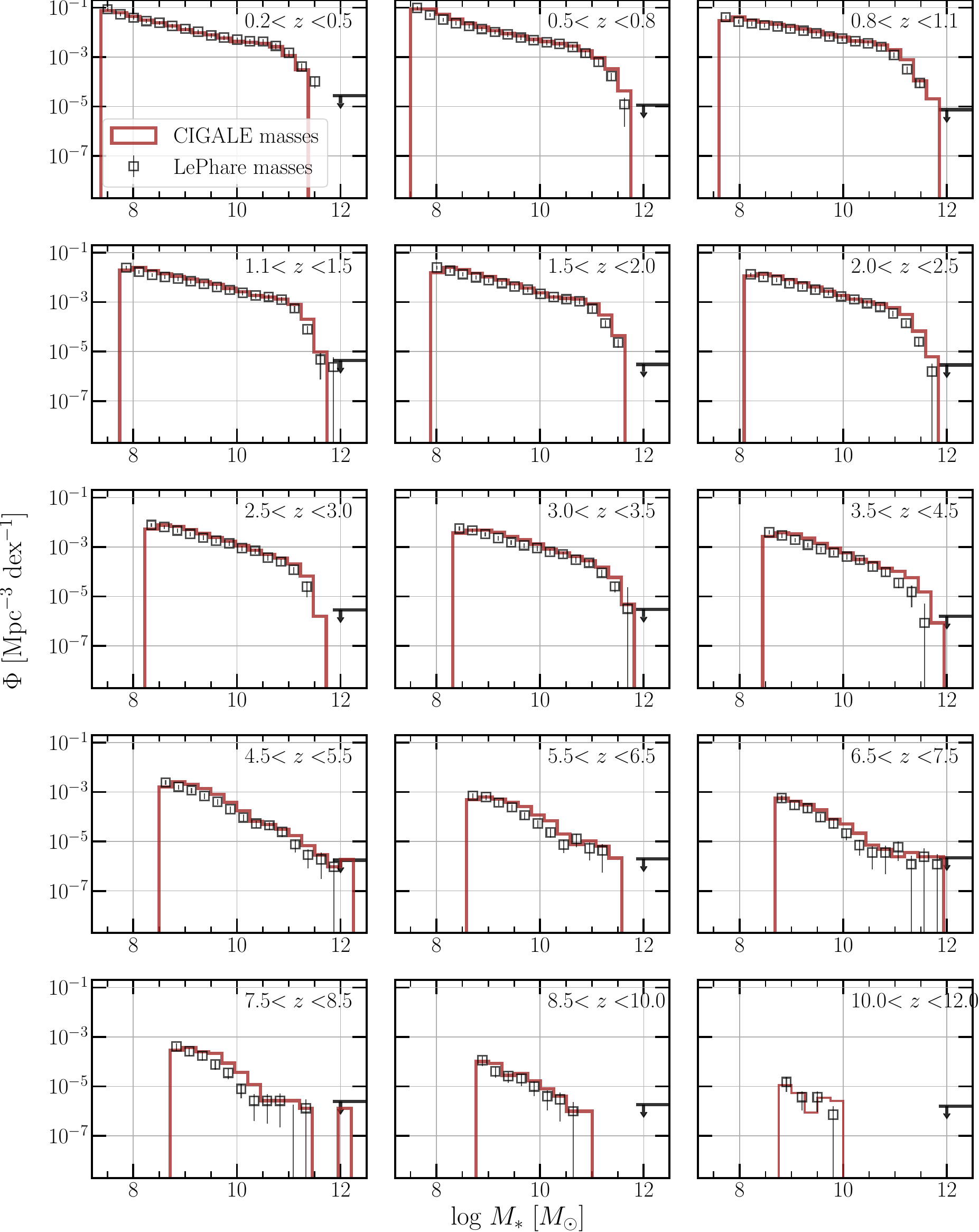}
 \caption{ The resulting SMF using \Cigale{} $M_{\star}$ (brown histogram), compared to the nominal SMF of this paper measured from \LePHARE{} masses (square with errorbars).
 }
  \label{fig:cigale-compar-smfs}
\end{figure}

To assess the robustness of our results, we ran \Cigale{}\footnote{\url{https://cigale.lam.fr/}} \citep{Boquien19} to estimate the physical parameters independently. \Cigale{} is a Bayesian-like analysis code that models the entire SED of galaxies from the X-rays to radio by adopting the energy budget balance between the UV-optical emission from young stars and absorption and re-emission by dust in IR. The SED modelling with \Cigale{} and the resulting performance is described in detail in Arango-Toro et al. (2024). Briefly, \Cigale{} is run to fit only for physical parameters by fixing the redshift on the $z_{\rm phot}$ solution from \LePHARE{}. It implements the \texttt{sfhNlevels} non-parametric SFH module with a bursty continuity prior (presented and tested in \citealt{Ciesla23}, see also \citealt{Arango23}), the stellar population models (SSPs) of \citet{BruzualCharlot03} and the \cite{Dale14} dust emission library. The dust attenuation is a modified \citet{Calzetti00} law, which implements a modified starburst law with the continuum attenuated with the \cite{Calzetti00} curve and the lines extincted with a Milky Way or a Magellanic Cloud attenuation curve; the allowed $E(B-V)$ range is  $0-1.8$. We note that the attenuation modeling is different than our nominal \LePHARE{} run that uses three different attenuation curves but with a more limited $E(B-V)$ range of $0-1.2$.

This SED modelling, most notably the dust attenuation and the non-parametric SFH, allows us to estimate how our results on the measurements of $M_{\star}$ and the SMF would be affected by SED modelling. However, we note that the fact that $\zphot$ is fixed at the solution from \LePHARE{} limits the full and independent exploration of SED solutions by \Cigale{}. 

Figure~\ref{fig:cigale-compar-masses} compares $M_{\star}$ by the two codes in the 15 redshift bins. This shows a constant bias towards higher $M_{\star}$ inferred by \Cigale{} that is dependent on both redshift and mass. The difference is about 0.1 at $z<0.5$ and $\sim 0.2-0.3$ at higher redshifts. With respect to the mass, the difference is higher at the low-mass end and decreases with mass out to $z\sim 5.5$. At $z>5.5$ this trend reverses and \Cigale{} results in higher masses with increasing \LePHARE{} mass by about $0.1-0.3$ dex.

Figure~\ref{fig:cigale-compar-smfs} shows how these differences propagate into the SMF and capture the variance and uncertainty arising from different SED modelling assumptions. Interestingly, however, our findings of increased abundances of massive galaxies at $z\gtrsim5$ are corroborated by the SED modelling by \Cigale{}.

\section{Fitting results}
Table \ref{table:fit_total} presents the parameter values of the double/single Schechter and double power law models, derived as the median, 16th and 84th percentiles of the posterior distribution.

\begin{table*}
\centering
\caption{Values of the double/single Schechter and double power law model parameters. At $z>3.5$ we show the parameters for both single Schechter and DPL.}
\begin{threeparttable}
\begin{tabular}{ccccccccccccc}
\hline
\hline
$z$-bin & $N_{\rm gal}$ & ${\rm Log}_{\rm 10}\,{M}^*$ & log$\,\Phi_1$ & $\alpha_1$ & log$\,\Phi_2$ & $\alpha_2$ & BIC & log$(\rho_{\star})$ & \\ 
& & ($M_\odot$) & ($\mathrm{Mpc}^{-3}\,\mathrm{dex}^{-1}$)  &  &  ($\mathrm{Mpc}^{-3}\,\mathrm{dex}^{-1}$) & & & ($M_{\odot} \, \mathrm{Mpc}^{-3}$) \\
\hline

$0.2 < z < 0.5$ & 19092 &  &  &  &  &  &  & \\
Double Schechter & & $10.78^{+0.12}_{-0.15}$ & $-3.19^{+0.19}_{-0.30}$ & $-1.52^{+0.06}_{-0.08}$ & $-2.67^{+0.16}_{-0.21}$ & $-0.52^{+0.62}_{-0.34}$ & $10.73$ & $8.27^{+0.07}_{-0.08}$ \\
 \hline 
$0.5 < z < 0.8$ & 40107 &  &  &  &  &  &  & \\
Double Schechter & & $10.73^{+0.07}_{-0.10}$ & $-3.43^{+0.26}_{-0.33}$ & $-1.62^{+0.09}_{-0.11}$ & $-2.70^{+0.11}_{-0.14}$ & $-0.70^{+0.45}_{-0.22}$ & $7.77$ & $8.15^{+0.06}_{-0.07}$ \\
 \hline 
$0.8 < z < 1.1$ & 44742 &  &  &  &  &  &  & \\
Double Schechter & & $10.73^{+0.09}_{-0.13}$ & $-2.99^{+0.16}_{-0.22}$ & $-1.40^{+0.05}_{-0.07}$ & $-2.76^{+0.14}_{-0.16}$ & $-0.60^{+0.62}_{-0.31}$ & $8.54$ & $8.21^{+0.06}_{-0.06}$ \\
 \hline 
$1.1 < z < 1.5$ & 40801 & &  &  &  &  &  & \\
Double Schechter & & $10.58^{+0.12}_{-0.10}$ & $-3.16^{+0.11}_{-0.18}$ & $-1.42^{+0.04}_{-0.06}$ & $-2.93^{+0.11}_{-0.10}$ & $0.20^{+0.55}_{-0.65}$ & $9.22$ & $7.96^{+0.06}_{-0.07}$ \\
 \hline 
$1.5 < z < 2.0$ & 55635 &  &  &  &  &  &  & \\
Double Schechter & & $10.66^{+0.13}_{-0.13}$ & $-3.37^{+0.14}_{-0.21}$ & $-1.53^{+0.05}_{-0.07}$ & $-3.10^{+0.13}_{-0.15}$ & $-0.17^{+0.61}_{-0.57}$ & $8.59$ & $7.85^{+0.06}_{-0.07}$ \\
 \hline 
$2.0 < z < 2.5$ & 32602 &  &  &  &  &  &  & \\
Double Schechter & & $10.73^{+0.09}_{-0.11}$ & $-3.58^{+0.14}_{-0.18}$ & $-1.54^{+0.05}_{-0.07}$ & $-3.34^{+0.13}_{-0.13}$ & $-0.44^{+0.53}_{-0.39}$ & $10.09$ & $7.68^{+0.06}_{-0.06}$ \\
 \hline 
$2.5 < z < 3.0$ & 18886 &  &  &  &  &  &  & \\
Double Schechter & & $10.67^{+0.15}_{-0.16}$ & $-3.80^{+0.19}_{-0.26}$ & $-1.58^{+0.06}_{-0.09}$ & $-3.56^{+0.17}_{-0.19}$ & $-0.45^{+0.65}_{-0.44}$ & $8.61$ & $7.43^{+0.07}_{-0.09}$ \\
 \hline 
$3.0 < z < 3.5$ & 13059 &  &  &  &  &  &  & \\
Double Schechter & & $10.75^{+0.17}_{-0.21}$ & $-4.02^{+0.22}_{-0.8}$ & $-1.59^{+0.08}_{-0.14}$ & $-3.62^{+0.31}_{-0.22}$ & $-0.51^{+0.70}_{-0.45}$ & $4.75$ & $7.39^{+0.09}_{-0.12}$ \\
 \hline 
$3.5 < z < 4.5$ & 14281 &  &  &  &  &  &  & \\
Single Schechter & & $10.99^{+0.18}_{-0.18}$ & $-4.31^{+0.19}_{-0.23}$ & $-1.63^{+0.07}_{-0.08}$ & -- & -- & $5.45$ & $7.03^{+0.07}_{-0.09}$ \\
Double power law & & $11.00^{+0.14}_{-0.16}$ & $-4.06^{+0.20}_{-0.19}$ & $-1.68^{+0.06}_{-0.06}$ & -- & $-5.25^{+2.48}_{-2.48}$ & $7.94$ &  \\
 \hline 
$4.5 < z < 5.5$ & 6700 &  &  &  &  &  &  & \\
Single Schechter & & $11.47^{+0.50}_{-0.35}$ & $-5.59^{+0.44}_{-0.63}$ & $-1.93^{+0.07}_{-0.06}$ & -- & -- & $2.04$ & $6.62^{+0.14}_{-0.11}$ \\
Double power law & & $11.03^{+0.49}_{-1.51}$ & $-4.77^{+1.69}_{-0.61}$ & $-1.89^{+0.29}_{-0.09}$ & -- & $-2.91^{+3.41}_{-3.41}$ & $3.32$ &  \\
 \hline 
$5.5 < z < 6.5$ & 1974 &  &  &  &  &  &  & \\
Single Schechter & & $10.93^{+0.87}_{-0.50}$ & $-5.64^{+0.59}_{-0.92}$ & $-2.00$ & -- & -- & $-2.04$ & $6.13^{+0.17}_{-0.15}$ \\
Double power law & & $10.53^{+0.97}_{-0.50}$ & $-4.83^{+0.61}_{-1.04}$ & $-2.00$ & -- & $-3.35^{+3.58}_{-3.58}$ & $-14.24$ &  \\
 \hline 
$6.5 < z < 7.5$ & 1022 &  &  &  &  &  &  & \\
Single Schechter & & $10.64^{+0.94}_{-0.39}$ & $-5.53^{+0.47}_{-1.02}$ & $-2.00$ & -- & -- & $0.59$ & $5.87^{+0.18}_{-0.15}$ \\
Double power law & & $10.34^{+0.97}_{-0.41}$ & $-4.85^{+0.52}_{-1.06}$ & $-2.00$ & -- & $-3.68^{+3.20}_{-3.20}$ & $-8.71$ &  \\
 \hline 
$7.5 < z < 8.5$ & 704 &  &  &  &  &  &  & \\
Single Schechter & & $10.01^{+0.27}_{-0.22}$ & $-4.92^{+0.29}_{-0.38}$ & $-2.00$ & -- & -- & $0.75$ & $5.67^{+0.12}_{-0.14}$ \\
Double power law & & $9.86^{+0.21}_{-0.23}$ & $-4.46^{+0.31}_{-0.30}$ & $-2.00$ & -- & $-4.88^{+2.79}_{-2.79}$ & $-5.57$ &  \\
 \hline 
$8.5 < z < 10.0$ & 209 &  &  &  &  &  &  & \\
Single Schechter & & $11.37^{+0.71}_{-0.81}$ & $-6.97^{+0.84}_{-0.74}$ & $-2.00$ & -- & -- & $4.57$ & $5.23^{+0.18}_{-0.19}$ \\
Double power law & & $10.90^{+0.75}_{-0.71}$ & $-6.13^{+0.78}_{-0.76}$ & $-2.00$ & -- & $-5.19^{+2.66}_{-2.66}$ & $-0.59$ &  \\
 \hline 
$10.0 < z < 12.0$ & 27 &  &  &  &  &  &  & \\
Single Schechter & & $10.79^{+1.12}_{-1.14}$ & $-7.62^{+1.39}_{-1.17}$ & $-2.00$ & -- & -- & $1.33$ & $4.10^{+0.35}_{-0.83}$  \\
Double power law & & $10.60^{+0.94}_{-0.99}$ & $-7.10^{+1.15}_{-0.99}$ & $-2.00$ & -- & $-5.52^{+2.53}_{-2.53}$ & $1.05$ &  \\
 \hline 

\hline
\end{tabular}
\end{threeparttable}
\label{table:fit_total}
\end{table*}

\end{document}